\documentclass[deluxetables]{aastex631}
\usepackage{amsmath}
\usepackage{bm}
\usepackage[caption=false]{subfig}
\usepackage{multirow}
\usepackage{graphicx}
\usepackage{subfig}
\usepackage{comment}

\newcommand{\squishlist}{
 \begin{list}{$\bullet$}
  { \setlength{\itemsep}{0pt}
     \setlength{\parsep}{3pt}
     \setlength{\topsep}{3pt}
     \setlength{\partopsep}{0pt}
     \setlength{\leftmargin}{1.5em}
     \setlength{\labelwidth}{1em}
     \setlength{\labelsep}{0.5em} } }

\newcommand{\squishlisttwo}{
 \begin{list}{$\bullet$}
  { \setlength{\itemsep}{0pt}
     \setlength{\parsep}{0pt}
    \setlength{\topsep}{0pt}
    \setlength{\partopsep}{0pt}
    \setlength{\leftmargin}{2em}
    \setlength{\labelwidth}{1.5em}
    \setlength{\labelsep}{0.5em} } }

\newcommand{\squishend}{
  \end{list}  }

\def\etal{{et al.\thinspace}}

\def\eg{{\em e.g.,\ }}

\def\ie{{\em i.e.,\ }}
\def\spose#1{\hbox to 0pt{#1\hss}}

\def\multleft#1{\hbox to size{\vbox {\halign {\lft{##}\cr #1}}\hfill}\par}
\def\multright#1{\hbox to size{\vbox {\halign {\rt{##}\cr #1}}\hfill}\par}

\def\degmark{^\circ}
\def\boxit#1{\vbox{\hrule\hbox{\vrule\kern3pt\vbox{\kern3pt
          #1 \kern3pt}\kern3pt\vrule}\hrule}}

\def\cm{{\rm\thinspace cm}}

\def\erg{{\rm\thinspace erg}}
\def\eV{{\rm\thinspace eV}}

\def\keV{{\rm\thinspace keV}}
\def\km{{\rm\thinspace km}}

\def\ph{{\rm\thinspace ph}}
\def\s{{\rm\thinspace s}}
\def\ks{{\rm\thinspace ks}}

\def\cmsq{\hbox{$\cm^2\,$}}

\def\pcmcu{\hbox{$\cm^{-3}\,$}}

\def\ergcmps{\hbox{$\erg\cm\s^{-1}\,$}}

\def\ergpcmsqps{\hbox{$\erg\cm^{-2}\s^{-1}\,$}}

\def\ergps{\hbox{$\erg\s^{-1}\,$}}

\def\kmps{\hbox{$\km\s^{-1}\,$}}

\def\pcmsq{\hbox{$\cm^{-2}\,$}}
\def\pcmcu{\hbox{$\cm^{-3}\,$}}

\def\phpcmsqps{\hbox{$\ph\cm^{-2}\s^{-1}\,$}}

\graphicspath{{./}{figures/}}

\shorttitle{The X-ray Spectrum of MCG--6-30-15 with
  XRISM}
\shortauthors{L.~W.~Brenneman}

\DeclareUnicodeCharacter{2212}{-}
\begin{document}

\title{A Sharper View of the X-ray Spectrum of MCG--6-30-15 with
  XRISM, XMM-Newton and NuSTAR}

\author[0000-0003-2663-1954]{Laura~W.~Brenneman}
\affiliation{Center for Astrophysics $|$ Harvard~\&~Smithsonian, 60 Garden St., Cambridge, MA 02138, USA}

\author[0000-0002-4794-5998]{Daniel~R.~Wilkins}
\affiliation{The Ohio State University, Department of Astronomy, 4006 McPherson Laboratory, 140 W 18th Ave, Columbus, OH 43210, USA}

\author[0000-0003-4504-2557]{Anna~Ogorza{\l}ek}
\affiliation{Department of Astronomy, University of Maryland, College Park, MD 20742, USA} 
\affiliation{NASA / Goddard Space Flight Center, Greenbelt, MD 20771, USA}
\affiliation{Center for Research and Exploration in Space Science and Technology, NASA / GSFC (CRESST II), Greenbelt, MD 20771, USA}

\author[0000-0002-0786-7307]{Daniele~Rogantini}
\affiliation{Department of Astronomy and Astrophysics, University of Chicago, 5640 S Ellis Ave, Chicago, IL 60637, USA}

\author[0000-0002-9378-4072]{Andrew~C.~Fabian}
\affiliation{Institute of Astronomy, Cambridge University, Madingley Road, Cambridge CB3 0HA, UK}

\author[0000-0003-3828-2448]{Javier~A.~Garc\'ia}
\affiliation{NASA / Goddard Space Flight Center, Greenbelt, MD 20771, USA}
\affiliation{California Institute of Technology, Pasadena, CA 91125, USA}

\author[0000-0002-7292-6852]{Anna~Jur\'{a}\v{n}ov\'{a}}
\affiliation{Kavli Institute for Astrophysics and Space Research, Massachusetts Institute of Technology, MA 02139, USA}

\author[0000-0003-2161-0361]{Misaki~Mizumoto}
\affiliation{Science Research Education Unit, University of Teacher Education Fukuoka, Fukuoka 811-4192, Japan} 

\author[0000-0001-6020-517X]{Hirofumi~Noda}
\affiliation{Astronomical Institute, Tohoku University, Miyagi 980-8578, Japan} 

\author[0000-0001-9735-4873]{Ehud~Behar}
\affiliation{Department of Physics, Technion, Technion City, Haifa 3200003, Israel} 

\author[0000-0003-2704-599X]{Rozenn~Boissay-Malaquin}
\affiliation{Center for Space Sciences and Technology, University of Maryland, Baltimore County (UMBC), Baltimore, MD, 21250 USA}
\affiliation{NASA / Goddard Space Flight Center, Greenbelt, MD 20771, USA}
\affiliation{Center for Research and Exploration in Space Science and Technology, NASA / GSFC (CRESST II), Greenbelt, MD 20771, USA}

\author[0000-0002-1094-3147]{Matteo~Guainazzi}
\affiliation{European Space Agency (ESA), European Space Research and Technology Centre (ESTEC), 2200 AG Noordwijk, The Netherlands} 

\author[0000-0002-6054-3432]{Takashi~Okajima}
\affiliation{NASA / Goddard Space Flight Center, Greenbelt, MD 20771, USA}

\author[0000-0003-3496-8928]{Erika~Hoffman}
\affiliation{Department of Astronomy, University of Maryland, College Park, MD 20742, USA}

\author[0000-0003-2535-6436]{Noa~Keshet}
\affiliation{Department of Physics, Technion, Technion City, Haifa 3200003, Israel}

\author[0000-0001-5540-2822]{Jelle~Kaastra}
\affiliation{SRON Netherlands Institute for Space Research, Leiden, The Netherlands} 
\affiliation{Leiden Observatory, University of Leiden, P.O. Box 9513, NL-2300 RA, Leiden, The Netherlands} 

\author[0000-0003-0172-0854]{Erin~Kara}
\affiliation{Kavli Institute for Astrophysics and Space Research, Massachusetts Institute of Technology, MA 02139, USA} 

\author[0000-0003-1100-1423]{Makoto~Yamauchi}
\affiliation{Faculty of Engineering, University of Miyazaki, 1-1 Gakuen-Kibanadai-Nishi, Miyazaki, Miyazaki 889-2192, Japan}

\begin{abstract}
We present a time-averaged spectral analysis of the 2024 {\it XRISM} observation of the narrow-line Seyfert-1 galaxy MCG--6-30-15, taken contemporaneously with {\it XMM-Newton} and {\it NuSTAR}.  Our analysis leverages a unique combination of broadband and high-resolution X-ray spectroscopy to definitively isolate and characterize both broad and narrow emission and absorption features in this source.  The best-fitting model for the joint spectral analysis is very well described by reflection from the inner accretion disk illuminated by a compact corona, modified by multi-zone ionized absorption from an outflowing wind along the line of sight.  The {\it XRISM}/Resolve data confirm that a strong, relativistically-broadened Fe K$\alpha$ emission line is required in order to obtain an adequate model fit.  The Resolve data additionally verify the presence of a $v_{\rm out} \sim 2300 \kmps$ component of this outflowing wind, find tentative evidence for a $v_{\rm out} \sim 20,000 \kmps$ wind component, and indicate that the reflection from distant, neutral material may originate in a non-uniform structure rather than the traditional torus of AGN unification schemes.  Though a rapid prograde black hole spin is statistically preferred by the best-fitting model, consistent with previous results, the AGN flux variability over the course of the observation complicates the interpretation of the time-averaged spectra.  This insight, clarified by the combination of high signal-to-noise and high spectral resolution in the joint dataset, emphasizes the importance of time-resolved, high-resolution spectral analysis in unambiguously measuring the physical properties of variable AGN.  
\end{abstract}


\section{Introduction}
\label{sec:intro}

Over the past three decades, considerable literature has been published on the detection of ``relativistic reflection" features emitted from the innermost accretion disk in active galactic nuclei (AGN).  This term describes the reprocessed spectrum created when a fraction of the thermal disk photons that inverse-Compton scatter off of electrons in a compact corona \citep[\ie $\leq10\,r_{\rm g}$;][]{Chen2011} irradiate the optically-thick portion of the accretion disk down to radii at or near the innermost stable circular orbit (ISCO).  Because relativistic effects from the black hole are significant in the innermost disk, these ``reflected" features appear highly distorted \citep[\eg][]{Fabian1989,RN2003}.  The most prominent characteristics of this relativistic reflection are in the X-ray band: the broad, asymmetric profile produced by the Fe K$\alpha$ line at $6.4-6.9 \keV$, the corresponding Fe K edge in absorption at ${\sim} 7-10\,\keV$ and the ``Compton hump" peaking at $20-30 \keV$.  These features are of interest primarily because their morphology encodes information about the supermassive black hole (SMBH) and inner accretion flow that can be extremely difficult to obtain through other observational means: the extent of the red wing of the Fe K$\alpha$ line is set by the SMBH spin, while the sharp drop of the blue wing of the line is defined by the inclination angle of the innermost disk to the line of sight \citep[\eg][ and references therein]{BR2006}.
Understanding how accretion fuels the co-evolution of SMBHs and their host galaxies ---catalyzing both SMBH growth and wind-driven outflows through the transfer of mass and angular momentum--- remains one of the most pressing questions in astrophysics. 

From a theoretical standpoint, relativistic reflection is an expected byproduct of radiatively-efficient accretion onto a SMBH.  \citet{Thorne1974} showed that for a Novikov-Thorne accretion disk, $50\%$ of the flux going to infinity originates from ${<}5\,r_{\rm g}$ for near-maximal prograde BH spins and from ${<}30\,r_{\rm g}$ for non-spinning BHs.  We should therefore expect that relativistic effects become increasingly important for spectra from more rapidly spinning BHs.  Yet isolating and modeling these features is often challenging: doing so requires high signal-to-noise (S/N) X-ray spectra that demand long observation times (${>}100 \ks$ with present day instrumentation, even for the brightest sources), as well as data sensitive enough to disentangle the broad reflection signatures from spectral curvature due to complex absorption along the line of sight from wind-driven outflows, which are present in a majority of AGN \citep{Reynolds1997,Brenneman2013}.  

In spite of these challenges, signatures of relativistic reflection have been reported to be robustly detected in several dozen AGN, resulting in a growing population of SMBH spin measurements \citep{Brenneman2013,Reynolds2021,Bambi2021}.  The first broad Fe K$\alpha$ line observed in an AGN was found by {\it ASCA} in the narrow-line Seyfert 1.2 galaxy MCG--6-30-15 \citep{Tanaka1995,Iwasawa1996,Dabrowski1997}.  Since then, this source (MCG-6 hereafter) has been one of the most extensively studied AGN, having been the subject of long observations with {\it BeppoSAX} \citep{Guainazzi1999}, {\it RXTE} \citep{Lee1999,Lee2000}, {\it Chandra} \citep{Lee2001,Lee2002,Young2005}, {\it XMM-Newton} \citep{Wilms2001,Fabian2002,Reynolds2004}, {\it Suzaku} \citep{Miniutti2007,Chiang2011} and {\it NuSTAR} \citep{Marinucci2014}.  Results from all of these datasets report a broad Fe K$\alpha$ emission feature that is consistent with a highly redshifted line from the inner parts of an accretion disk, in spite of the clear presence of complex, multi-zone absorption from outflowing winds \citep[\eg][]{Lee2002,Young2005}.  \citet{BR2006} reported one of the first spin measurements of this SMBH using {\it XMM} data: $a=0.989^{+0.009}_{-0.002}$, where $a \equiv cJ/GM^2$ in dimensionless units.  

Because of the prominence of its broad iron line and rich archive of X-ray observations, MCG-6 has become the archetype of relativistic reflection studies in AGN.  It has also become a lightning rod for controversy, however: although the X-ray collecting area of most of the instruments listed above have provided the necessary S/N to detect broad Fe K lines, their limited spectral resolution has hampered the effort to break modeling degeneracies between inner disk reflection and complex absorption.  For example, an alternative absorption-dominated model has also been proposed to explain the spectral shape and flux variability of MCG−6 in the observations above \citep{Miller2008,Miller2009,Miyakawa2012}.  In this model, the red wing of the line is not due to strong relativistic effects but to a complex structure of absorbers along the line of sight that partially cover the primary X-ray source (\ie the AGN corona).  These complex absorbing structures are proposed to be Compton-thick clouds at or within the broad line region (BLR), partially covering the X-ray source, that can produce absorbed continua mimicking the shape of the Fe K$\alpha$ emission line and Compton hump \citep{Tatum2013}.

Progress has been made in distinguishing between the inner disk reflection and absorption-dominated models by expanding the X-ray bandpass of AGN observations.  The launch of {\it NuSTAR} in 2012 has enabled broadband X-ray spectral fitting out to ${\sim} 80 \keV$ in many AGN, helping to differentiate between inner disk reflection and complex absorption, particularly when utilized in coordinated campaigns with {\it XMM} \citep[\eg NGC~1365;][]{Walton2014}.  The increased bandpass allows the broad Fe K$\alpha$ line and Compton hump to be holistically modeled for the first time, and, in many cases, for the direct (corona) and reflected continuum components to be more cleanly differentiated as well \citep[\eg][]{Fabian2015, Wilkins2015}.  This approach was employed on a joint {\it XMM}+{\it NuSTAR} observing campaign on MCG-6 by \citet{Marinucci2014}, who found that the inner disk reflection model was favored over the absorption-dominated model, but noted that the latter could not be statistically ruled out.   

The launch of {\it XRISM} in 2023 has finally brought a high-resolution X-ray micro-calorimeter spectrometer to bear in the effort to understand the innermost regions of AGN, including the true properties of complex absorbers, the geometry of the inner accretion flow, and the role of relativistic reflection.  The {\it XRISM}/Resolve instrument combines the effective area of typical X-ray CCDs with ${<}5$-eV spectral resolution, ${\sim}7\times$ better than {\it Chandra}/HETG in the Fe K band with an effective area ${\sim}100\times$ larger.  This sensitivity allows intrinsically narrow emission lines and the discrete lines of absorption from highly ionized outflowing winds to be definitively resolved for the first time, rather than blended together.  With its unique ability to accurately characterize the narrow signatures of both absorption and emission in the Fe K band, thereby disentangling them from the broader features of inner disk reflection, {\it XRISM}/Resolve is now an indispensable tool in the effort to properly characterize the spectra of MCG-6 and other AGN reported to have relativistic reflection features and measured SMBH spins. 

With this goal in mind, we have obtained a long, contemporaneous observation of MCG-6 with {\it XRISM}, {\it XMM} and {\it NuSTAR} in 2024, during the {\it XRISM} performance verification (PV) phase.  In this first paper from the MCG-6
campaign, we utilize the highest S/N broadband, time-averaged data
from each telescope to establish the basic X-ray continuum, reflection
and absorption properties of the AGN.  These datasets include {\it
  XRISM}/Xtend  \citep{Noda2025}, {\it XMM-Newton}/EPIC-pn
\citep{Struder2001} and {\it NuSTAR}/FPMA and
FPMB \citep{Harrison2013}.  We then use this model
framework to scaffold the first-ever spectral decomposition of the Fe K
energy band in MCG-6 at ${<}5$-eV resolution with {\it XRISM}/Resolve \citep{Ishisaki2022}.  (We note that data from the {\it
  XMM}/RGS and OM instruments will be presented in subsequent papers,
as well as detailed, time-resolved spectral analyses of data from all instruments.)  Finally, we jointly fit all of the datasets to holistically model both the reflection and absorption in MCG-6, obtaining the most accurate characterization of its X-ray spectrum ever achieved.  The Resolve data allow us to correctly model the narrow spectral features in emission and absorption, while the Xtend, {\it XMM} and {\it NuSTAR} data provide the S/N and bandpass necessary to correctly model the continuum and the broader, more subtle features of inner disk reflection.  

The remainder of the paper is structured as follows: observations and
data reduction for each instrument are outlined in \S\ref{sec:data};
our timing and spectral analyses and results are detailed in \S\ref{sec:analysis};
implications of our findings are discussed in \S\ref{sec:discussion},
and our summary and conclusions are presented in \S\ref{sec:conclusions}.

\section{Observations and Data Reduction}
\label{sec:data}

MCG--6-30-15 was observed by {\it XRISM, XMM-Newton}
and {\it NuSTAR} in a coordinated campaign during 2024 February 5-8.  The data from each
observatory were processed separately as described in the following
subsections.  Before filtering out Earth occultations, incidents of
high background and other events, the {\it XRISM} observation totaled ${\sim}213 \ks$ of
elapsed time.
The {\it XMM} and {\it NuSTAR} observations totaled ${\sim}135 \ks$ and
${\sim}269 \ks$ of elapsed time, respectively.  Here we utilize only the data from
{\it XMM} and {\it NuSTAR} that falls within the {\it XRISM} observation, and within
the energy band where the source is brighter than the background.
Count rates, total counts, net exposure times and
S/N ratios for the data from each instrument used in our analysis are
listed in Table~\ref{tab:obs}.  

\begin{table}
\begin{center}
\begin{tabular}{|cccccc|} 
\hline\hline
{\bf Instrument} & {\bf Energy Band (keV)} & {\bf Net Exposure Time (ks)} &
{\bf Count Rate (cts/s)} & {\bf Total
  Counts} & {\bf S/N} \\
\hline
{\it XRISM}/Resolve & $2.0-11.0$ & $142$ & $0.704 \pm 0.002$ & $99,733$ & $28$\\
{\it XRISM}/Xtend & $0.3-12.0$ & $112$ & $6.562 \pm 0.008$ & $765,831$ & $88$\\
{\it XMM}/pn & $0.5-10.0$ & $44.7$ & $27.24 \pm 0.03$ & $1,233,210$ & $699$\\
{\it NuSTAR}/FPMA  & $3.0-55.0$ & $101$ & $1.193 \pm 0.003$ & $160,529$ & $39$\\
{\it NuSTAR}/FPMB  & $3.0-55.0$ & $99.6$ & $1.101 \pm 0.003$ & $146,578$ & $41$\\
\hline\hline
\end{tabular}
\end{center}
\caption{\small{Observation details for the 2024 {\it XRISM, XMM} and {\it NuSTAR}
    campaign on MCG--6-30-15.  Energy ranges are those for which the
    source is dominant over the background for each instrument.  Net exposure times account for filtering
    and matching to the {\it XRISM}/Resolve GTIs.  Count rates and total counts are
    taken from the background-subtracted data in the given energy
    band.  S/N is
    calculated as an average source-to-background ratio for the
    unbinned spectra over the given energy bands for all instruments using the source-dominated
    case described in \citet{Longair2011}.}}
\label{tab:obs}
\end{table}

\subsection{XRISM}
\label{sec:xrism_obs}

The {\it XRISM} observation (OBSID 000161000) began on 2024 February 5 UTC 23:13 and ended on 2024 February 8 UTC 10:45.
MCG--6-30-15 was observed without a filter for the
Resolve micro-calorimeter and using $1/8$-window mode for the Xtend
CCD instrument to limit photon pile-up.  We reduced these data
using the first public release of the {\it XRISM} software within
HEASoft 6.34 in conjunction with version 11 of the {\it
  XRISM} CALDB.  Routines used to conduct the data reduction and
products extraction followed
the steps presented in 
the {\it XRISM} ABC Guide\footnote{https://heasarc.gsfc.nasa.gov/docs/xrism/analysis/abc\_guide/Contents.html}.

The Resolve instrument data reduction included additional
screening to 
ensure pulse shape validity and to avoid pulse rise-time and
frame-time coincidences.  Time filtering was also performed to avoid
periods corresponding to South Atlantic Anomaly (SAA) passage, Earth’s sunlit limb,
and the instrument's $4300$-s adiabatic demagnetization refrigerator (ADR)
recycling intervals.  Once these steps were performed, we extracted
light curves and spectra over the whole $6\times~6$-pixel array using
{\sc xselect} (data from pixels 12 and 27 were not used: pixel 12 is the calibration pixel, and pixel 27 shows anomalous gain variations). 
For our observation, $95\%$ of the Resolve events are classified as
high-resoultion (\ie Hp), and the $^{55}\,{\rm Fe}$ calibration pixel indicates a
FWHM of $4.50 \pm 0.03 \eV$ for the Mn K$\alpha$ emission
line, with a gain uncertainty less than $1 \eV$ across the bandpass used in our analysis.  As the Hp events
provide the finest spectral spectral resolution for Resolve, we
restrict our data analysis to these events.
We generated response files using the {\it XRISM} RMF and ARF generation tools,
``rslmkrmf'' and ``xaarfgen,'' assuming a point-like source for the ARF.  We
utilize the ``XL'' RMF in our spectral analysis, as this
is the full response file that accounts for all energy redistribution processes.  Due to
the size and shape of a point source PSF on the
$6\times6$-pixel array, one cannot simply identify a background region
that is uncontaminated by source photons and extract the spectrum.  While the sky background is assumed to be negligible in the Resolve bandpass ${\geq}2 \keV$, the non-X-ray background (NXB) must be based on an empirical model and drawn from a database using the ``rslnxbgen'' tool.  The
NXB model includes Gaussian emission lines from Al, Cr, Mn,
Fe, Ni, and Cu.  For this observation the NXB is a factor of ${\sim}100\times$
weaker than the source, so the Gaussian line approximation is adequate
and the NXB is effectively negligible on our spectral modeling.  Lastly, we rebinned the spectra using the ``ftgrouppha''
task with the optimal method described in \citet{Kaastra2016} in order
to optimally sample the instrumental response.

The Xtend instrument data reduction via the ABC Guide includes an option to run the
``searchflickpix'' tool in order to determine whether 
flickering pixels in the CCD are present and problematic for a given observation.  Upon
inspection of the Xtend image, we were able to
establish source and background regions of the recommended size that
did not include any anomalous pixels, thereby negating the need to run
``searchflickpix.''  Per guidelines for observations taken in
$1/8$-window mode, the source region is $5 \arcmin$ long and $3.65
\arcmin$ wide (\ie the width of the frame), centered on the point
source.  The background region is the same size, taken from a spot on
the same chip as far from the source as possible while avoiding
anomalous pixels.  We extracted light curves and spectra using {\sc
  xselect}, then created RMF and ARF files using the
response generation tools, ``xtdrmf'' and ``xaarfgen,'' with the
settings as described in the ABC Guide.  As with Resolve, we rebinned the spectra using the ``ftgrouppha'' task.

\subsection{XMM-Newton}
\label{sec:xmm_obs}

{\it XMM-Newton} observed MCG-6 during revolution 4425 from 2024 February 6 UTC 00:56 to
2024 February 7 UTC 12:33 with OBSID 0921420101.  The EPIC/pn observation was
taken in Prime Small Window mode with the medium optical blocking filter.
The data were reprocessed and reduced using the Science
Analysis System (SAS) software version 21.0 and the CALDB release of
April 24, 2024, per instructions in the
{\it XMM-Newton} ABC
Guide\footnote{https://heasarc.gsfc.nasa.gov/docs/xmm/abc/abc.html}.

After reprocessing via ``epproc'' and performing standard filtering
within ``evselect,'' we established source and background
regions within CCD-4.  The source region is the recommended
circle $40 \arcsec$ in radius, centered on the source.  The background region
consists of two boxes along the top edge of the same chip that are as large
as possible while still avoiding the
PSF diffraction spikes from the source.  These boxes are both $41
\arcsec$ in width, and are $131 \arcsec$ and $89 \arcsec$ in length,
respectively.  After determining that pile-up is negligible in these
data using ``epatplot,'' we extracted source and background light
curves and spectra for the whole observation using ``evselect,''
correcting the light curves and subtracting backgrounds via
``epiclccor.''  We then
generated RMF and ARF files using ``rmfgen'' and ``arfgen.''  We 
applied the {\it XRISM}/Resolve GTIs as time filters and extracted
strictly simultaneous EPIC/pn data and response files for our spectral
analysis as well.  As for
the {\it XRISM} data, we rebinned the spectra using the ``ftgrouppha''
task, as above for the {\it XRISM} data. 

\subsection{NuSTAR}
\label{sec:nustar_obs}

{\it NuSTAR} observed MCG-6 from 2024 February 6
UTC 00:31 to 2024 February 9 UTC 03:41 with OBSID 60902004002.  The
data were reprocessed and reduced using the {\it NuSTAR} Data Analysis
System (NUSTARDAS) software version 2.1.2 and CALDB version 20240826
via HEASoft v.6.34, and following the instructions in the {\it NuSTAR}
data analysis software users' guide
v.1.9.7\footnote{https://heasarc.gsfc.nasa.gov/docs/nustar/analysis/nustar\_swguide.pdf}.

After reprocessing the data via the ``nupipeline'' task, we
established source and background regions for the FPMA and FPMB
detectors, each a circle with radius $1.5 \arcmin$.  We used the
``nuproducts'' task to extract light curves and spectra for the whole
observation from each
region, and to apply the {\it XRISM} GTIs as time filters in order to
extract strictly simultaneous spectra.  We note that, due to the
relatively steep spectral slope of MCG-6 ($\Gamma > 2$), the
background flux becomes comparable to the source flux at $E > 55
\keV$, so we limit our spectral analysis to energies below this
value.  As for
the {\it XRISM} and {\it XMM}/pn data, we rebinned the spectra using ``ftgrouppha''
task with the optimal method.  

\section{Data Analysis and Results}
\label{sec:analysis}

In the following section, we present our initial analyses
of the joint {\it XRISM, XMM} and {\it NuSTAR} data from this 2024
observing campaign on MCG--6-30-15.  After providing a brief look at
the timing properties of these data, we move on the focus on
the analysis of the time-averaged spectra, with a particular emphasis
on the Fe K region and {\it XRISM}/Resolve's unique ability to
definitively disentangle features corresponding to distant reflection,
relativistic reflection and absorption.  

For all the analyses presented here, we used {\sc xronos} version 5.22 and
{\sc xspec} version 12.14.1d, along with other {\sc ftools} packages within
HEASoft 6.34.  Spectral analysis employed Cash statistics \citep{Cash1979}, Verner
cross-sections \citep{Verner1995,Verner1996} and Wilms abundances \citep{Wilms2000}, and included Galactic
photoabsorption via the {\tt TBabs} model.  Errors quoted within the text are at the 1$\sigma$ level of
confidence unless otherwise specified, while those in tables are quoted at
$90\%$ confidence.  All plots are shifted to the cosmological rest
frame of MCG-6: $z=0.00775$ (an average of the values listed in NED, the NASA/IPAC extragalactic
database\footnote{https://ned.ipac.caltech.edu}).

\subsection{Timing Behavior}
\label{sec:timing}

\begin{figure}
\centering
\subfloat{%
    \includegraphics[width=0.8\textwidth, trim = 0cm 1.5cm 0cm 0cm, clip]{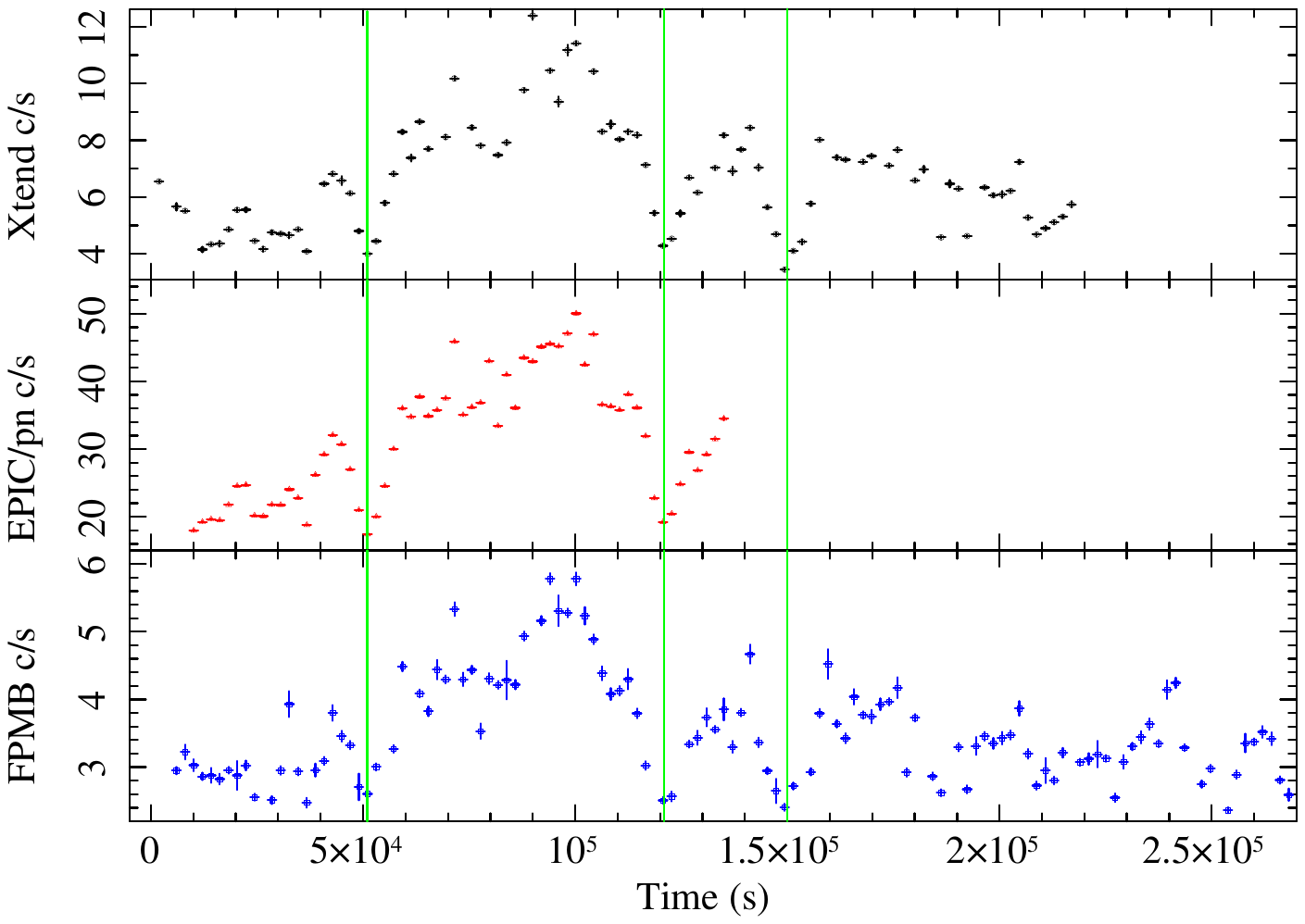}
}
\\
\subfloat{%
    \includegraphics[width=0.8\textwidth, trim = 0cm 1cm 0cm 2.5cm, clip]{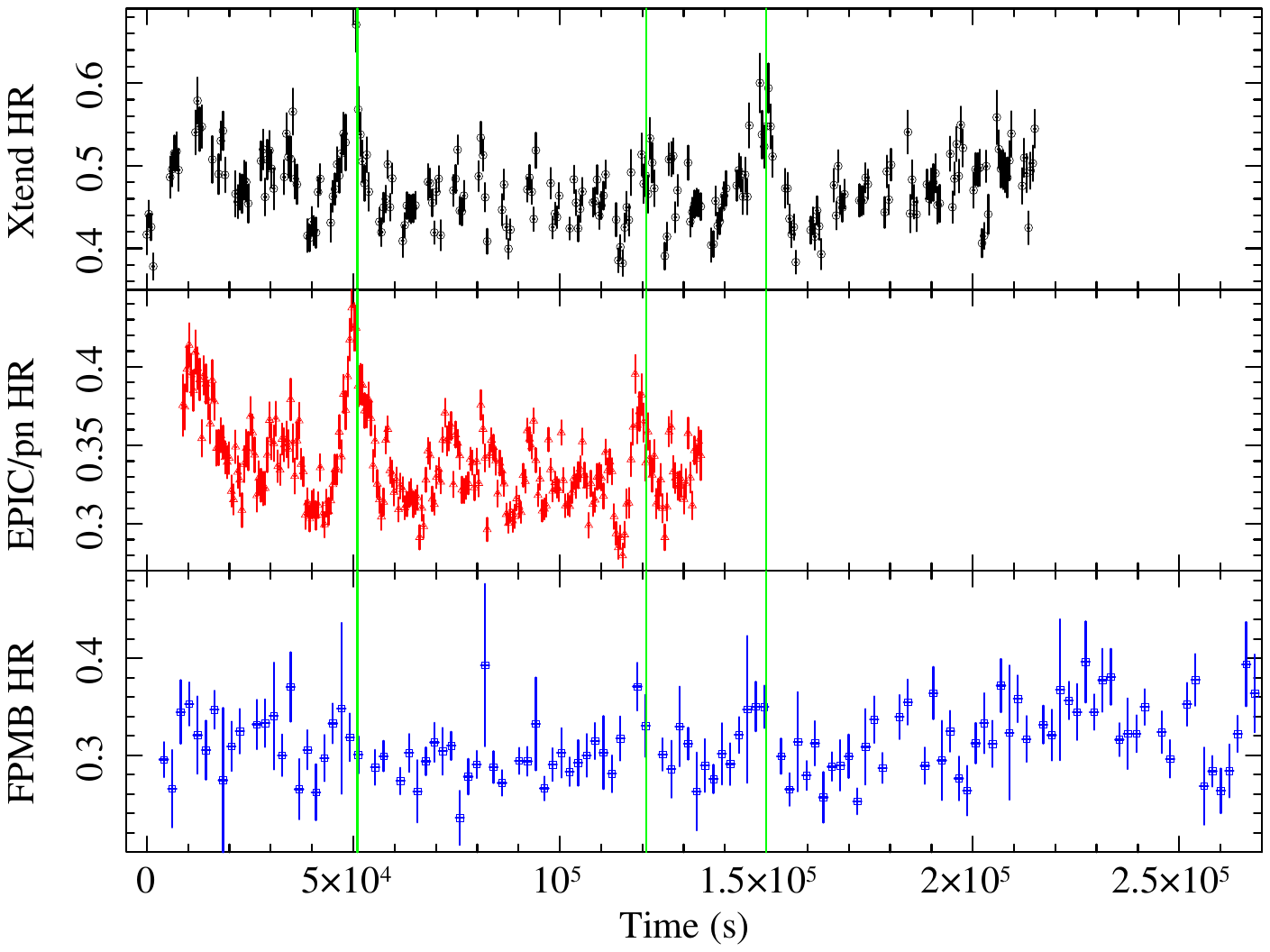}
}
\caption{{\small {\it Top three panels:} Background-subtracted light curves for the 2024 X-ray
campaign on MCG-6.  From top to bottom: {\it XRISM}/Xtend (black,
$0.3-12 \keV$), {\it
  XMM}/EPIC-pn (red, $0.3-12 \keV$) and {\it NuSTAR}/FPMA+FPMB (blue, $3-78
\keV$).  All are shown at a time resolution of $2048 \s$ per bin.  {\it Bottom three panels:} Hardness ratio vs. time for the Xtend (black
    circles; top), pn (red triangles; middle) and {\it NuSTAR}/FPMB
    (blue squares; bottom) instruments.  The HR is calculated
    by dividing the background-subtracted light curves in the hard
    and soft bands ($H = 2-10 \keV$ for Xtend and pn, $10-50 \keV$ for
    FPMB; $S = 0.5-2 \keV$ for Xtend and pn, $3-10 \keV$ for FPMB).  The Xtend and pn HRs have time bins of $512 \s$, while the FPMB is binned to $2048 \s$ for clarity.}  Note the green vertical lines marking dips in the light curves that correspond to spikes in the HRs.}
\label{fig:hrlcs}
\end{figure}

As a low-mass narrow-line Seyfert-1 (NLSy1) AGN, MCG--6-30-15 has typically been quite changeable
in flux both within and between observations.  Historically, the
source has displayed a $2-10 \keV$ flux between
$F_{2-10} \sim 2-7 \times 10^{-11} \ergpcmsqps$ (NED).  The 2024 X-ray
observing campaign caught the source in a relatively bright state,
resembling those reported in the {\it XMM} observation of \citet{Fabian2002} and
the joint {\it XMM} and {\it NuSTAR} campaign of \citet{Marinucci2014}: $F_{2-10}=5.1 \times
10^{-11} \ergpcmsqps$ ($L_{2-10} = 6.82 \times 10^{42} \ergps$).  MCG-6 varied by a factor of
${\sim}3\times$ in flux over the course of the 2024 observations,
which is evident in both the Xtend and pn bandpasses as well as that of
{\it NuSTAR}.  Significant flux
variations are apparent on timescales as small as ${\sim}2000 \s$.  The
background-subtracted light curves from each of the low-resolution detectors used
in this work are shown in the top three panels of Fig.~\ref{fig:hrlcs}.  {\it XRISM}/Xtend is
shown because its CCD collects more
photons than the Resolve micro-calorimeter, but the light curves from
the two instruments show identical behavior. 

The hardness ratio (HR) between the hard and soft energy bands is
often used as an indicator of spectral shape change
\citep[see, \eg][]{Marinucci2014}.  The HR vs. time for
each of the low-resolution detectors is in our campaign is shown in the bottom three panels of
Fig.~\ref{fig:hrlcs} (only the FPMB detector data for {\it NuSTAR} is shown, as it has the higher S/N).  The HRs for all
instruments are overall relatively constant, with Xtend
averaging ${\rm HR} = 0.47 \pm 0.04$, the pn averaging
${\rm HR} = 0.34 \pm 0.03$, and FPMB averaging ${\rm HR} = 0.32 \pm 0.07$ (mean $\pm$ standard deviation, which we denote $\sigma$).
However, there are two noteworthy spikes seen in the
HRs: one at ${\sim}50 \ks$ of elapsed time lasting ${\sim}10 \ks$ (a ${\sim}5\sigma$ increase for
Xtend, ${\sim}3\sigma$ for pn, marginal in FPMB), and one at ${\sim}150 \ks$ lasting
${\sim}15 \ks$ (a ${\sim}3\sigma$ increase for
Xtend, ${\sim}1\sigma$ for FPMB), which occurred after the pn observation had ended.  Another
spike is seen at ${\sim}120 \ks$ lasting
${\sim}10 \ks$, though statistically this is only a $2\sigma$
detection in Xtend and pn, and $1\sigma$ in FPMB.  All three HR spikes correspond to prominent (${\sim}2\times$) flux dips at the
same times in the Xtend and pn
light curves; these are less prominent, but also noticeable, in the FPMB
light curve (Fig.~\ref{fig:hrlcs}).  Taken together, this information indicates that while changes in the
absorbing column may play a role, some of
this spectral shape variability must be broad-band in nature, \ie
changes in the intrinsic continuum of the source.

Given the relatively short duration of the HR changes and their
correspondence to the periods of lowest flux in our observations, we
turn our focus in subsequent sections to the time-averaged spectral
modeling of the data, with the goal of characterizing all of the spectral components.  We leave a
detailed exploration of the spectral variability to subsequent papers (\eg Wilkins \etal, in prep.).

\subsection{Time-averaged Spectral Analysis}
\label{sec:spectral}

Our principal goal in this work is to ascertain
whether relativistic reflection is required in order to fit the
spectrum of MCG-6, given the complexity of the Fe K band in this AGN and the
unprecedented spectral resolution of Resolve.  If so, the combination of broadband CCD-resolution data and Resolve data will facilitate
characterization of this reflection with the most up-to-date spectral models
for comparison with previous work on MCG-6 \citep[\eg][]{BR2006,Marinucci2014}.
We begin our analysis of
the time-averaged spectrum of MCG-6 with
a characterization of the broadband continuum shape, laying the foundation for modeling discrete spectral features in the Fe K band of {\it XRISM}/Resolve. 
This broadband analysis constrains the properties of the corona and yields a preliminary
assessment of the contributions of complex absorption intrinsic to the
AGN, distant reflection from the outer disk and/or torus, and
relativistic inner disk reflection.  

\subsubsection{Modeling the Broadband X-ray Continuum, Reflection and Absorption}
\label{sec:broadband}

\begin{figure}
\includegraphics[width=1.0\textwidth]{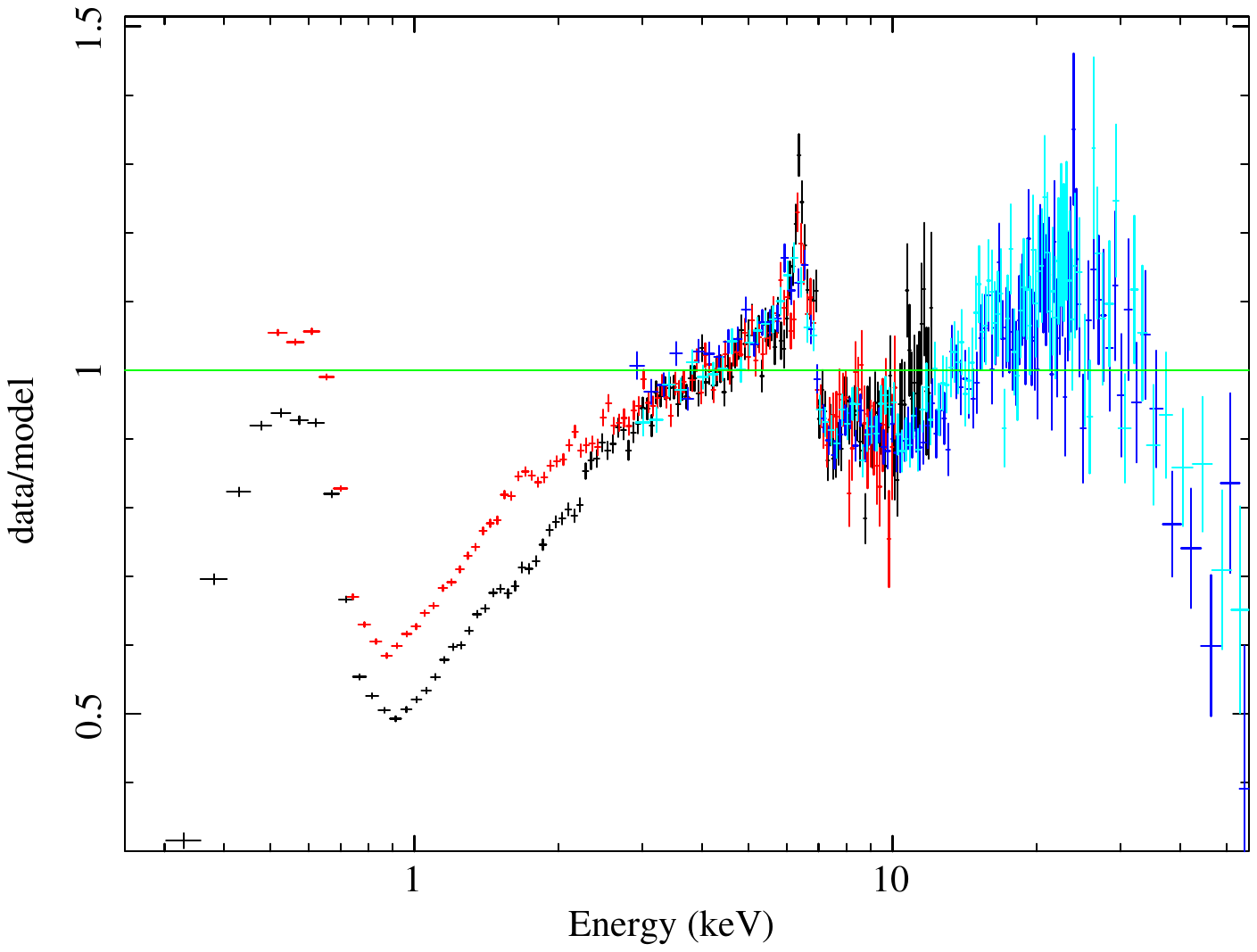}
\caption{{\small Ratio of the time-averaged broadband data from the CCD-resolution detectors to a simple power-law modified by Galactic photoabsorption.  {\it XRISM}/Xtend is in black, {\it XMM}/pn is in red, {\it
  NuSTAR}/FPMA is in dark blue, and {\it NuSTAR}/FPMB is in light
    blue.  Note the curvature due to low-ionization absorption below
    ${\sim}3 \keV$, the soft excess at the lowest energies, and the
    prominence of reflection in the Fe K band and above $10 \keV$.
    Some complexity due to ionized absorption is also evident between
    ${\sim}6-7 \keV$.  The difference between the pn and Xtend is due to their different mirror and detector responses (see text for details)}.}
\label{fig:phpo}
\end{figure}

We begin by modeling the broadband X-ray
continuum of MCG-6 using the data from {\it XRISM}/Xtend, {\it XMM}/pn and {\it NuSTAR}.  As with most radio-quiet AGN, the continuum shape
attributable to the corona in this source can be
phenomenologically represented by a single power-law, often with a
high-energy cutoff.  We begin with the assumptions that the
power-law slope is similar to that found in previous observations \citep[$\Gamma \sim 1.9-2.2$;][]{Fabian2002,BR2006,Marinucci2014}, and that this component
of the spectrum dominates over the energy ranges that do not
prominently feature reflection or absorption, \ie $3-4 \keV$,
$8-10 \keV$ and $40-55 \keV$.  
Fitting a simple photoabsorbed power-law over these energies returns an initial fit of
${\rm C/dof} = 240/154$, with $\Gamma = 2.05 \pm 0.01$.  A cross-normalization constant is also
employed to account for differences in flux calibration between the
different detectors.  We show the data-to-model
ratio over the entire $0.3-55 \keV$ range in Fig.~\ref{fig:phpo}, which illustrates the significant deviations from a simple power-law continuum due to unmodeled emission and absorption.  Note that, in spite of the cross-normalization constant, the Xtend and pn instruments show an offset of ${\sim}20\%$ at energies ${<}3 \keV$.  Similar discrepancies between the two detectors have been noted in other sources as well (\eg NGC~3783; Kaastra \etal, in prep.).  This offset persists even when precise overlap between the good time intervals of the two instruments is enforced, meaning that we can rule out source variability as the cause.  We attribute this offset to differences in the effective area calibration between these two instruments, which are currently being investigated.  We therefore limit the Xtend data to the $3-12 \keV$ energy range hereafter, where there is excellent agreement between the two detectors, including in the crucial Fe K band.

With the power-law continuum accounted for, the spectral shape below ${\sim}3 \keV$ is dominated by
absorption intrinsic to the AGN along with a soft excess at energies ${<}1
\keV$.  Focusing on the $3-7 \keV$ band,
narrow and broad Fe K$\alpha$ emission lines residuals are present, as are discrete lines from ionized absorption above ${\sim}6.5 \keV$.  Finally, ${>}10 \keV$ shows a broad, hump-like feature peaking at ${\sim}20-30 \keV$ that likely represents a combination of coronal curvature and the Compton hump characteristic of reflection.

\begin{figure}
\centering
\subfloat{%
    \includegraphics[width=0.55\textwidth, trim = 0cm 10.5cm 0cm 2.5cm, clip]{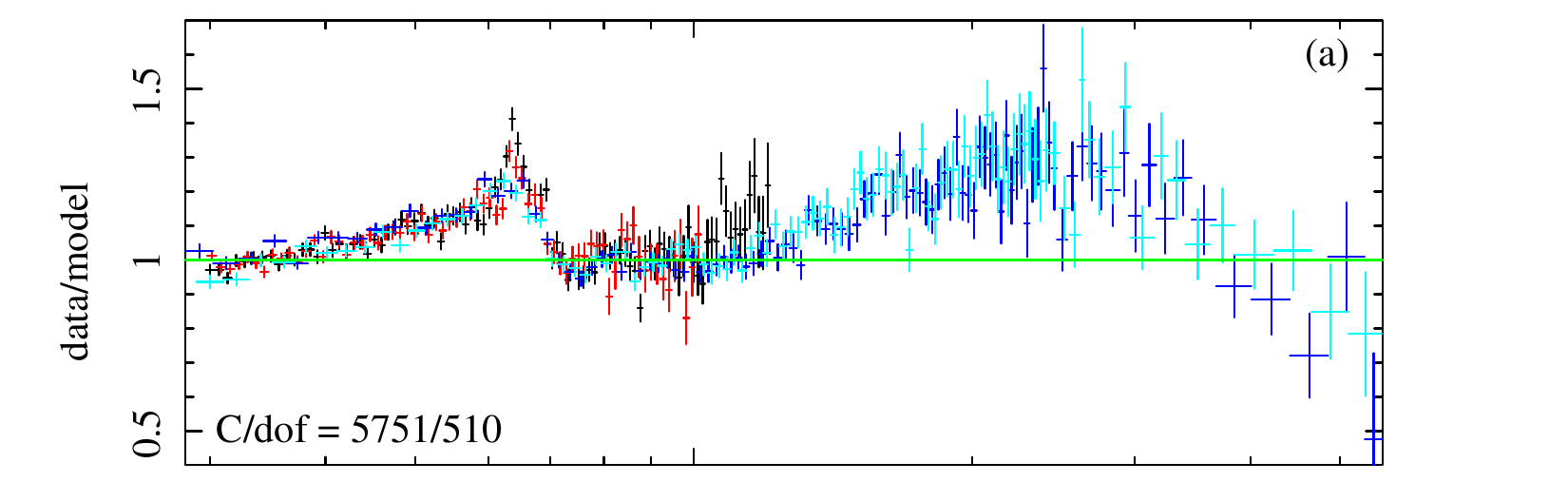}
}
\\
\subfloat{%
    \includegraphics[width=0.55\textwidth, trim = 0cm 10.5cm 0cm 2.8cm, clip]{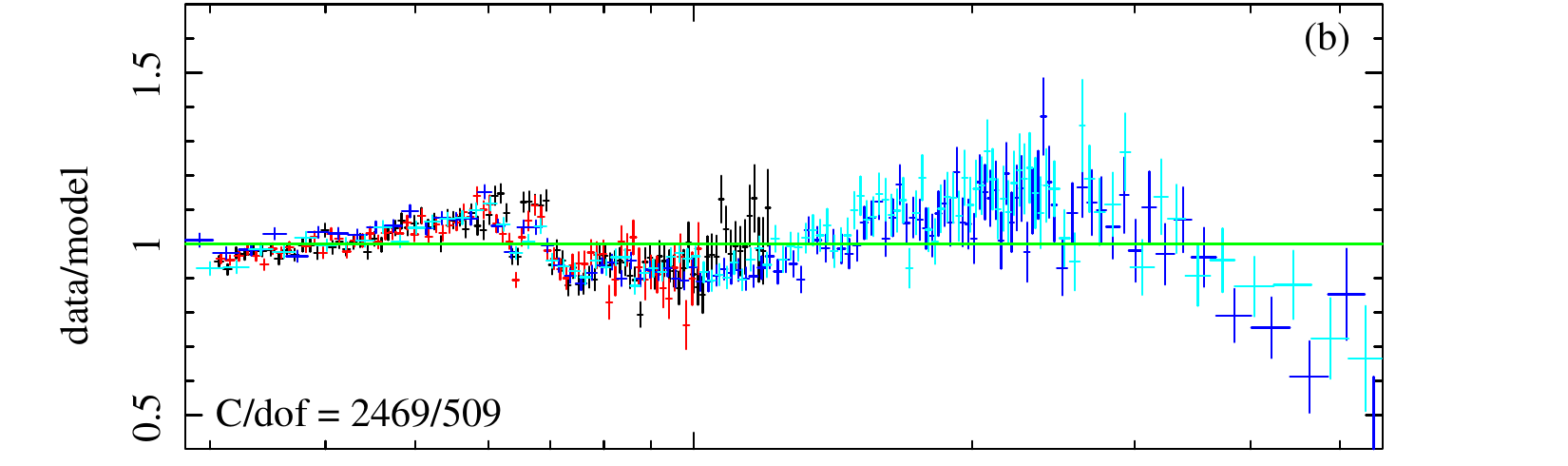}
}
\\
\subfloat{%
    \includegraphics[width=0.55\textwidth, trim = 0cm 10.5cm 0cm 2.8cm, clip]{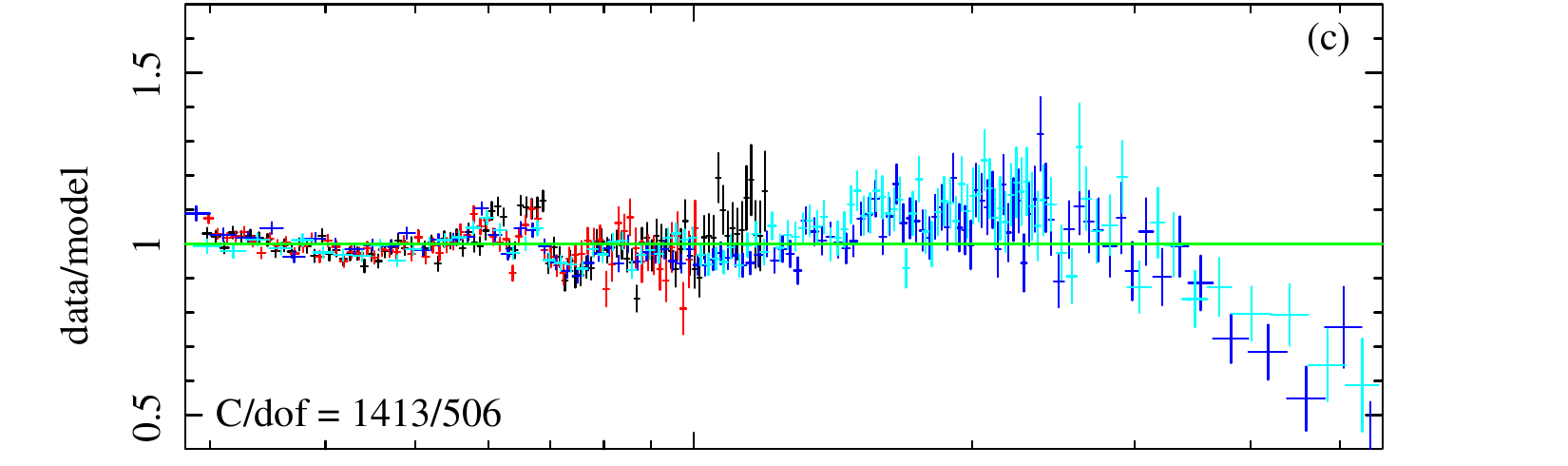}
}
\\
\subfloat{%
    \includegraphics[width=0.55\textwidth, trim = 0cm 10.5cm 0cm 2.8cm, clip]{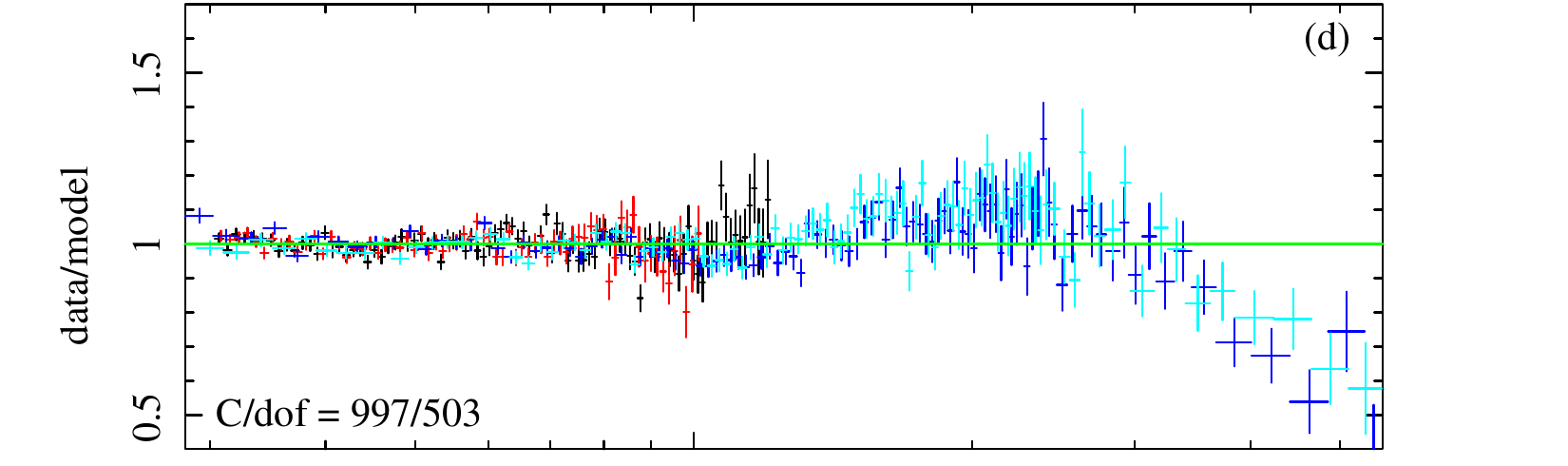}
}
\\
\subfloat{%
    \includegraphics[width=0.55\textwidth, trim = 0cm 10.5cm 0cm 2.8cm, clip]{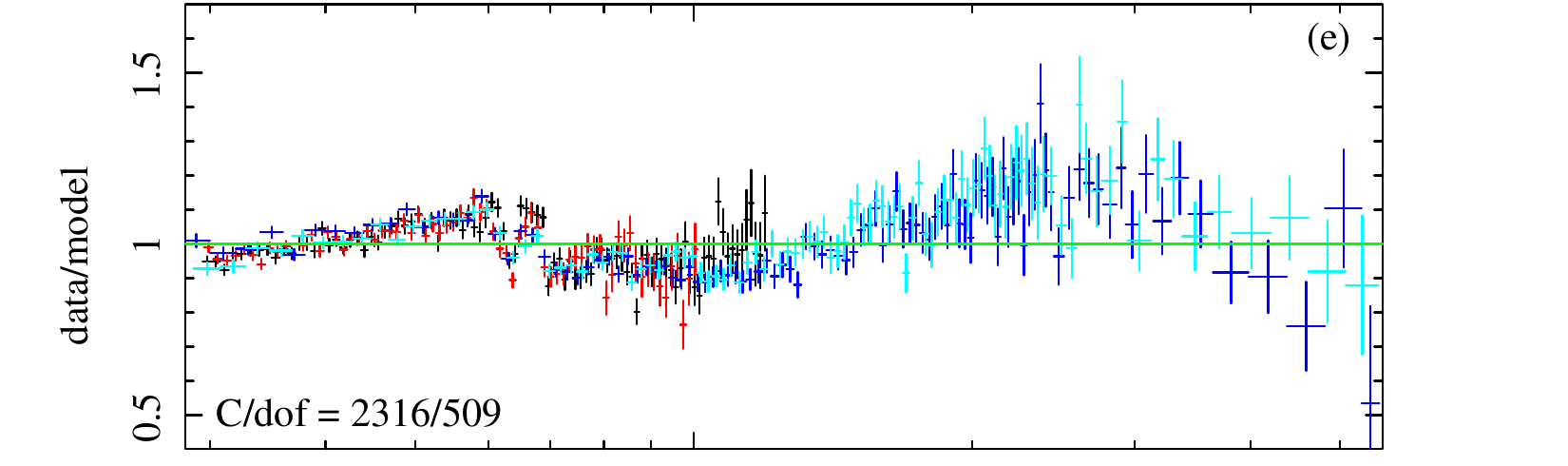}
}
\\
\subfloat{%
    \includegraphics[width=0.55\textwidth, trim = 0cm 8.5cm 0cm 2.8cm, clip]{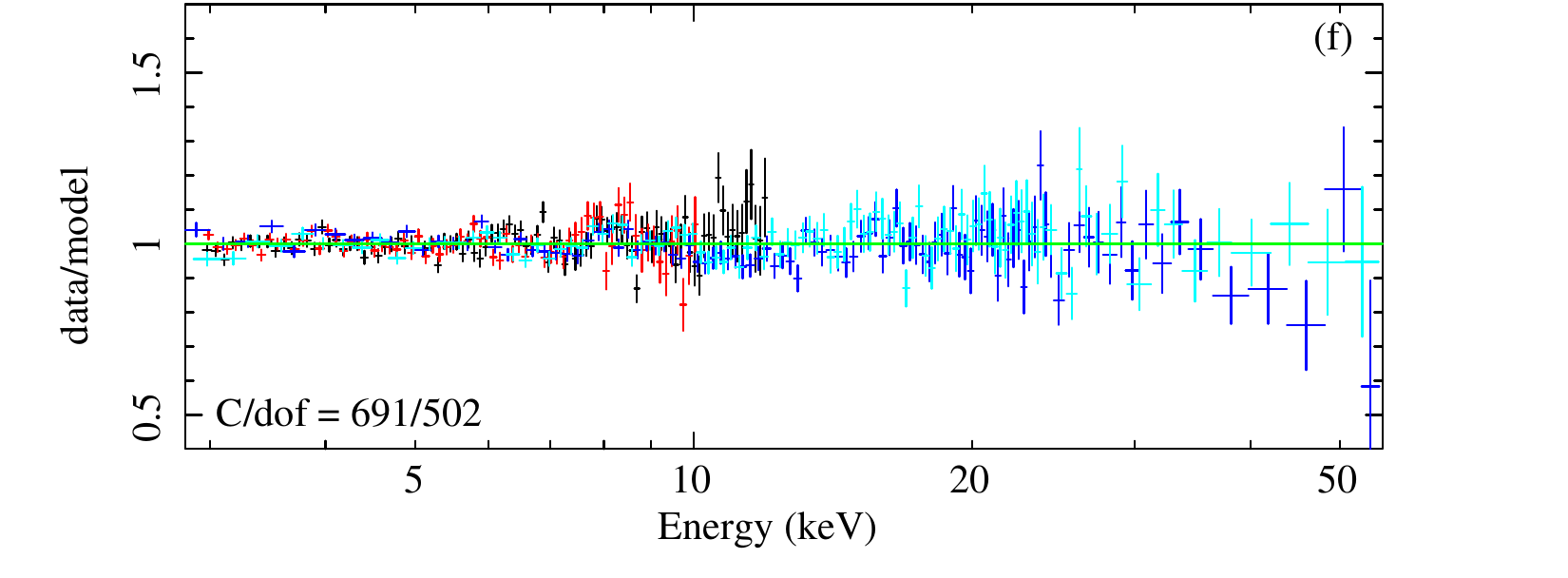}
}
\caption{{\small Ratios of the $3-55 \keV$ time-averaged broadband data to a series of
    models (see text for details): (a) a simple photoabsorbed power-law; (b) power-law plus
    narrow Gaussian emission line at $6.4 \keV$; (c) power-law plus
    narrow and broad Gaussian lines at $6.4$ and $3.29 \keV$,
    respectively; (d)
power-law plus narrow Gaussian and broad relativistic emission lines
at $6.4 \keV$; (e) cutoff power-law plus {\tt MYtorus} distant
reflection; (f) Comptonized coronal emission ({\tt nthcomp}) plus {\tt MYtorus} distant and {\tt
  relxilllpCp} relativistic reflection.  {\it XRISM}/Xtend is in black, {\it XMM}/pn is in red, {\it
  NuSTAR}/FPMA is in dark blue, and {\it NuSTAR}/FPMB is in light
blue.}}
\label{fig:ratios}
\end{figure}

As is common practice in fitting the X-ray spectra of AGN \citep[\eg][]{Brenneman2013}, we take a sequential approach to modeling these residual features, starting with simple phenomenological models and progressing to more complex physical models.  The results are shown in Fig.~\ref{fig:ratios}, with Fig.~\ref{fig:ratios}a showing the $3-55 \keV$ data ratioed against the power-law continuum described above.

To assess the importance of distant, neutral reflection from the putative torus of AGN unification schemes \citep{Antonucci1993} in the $3-55 \keV$ data, we
add
a narrow Gaussian emission line with $E = 6.4 \keV$ and $\sigma = 10 \eV$
(unresolvable by all lower-resolution instruments).  This significantly improves the $3-55 \keV$ fit (Fig.~\ref{fig:ratios}b); however, a prominent broad residual emission feature still remains.
Including a second Gaussian emission line component with the centroid
energy and line width left free to vary yields a further significant
improvement in fit (Fig.~\ref{fig:ratios}c).  Requiring the centroid energy to be fixed at $6.4 \keV$ results in ${\rm \Delta C/ \Delta dof} = +522/+1$ relative to the fit with the centroid energy free to vary.  We note that in either case, the Gaussian line morphology does not adequately recover
the shape of the broad residual.

To get a preliminary sense of whether this broad residual represents a
relativistic Fe K$\alpha$ line (\ie originating from reflected emission off of the inner accretion disk and dominated by gravitational redshift and relativistic Doppler shifts), we replace the broad Gaussian with a {\tt relline}
component \citep{Dauser2010}.  This model parametrizes the illumination of a geometrically-thin, optically-thick
disk by the inverse-Compton-scattered X-ray photons from the corona, also including the dimensionless SMBH spin and disk inclination as free parameters.  As with the Gaussian used
to represent the narrow Fe K$\alpha$ line, we also fix the energy of the
relativistic line at $6.4 \keV$.  The inclusion of this feature
significantly improves the fit vs. the broad Gaussian (Fig.~\ref{fig:ratios}d) and yields
a dimensionless black hole spin $a \geq 0.97$ and disk inclination
to the line of sight $i = (41 \pm 1)\degmark$.  The broad
line has a strength of $EW = (323 \pm 32) \eV$, ${\sim}4\times$ stronger relative to the continuum than the narrow line component.  

We now move on to holistic, broadband reflection
models as opposed to only individual Fe K$\alpha$ lines.  The {\tt
  MYtorus} model of \citet{Murphy2009} parametrizes cold reflection of the coronal photons off of neutral material in the form of a torus, including transmitted and scattered continuum photons as well as fluorescence lines from Fe K$\alpha$ and K$\beta$.  The model has recently been updated to include Fe
K$\alpha$ and K$\beta$ line shapes that more accurately reflect laboratory data
\citep{Yaqoob2024}, and is ideal for the analysis of {\it
  XRISM}/Resolve data due to its ${<}5$-eV energy resolution.  Anticipating our incorporation of these data
once the broadband spectral properties of MCG-6 have been
established, we elect to use {\tt MYtorus} to represent the distant
reflection.  

Because curvature due to the high-energy cutoff seen in the coronal
power-law emission from an AGN can be degenerate with curvature due to
Compton downscattering, we also change our coronal model to a cutoff
power-law at this point.  The value of the cutoff is fixed in
the {\tt MYtorus} scattered continuum model at several values, ranging from $100-500 \keV$.
After trying out each value for the {\tt MYtorus} tables, we find the
$100 \keV$ cutoff provides the best fit to the {\it NuSTAR} data ${>}10
\keV$.  This is roughly consistent with the cutoff value constrained
by \citet{Marinucci2014} ($E_c > 110 \keV$).  As such, we initially fix the value of our
high-energy power-law cutoff to $100 \keV$ as well.  The fit is
insensitive to the column density and inclination of the torus, so we fix these
parameters at $N_{\rm H} = 10^{24} \pcmsq$ and $i = 30\degmark$
(standard assumptions for Sy-1 AGN).  This model (Fig.~\ref{fig:ratios}e), which does not include any contribution from inner disk reflection, does not result in as good a fit as the previous model (Fig.~\ref{fig:ratios}d), which does.  Clear
residuals corresponding to both the broad Fe K line and an additional
Compton hump component are present.

Finally, we add in the entire relativistic reflection spectrum via {\tt relxilllpCp}, the most tunable of the state-of-the-art {\tt relxill} 
family of models \citep{Garcia2013}.  The {\tt relxill} models have the advantage of capturing the reflected continuum at soft and hard energies (\ie the soft excess and Compton hump) and in the Fe K band, whereas {\tt relline} only models the broad Fe K line.  The {\tt relxilllpCp} model variant parametrizes the continuum with a Comptonization component, including the temperature of the coronal electrons ($kT_e$), which we allow to vary freely rather than a phenomenological cutoff power-law. It also allows the user to vary the density and ionization parameter of the gas in the disk, and to give the electron plasma of the corona a bulk outflow (as in the base of a jet).  We begin with the parameters measured using
{\tt relline}, linking the continuum photon index within {\tt relxilllpCp} to
that of the {\tt MYtorus} component, and allowing the reflection fraction (the ratio between the direct and reflected continuum emission) and normalization to fit freely.  We also allow the disk ionization, density and iron abundance to
vary freely, maintaining constant ionization across the disk.  We assume no outflow of the coronal plasma.  This model (Fig.~\ref{fig:ratios}f) yields a significantly improved fit vs. the {\tt MYtorus} model above (Fig.~\ref{fig:ratios}e), with {\tt relxill} parameters that echo those seen with {\tt relline} alone: $a \geq 0.97$ and $i = (38 \pm 1) \degmark$.  

The contemporaneous {\it NuSTAR} data $>10 \keV$ allow us to disentangle the contributions of the distant ({\tt MYtorus}) and inner disk ({\tt relxilllpCp}) reflection to the morphology of the Compton hump.  Surprisingly, we find that the fit is slightly improved (${\rm \Delta C/ \Delta dof} = -8/0$) when the scattered component of {\tt MYtorus} is removed (the degrees of freedom do not change, as the three components of the {\tt MYtorus} models are linked).  There is also visual indication of unmodeled discrete features
due to highly ionized absorption between ${\sim}6.5-8 \keV$, as seen
previously in {\it Chandra}/HETG data \citep{Young2005}.  We leave
further exploration of both of these issues for our {\it XRISM}/Resolve
analysis (\S\ref{sec:resolve}) and Discussion (\S\ref{sec:discussion}).  

MCG-6 has historically shown evidence for a multi-zone
warm absorber, primarily manifesting as the broad trough in the source flux seen below ${\sim}3 \keV$ (Fig.~\ref{fig:phpo}).  Gratings observations have shown evidence for 
discrete absorption features from low-ionization gas and dust seen at
these energies \citep{Lee2001,Marinucci2014} and from highly-ionized gas in the Fe K band
\citep{Young2005}.  Detailed analyses of these features will be
presented in future work incorporating the RGS data; here we adopt a simple, three-zone (two gas zones, one dust zone) representation
of the low-ionization absorption, following the example of
\citet[][see their Table~2]{Marinucci2014}.  This will ensure that our modeling of the
reflection ${>}3 \keV$ is not biased by ignoring curvature due to
absorption ${<}3 \keV$.

Including the {\it XMM}/pn data down to $0.5 \keV$, we apply {\sc cloudy} \citep{Ferland2013} table models to account for
the two zones of low-ionization gas absorption, generated using the SED from this 2024 campaign as input.  The SED was constructed from IR to X-rays using an unabsorbed continuum modeled with the 2024 {\it XMM}/EPIC spectra, a theoretical optical/UV blackbody shape for typical $\alpha_{OX}=1.5$ based on large quasar sample studies \citep[\eg][]{Chiaraluce2018}, and IR observations from {\it IRAS} \citep{Moshir1990} and the NASA Infrared Telescope Facility \citep{Ward1987}.  A more detailed, self-consistent treatment of the ionizing SED will be presented in follow-up papers focused on outflows in MCG-6 (Rogantini \etal, in prep.; Ogorza{\l}ek \etal,in prep.).
The {\sc cloudy} tables were calculated in $194$ energy bins over the $0.15-12.15 \keV$ band.  The gas density
is assumed to be $n = 10^5 \pcmcu$, and abundances are set to their
solar values.  Fit parameters are the column density,
ionization parameter, line broadening and outflow velocity of the
absorbing gas.  We also apply the model of \citet{Lee2001} using a {\tt TBvarabs} component
\citep{Wilms2006} to account for the known iron dust in the system.

\begin{figure}
\includegraphics[width=1.0\textwidth, trim = 0cm 1.5cm 0cm 0cm, clip]{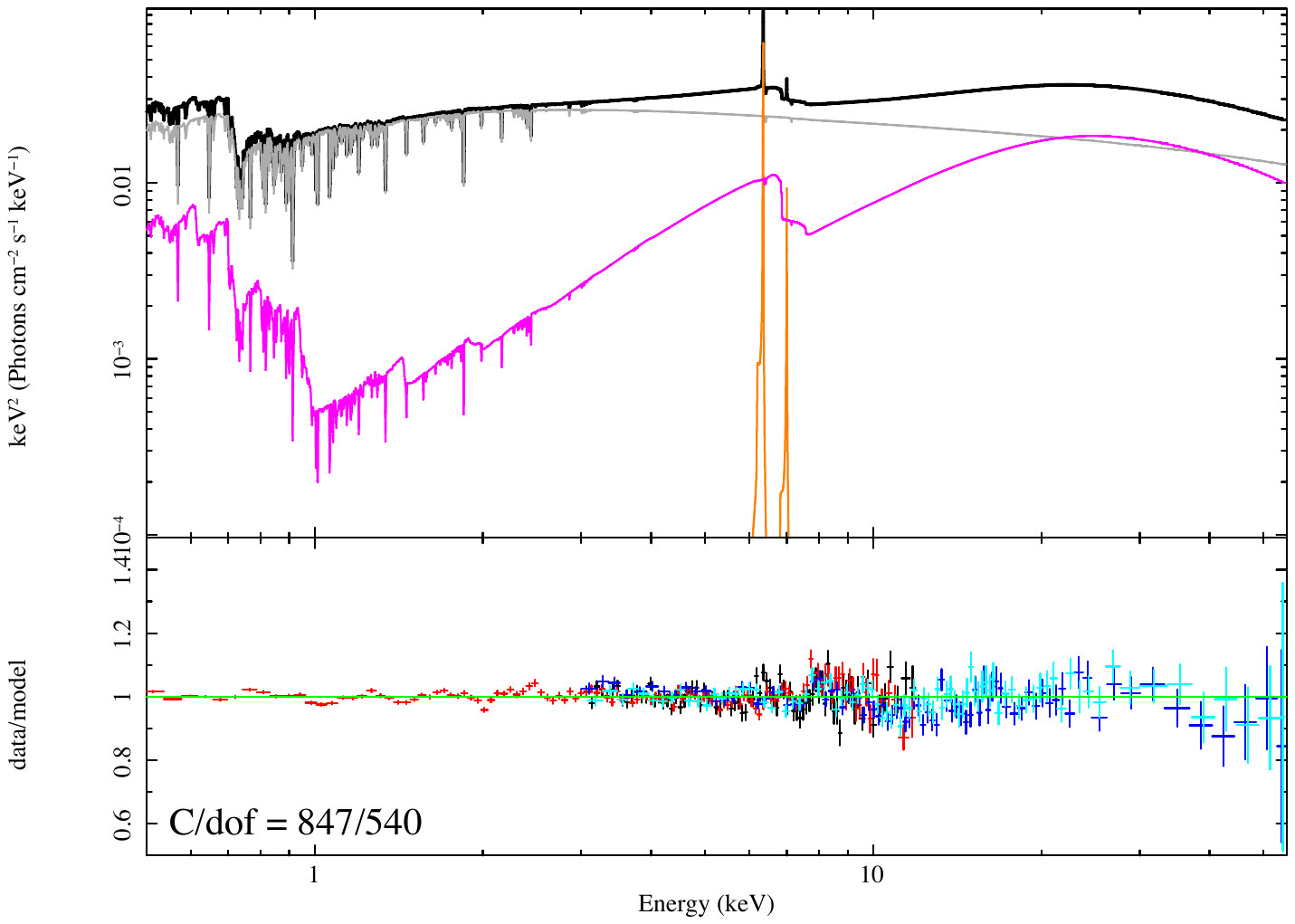}
\caption{{\small {\it Top:} Best-fit broadband model to the $0.5-55 \keV$ CCD
    data (total model is in black).  The model consists of a three-zone warm absorber modifying
    a coronal continuum via {\tt nthcomp} (plotted in grey, separately from the {\tt relxilllpCp} component for clarity), distant torus reflection via {\tt
      MYtorus} (orange; only the line component) and inner
    disk reflection via {\tt relxilllpCp} (magenta).  {\it Bottom:} Ratio of the best-fit
    broadband model to the data.  {\it XRISM}/Xtend is in black, {\it XMM}/pn is in red, {\it
  NuSTAR}/FPMA is in dark blue, and {\it NuSTAR}/FPMB is in light
    blue.}}
\label{fig:ccd_fit_ratio}
\end{figure}

The final CCD-resolution model syntax is as follows: {\tt constant $\times$ TBabs $\times$ TBvarabs $\times$ WA$_1$ $\times$ WA$_2$ $\times$ (MYtorus-lines + relxillllpCp).}  The best-fitting model to the $0.5-55 \keV$ data is shown
in Fig.~\ref{fig:ccd_fit_ratio}, and the corresponding parameter values and their $90\%$-confidence uncertainties
are listed in Table~\ref{tab:best_fit_tab}.  All remaining residual features in the ${<}3 \keV$ band
of the data-to-model ratio are at the ${<}5\%$ level, and
the parameter values of the continuum and reflection components remain
well constrained, with most similar to their values from the ${>}3 \keV$ fit.
Though the fit is not formally acceptable, mainly owing to unmodeled,
discrete features ${<}2 \keV$ which will be explored further in subsequent work
using the {\it XMM}/RGS data (Ogorza{\l}ek \etal, in prep.), it represents the major broadband spectral
characteristics accurately.  We therefore proceed to analyzing the
{\it XRISM}/Resolve spectrum, using our broadband best-fit model as a
baseline. 

\subsubsection{A New Look at the Fe K Band with {\em XRISM}/Resolve}
\label{sec:resolve}

With its ${<}5$-eV energy resolution, the {\it XRISM}/Resolve
micro-calorimeter provides the most detailed spectral view of the Fe K
band ever obtained for MCG--6-30-15.  As seen in other bright AGN \citep[\eg NGC~4151;][]{Miller2024}, this revolutionary instrument is unique in its
ability to definitively separate the different components that combine
to form the shapes of the spectral features in this bandpass.  Narrow emission
and absorption lines can easily be separated from broader features,
allowing us to accurately and precisely model the distant reflection,
inner disk reflection, and signatures of highly ionized outflowing
winds.  However, Resolve data are currently only recommended to be
analyzed between $2-11 \keV$ due to the gate valve remaining closed
and the status of the calibration.  This underscores the importance of the
contemporaneous {\it XMM} and {\it NuSTAR} observations in our 2024
campaign: these data are necessary in order to
scaffold and anchor the broadband continuum, allowing us to accurately model the discrete line features in the Resolve spectrum.

\begin{figure}
\hbox{
  \includegraphics[width=0.5\textwidth, trim = 0cm 1cm 7cm 1cm, clip]{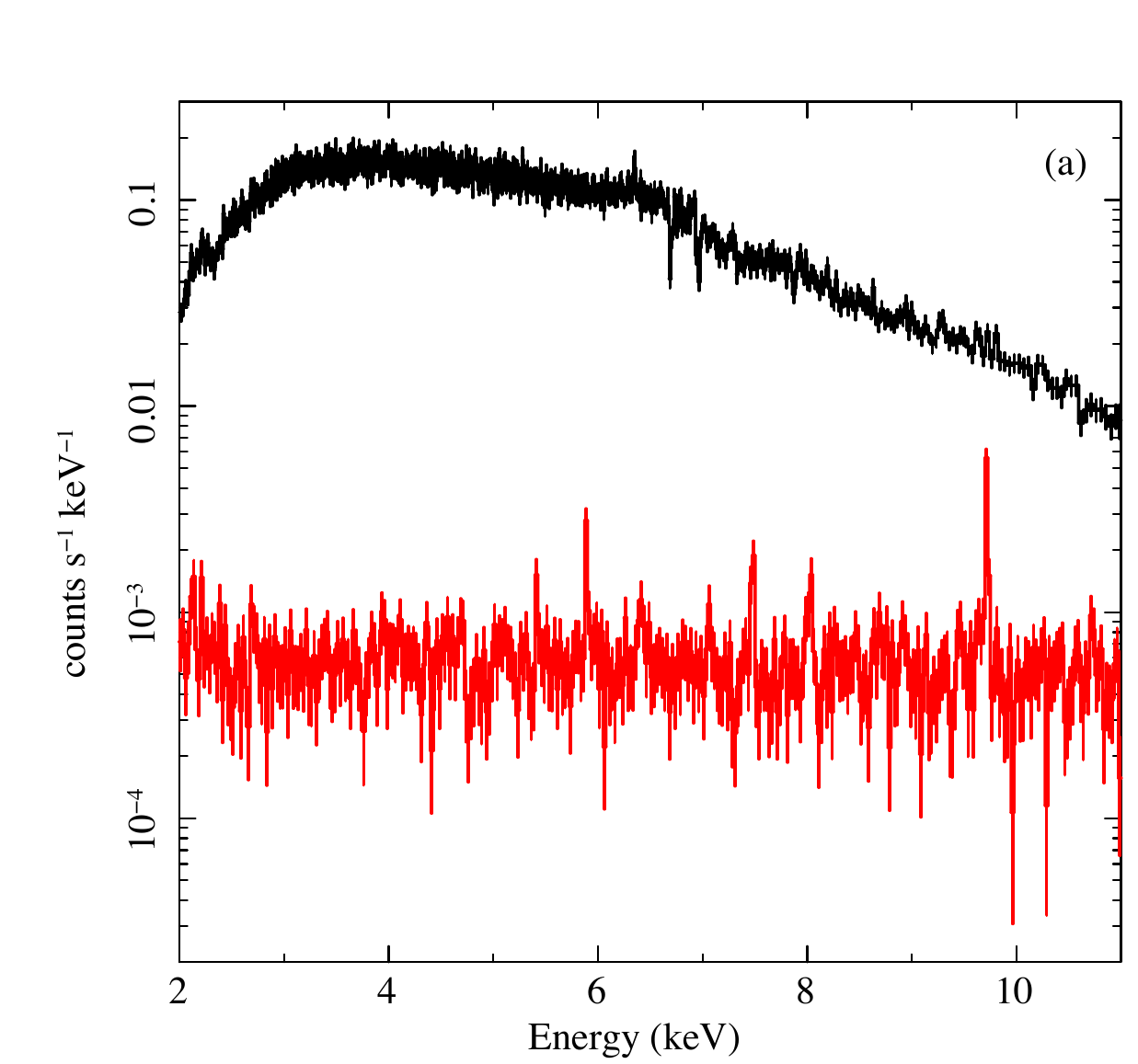}
  \includegraphics[width=0.5\textwidth, trim = 0cm 1cm 7cm 1cm, clip]{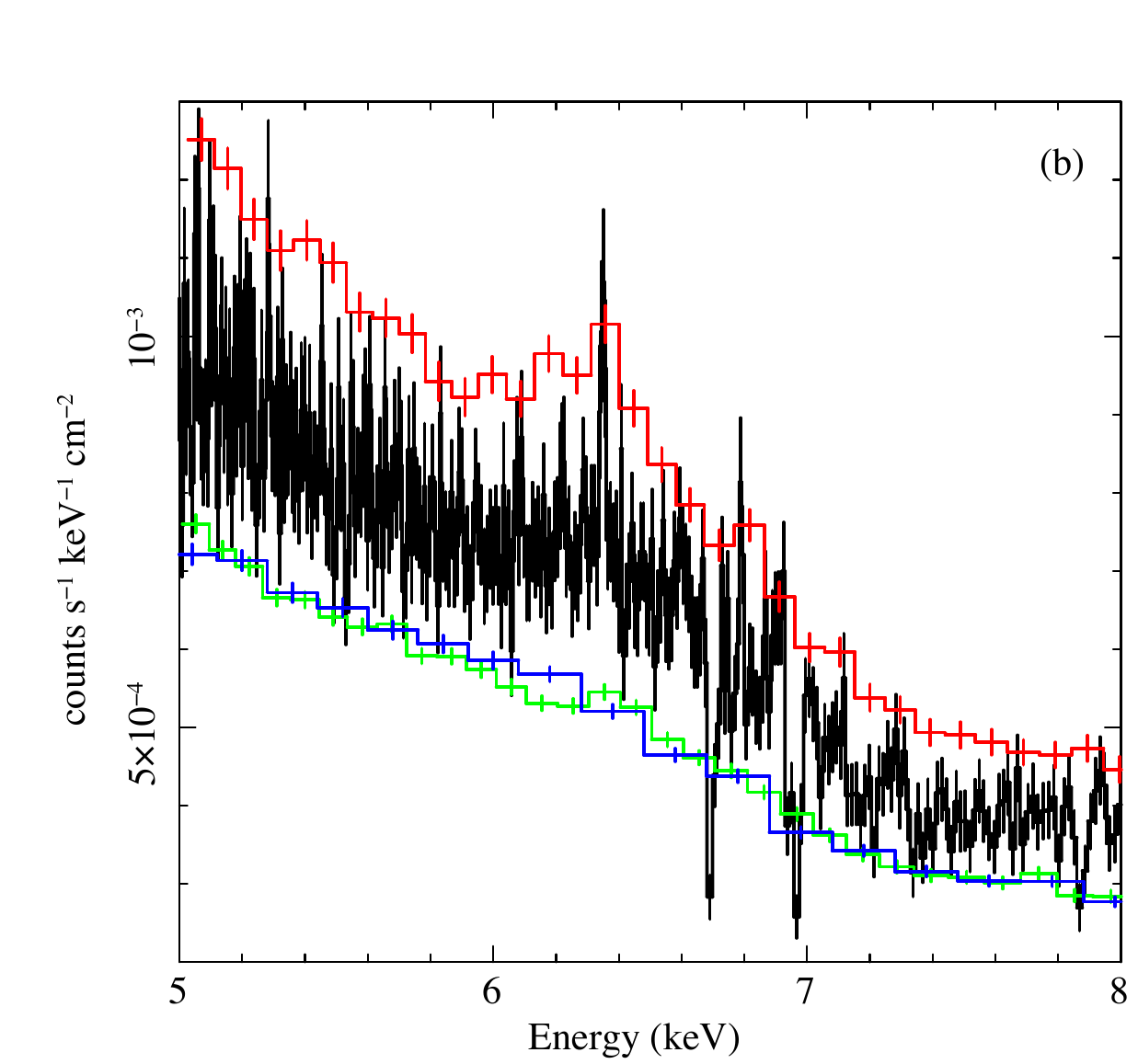}
}
\caption{{\small {\it Left:} Time-averaged {\it XRISM}/Resolve data
    (black) and background (red), both from the full array (minus
    pixels 12 and 27).  The background is negligible in this
    observation.  {\it Right:} Data divided by detector area
    for Resolve (black), Xtend (red), {\it
      XMM}/pn (green) and {\it NuSTAR}/FPMB (blue).  The offsets between detectors are
    indicative of current calibration differences.  No model is applied,
    but the data points are connected for visual clarity.}}
\label{fig:rsl_spec}
\end{figure}

As noted in \S\ref{sec:xrism_obs}, MCG-6 is detected at a level ${\sim}100\times$ higher than the
background across the majority of the Resolve bandpass.  We show the Resolve
source and background spectra in Fig.~\ref{fig:rsl_spec}a,
along with an Fe K band comparison of the Resolve data with those from the
{\it XRISM}/Xtend, {\it XMM}/pn and {\it NuSTAR} detectors in Fig.~\ref{fig:rsl_spec}b.  The advantage conferred by Resolve in spectral modeling of the narrow features in
the Fe K band is clear.

\begin{figure}
\hbox{
  \includegraphics[width=0.5\textwidth, trim = 0cm 1cm 7cm 1cm, clip]{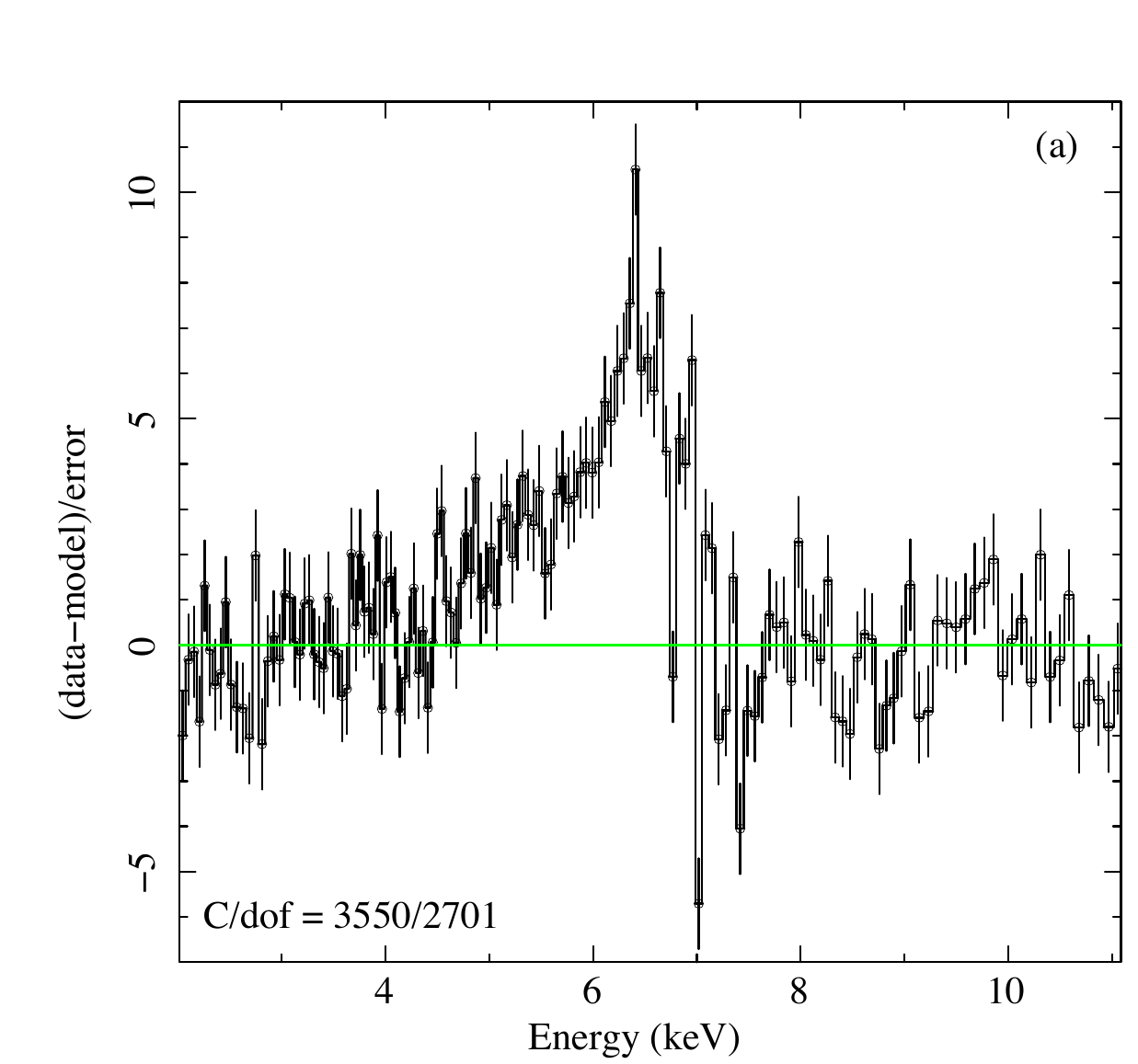}
  \includegraphics[width=0.5\textwidth, trim = 0cm 1cm 7cm 1cm, clip]{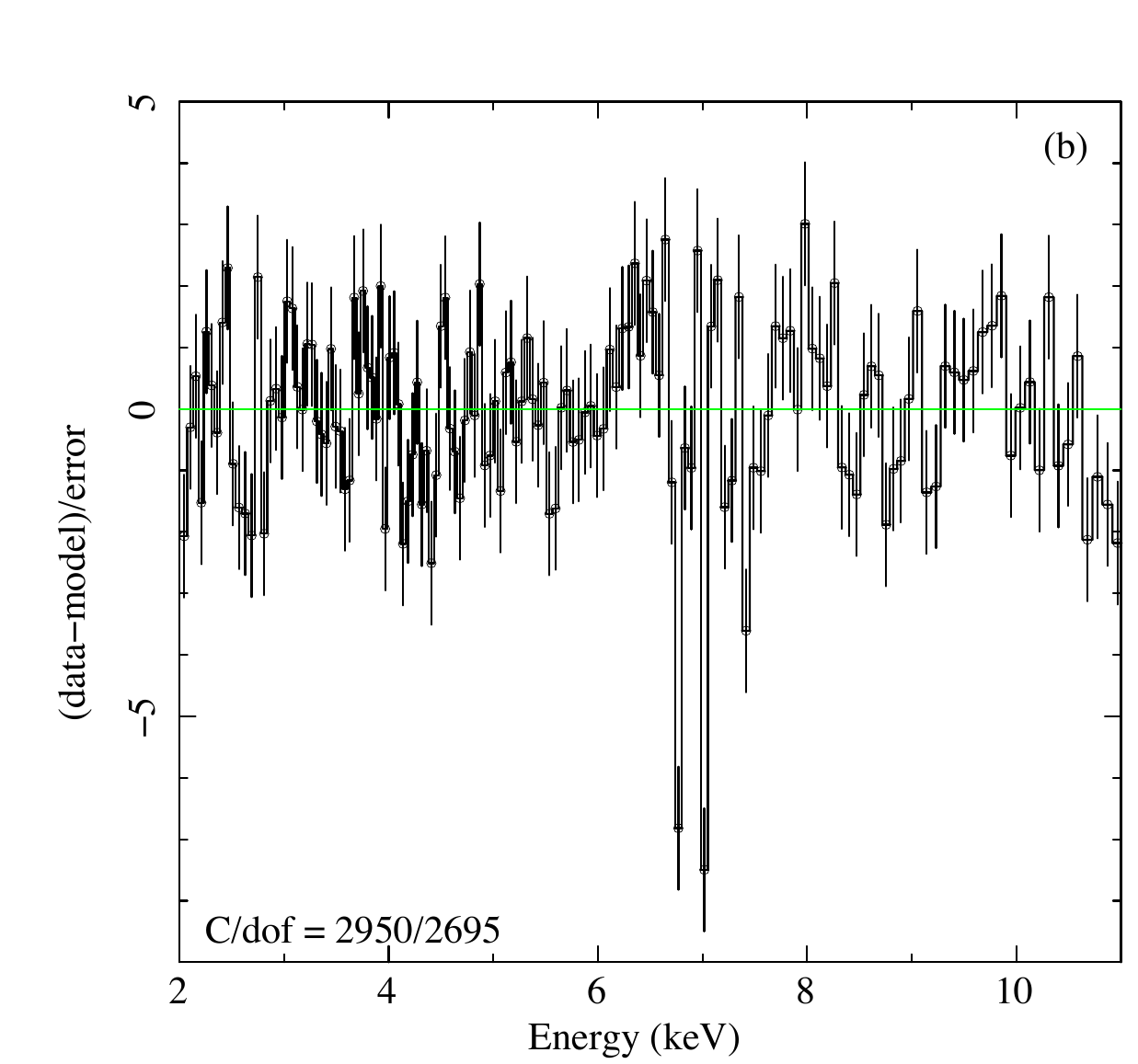}
}
\caption{{\small {\it Left:} The {\it XRISM}/Resolve data fit with a
    simple photoabsorbed power-law model.  Note the broad and
    narrow residuals that are apparent in emission and
    absorption in the Fe K band. {\it Right:} Residuals to the best-fit model for the
    broadband data (including reflection), refit to the Resolve spectrum.  Narrow residuals
    remain in the Fe K band, indicating the presence of highly-ionized gas, particularly in absorption.   Because we use Cash statistics
    to evaluate the goodness-of-fit,
error is calculated as the square root of the model-predicted number
of counts.}}
\label{fig:rsl_delchi}
\end{figure}

We begin our spectral modeling of the Resolve data by examining the
residual features from the same phenomenological photoabsorbed power-law fit initially applied to the broadband data (\S\ref{sec:broadband}).  As with the lower-resolution
detectors, we include a cross-normalization constant to account for
any differences in effective area calibration between Resolve and the
{\it XMM}/pn instrument, which we maintain as our baseline (\ie a
constant value fixed to $1.0$, as seen in Table~\ref{tab:best_fit_tab}).
Allowing the 
normalization of the power-law to vary freely results in an overall
goodness-of-fit of ${\rm C/dof} = 3550/2701$. 
Residuals from both broad and narrow features in emission and absorption are obvious in the Fe K
band of Resolve (Fig.~\ref{fig:rsl_delchi}a); this is in sharp contrast to how they are blurred together in lower-resolution data (Fig.~\ref{fig:rsl_spec}b).  Note that, due to the limited bandpass of Resolve, we see no 
curvature at low energies due to unmodeled absorption, or at high
energies due to unmodeled reflected continuum. 

We then apply the
best-fit model obtained for our broadband analysis in
\S\ref{sec:broadband} to the $2-11 \keV$ Resolve spectrum.  We fix the coronal power-law and low-ionization absorber parameter values at their broadband values, as they cannot viably be constrained across the narrow bandpass of Resolve.  Leaving all other parameters free to vary, the goodness-of-fit improves to ${\rm C/dof} = 2950/2695$.  Prominent, narrow absorption and emission features from ionized Fe\,{\sc xxv}
and Fe\,{\sc xxvi} between ${\sim}6.6 - 7 \keV$ (in the rest frame) are clearly present and unaccounted for by the model (Fig.~\ref{fig:rsl_delchi}b).  

Most of the parameter values are consistent with
those of the broadband fit within uncertainties, but there are some
notable exceptions within the {\tt relxilllpCp} model component: $R = 0.40 \pm 0.37$ (${\sim}3\times$ smaller than in the broadband fit), $i = (41 \pm 1) \degmark$ ($10\%$ larger than in the broadband fit), and ${\rm log}\,n_e = (19.02 \pm 1.97) \pcmcu$ (${>}100\times$ larger than in the broadband fit).  Additionally, the SMBH spin is only loosely constrained to $a \geq 0$, and each parameter of the {\tt relxilllpCp} component has a significantly larger uncertainty than in the broadband fit.  These differences in the inner disk reflection parameters may be attributable to the limited
bandpass and collecting area of Resolve compared to the lower-resolution instruments, making it
less effective at characterizing broadband reflection features on its own.  
Therefore, after using Resolve alone to account for the highly ionized
outflow in MCG-6, we will incorporate this data into the joint
fit with all instruments in order to finalize our modeling.

\begin{figure}
\hbox{
  \includegraphics[width=0.5\textwidth, trim = 0cm 1cm 7cm 1cm, clip]{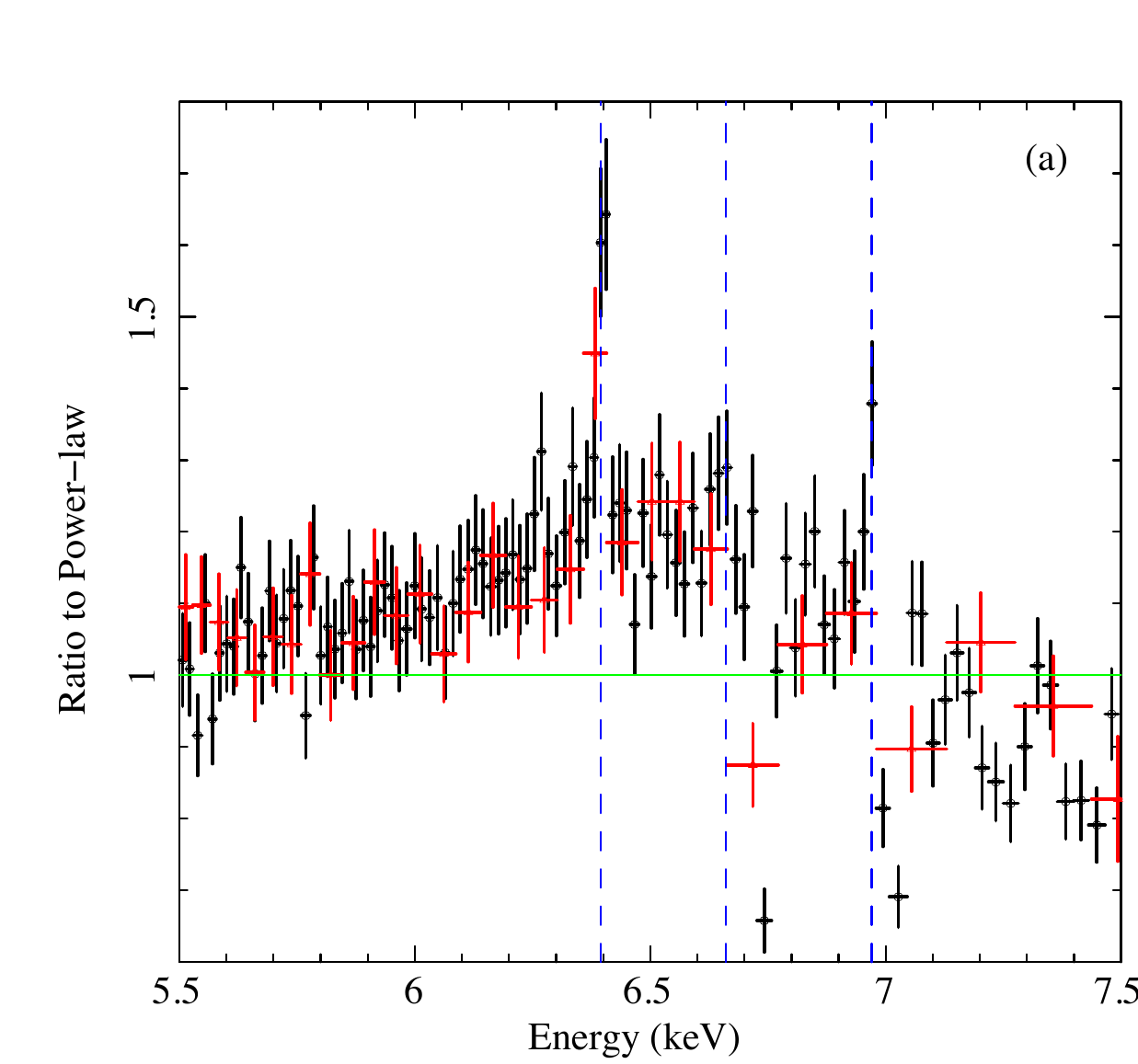}
  \includegraphics[width=0.5\textwidth, trim = 0cm 1cm 7cm 1cm, clip]{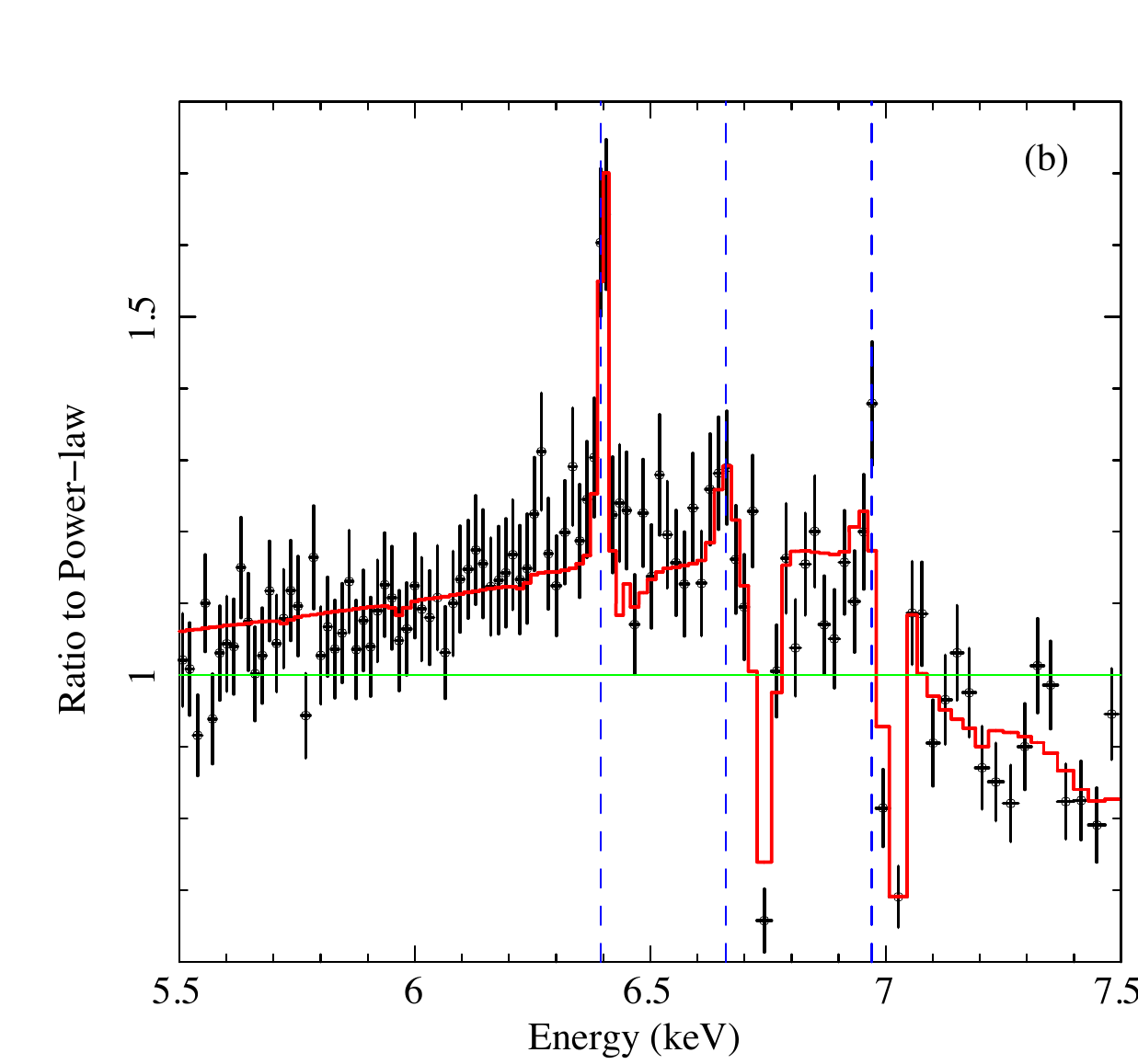}
}
\caption{{\small {\it Left:} Zoom-in of the Fe K region of the {\it XRISM}/Resolve spectrum divided by the power-law continuum (black points), plotted with the same ratio for the {\it Chandra}/HETG data from \citet{Young2005} (red points).  The spectra from both detectors are rebinned identically for comparison.  Though the HETG data show clear evidence for the strong, $v_{\rm out} \sim -2000 \kmps$ wind in the lines of Fe\,{\sc xxv} and Fe\,{\sc xxvi} also seen by Resolve, the diminishing effective area of the HETG above ${\sim}7 \keV$ renders it unable to detect the $v_{\rm out} \sim -20,000 \kmps$ wind component.  {\it Right:} The best-fitting model from \S\ref{sec:resolve} is plotted against the Resolve data shown in panel (a).  In both panels the blue hashed vertical lines show, from left to right, the rest-frame positions of the Fe K$\alpha$, Fe\,{\sc xxv} and Fe\,{\sc xxvi} lines.  Note that, while the emission features of Fe\,{\sc xxv} and Fe\,{\sc xxvi} appear at their rest-frame energies, the absorption components are blueshifted, indicating an origin in outflowing gas.}}
\label{fig:rsl_AY_ratplots}
\end{figure}

Though detailed modeling of
the highly ionized absorption signatures seen in the Resolve Fe
K band will be undertaken in a subsequent
paper, we make a first effort at characterizing the absorption
components here in order to investigate their basic properties and determine whether their presence impacts our reflection modeling.  \citet{Young2005} detected similar signatures of highly ionized Fe\,{\sc xxv} and {\sc xxvi} 
absorption within MCG-6 in the {\it Chandra}/HETG data (Fig.~\ref{fig:rsl_AY_ratplots}a).  However, the
resolving power of the HETG instrument is ${\sim}200$ at $6 \keV$
vs. ${\sim}1333$ for Resolve; the effective areas are
${\sim} 25 \cmsq$ and $130 \cmsq$,
respectively, and HETG's drops rapidly above $7 \keV$.  Resolve is therefore a much more powerful instrument for precisely characterizing narrow spectral lines in the Fe K band.  \citet{Young2005} fit the HETG absorption features
using an {\sc xstar} \citep{Kallman2001} table model, deriving gas parameters ${\rm log}\,N_{\rm H} = 23.2 \pcmsq$,
${\rm log}\,\xi = 3.6 \ergcmps$, line broadening $v_b = 100 \kmps$ and
outflow velocity $v_{\rm out} = -2000 \kmps$ relative to the MCG-6 rest
frame.  Applying this model to
the Resolve data without refitting improves the fit by ${\rm \Delta C/ \Delta dof} = -48/+3$ vs. the refitted broadband model, though several significant
residuals are visually evident around the Fe\,{\sc xxv} and Fe\,{\sc xxvi} lines.  

To more accurately model the highly-ionized outflow (HIO) seen in the 2024 Resolve data,
we replace the \citet{Young2005} model with the {\sc cloudy} table model referenced in \S\ref{sec:broadband}.  Incorporating one zone of this gas improves the
fit by ${\rm \Delta C/ \Delta dof} = -167/-4$ vs. the refitted broadband model (${\rm \Delta C/ \Delta dof} = -119/-7$ vs. the \citet{Young2005} absorber model), with initial parameters ${\rm log}\,N_{\rm H} = (22.65 \pm 0.12) \pcmsq$,
${\rm log}\,\xi = (4.95 \pm 0.11) \ergcmps$\footnote{We note that the reported parameters on the ionized absorbers here are best considered as upper limits, especially ionization parameters, given that they are driven by the fit to the Resolve Fe K band and are therefore blind to any soft features that Resolve cannot detect with the gate valve closed.}, line broadening
$v_b = (700 \pm 90) \kmps$
and outflow velocity $v_{\rm out} = (-2300 \pm 50) \kmps$.  This component
accurately accounts for the prominent Fe\,{\sc xxv} and Fe\,{\sc xxvi} absorption lines seen at
$E = 6.67 \keV$ and $E = 6.97 \keV$, respectively, in the rest frame of MCG-6, which have also been reported in several other AGN observations with {\it XRISM}/Resolve \citep[\eg NGC~3783 and NGC~4151;][]{Mehdipour2025,Xiang2025}.  Adding a second
component of absorbing gas further improves the fit by ${\rm \Delta C/ \Delta dof} = -13/-4$,
modeling the broader ${\sim}3\sigma$ residual absorption feature observed at $E \sim 7.2
\keV$.  This second component could not have been detected by the HETG owing to its very small effective area ${\geq}7 \keV$, but it has been hinted at previously through excess variance analysis \citep{Igo2020}.  The best fit to the Resolve data shows that the gas producing this feature has a column density and ionization
parameter roughly consistent with that of the gas in the first component, but with significantly
higher line broadening and outflow velocity: $v_b = (4800 \pm 1800) \kmps$ and $v_{\rm out} = (-20100 \pm 1200) \kmps$.  

Attempting to add
in a third absorber component in the Fe K band does not further improve the
fit.  However, we do note ${\sim}4\sigma$ emission features at
${\sim}6.67 \keV$ and ${\sim}6.97 \keV$ (Fe\,{\sc xxv} and {\sc xxvi} in the rest frame of MCG-6).  These are the corresponding emission features from the parts of the HIO not along our line of sight to the corona.  They are well
represented by Gaussian lines of $\sigma = (12 \pm 2) \eV$ with $EW_{\rm XXV} = (5 \pm 2) \eV$ and $EW_{\rm XXVI} = (3 \pm 2) \eV$, and their inclusion marginally improves the fit by ${\rm \Delta C/\Delta dof} = -7/-5$, if we assume the same line width for both features.
No additional improvement is found when attempting to add in any other
line-like components.  The fit of this model to the $5.5-7.5 \keV$ data is shown in Fig.~\ref{fig:rsl_AY_ratplots}b. 

\begin{figure}
    \begin{center}
    \subfloat{%
        \hbox{
            \includegraphics[width=0.5\textwidth, trim = 0cm 1cm 7cm 1cm, clip]{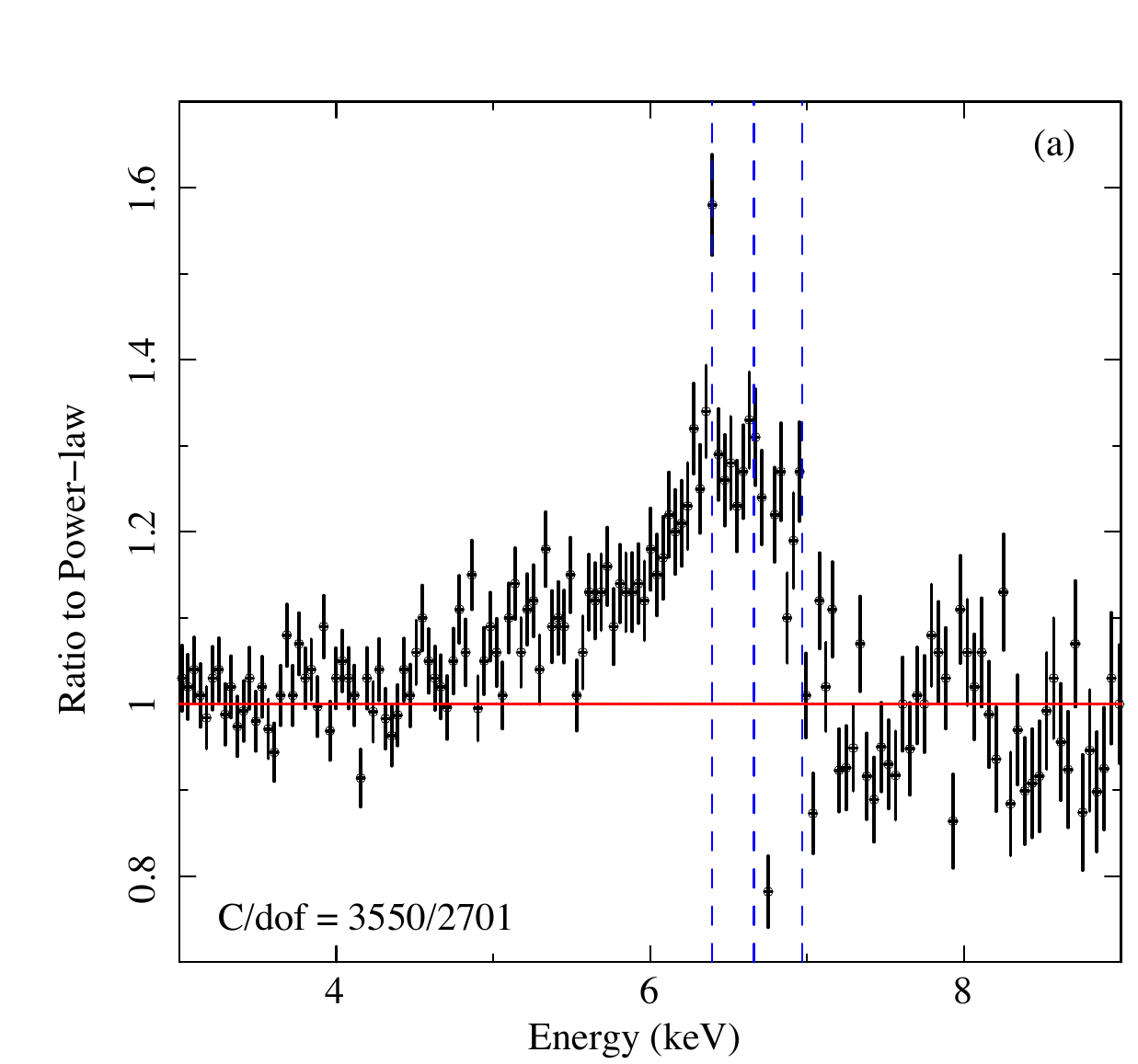}
            \includegraphics[width=0.5\textwidth, trim = 0cm 1cm 7cm 1cm, clip]{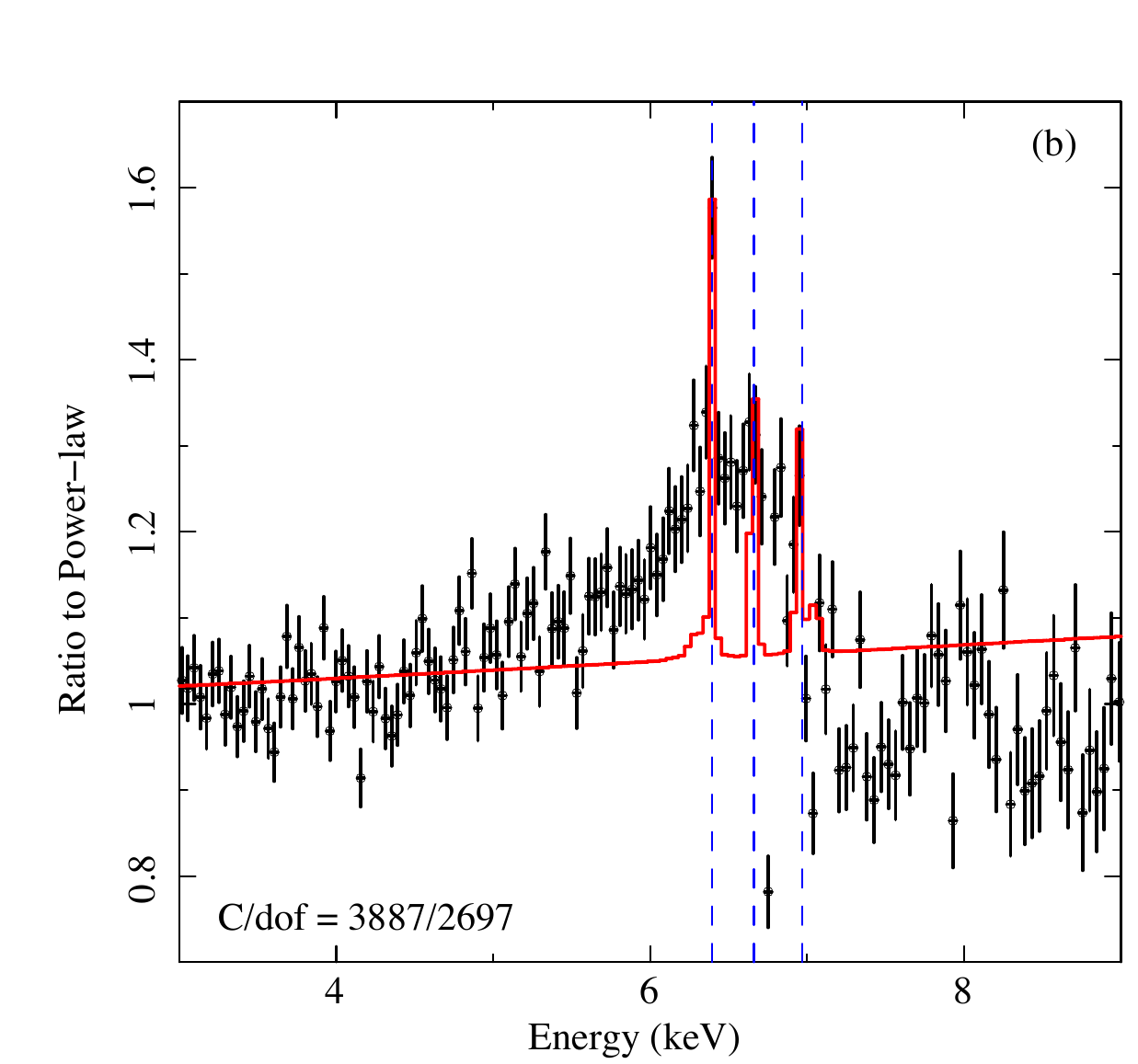}
        }
    }
    \\
    \subfloat{%
        \hbox{
            \includegraphics[width=0.5\textwidth, trim = 0cm 1cm 7cm 1cm, clip]{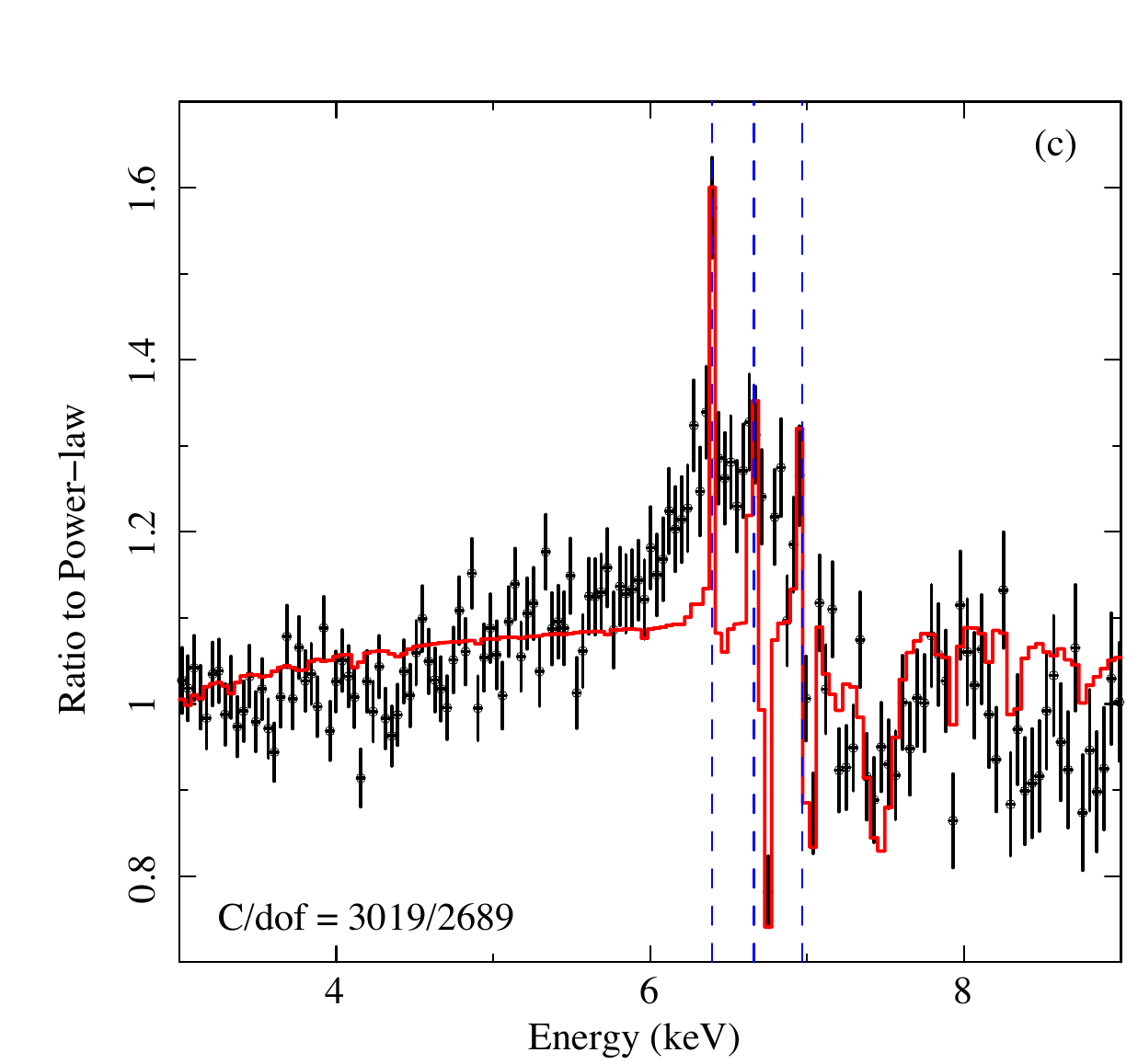}
            \includegraphics[width=0.5\textwidth, trim = 0cm 1cm 7cm 1cm, clip]{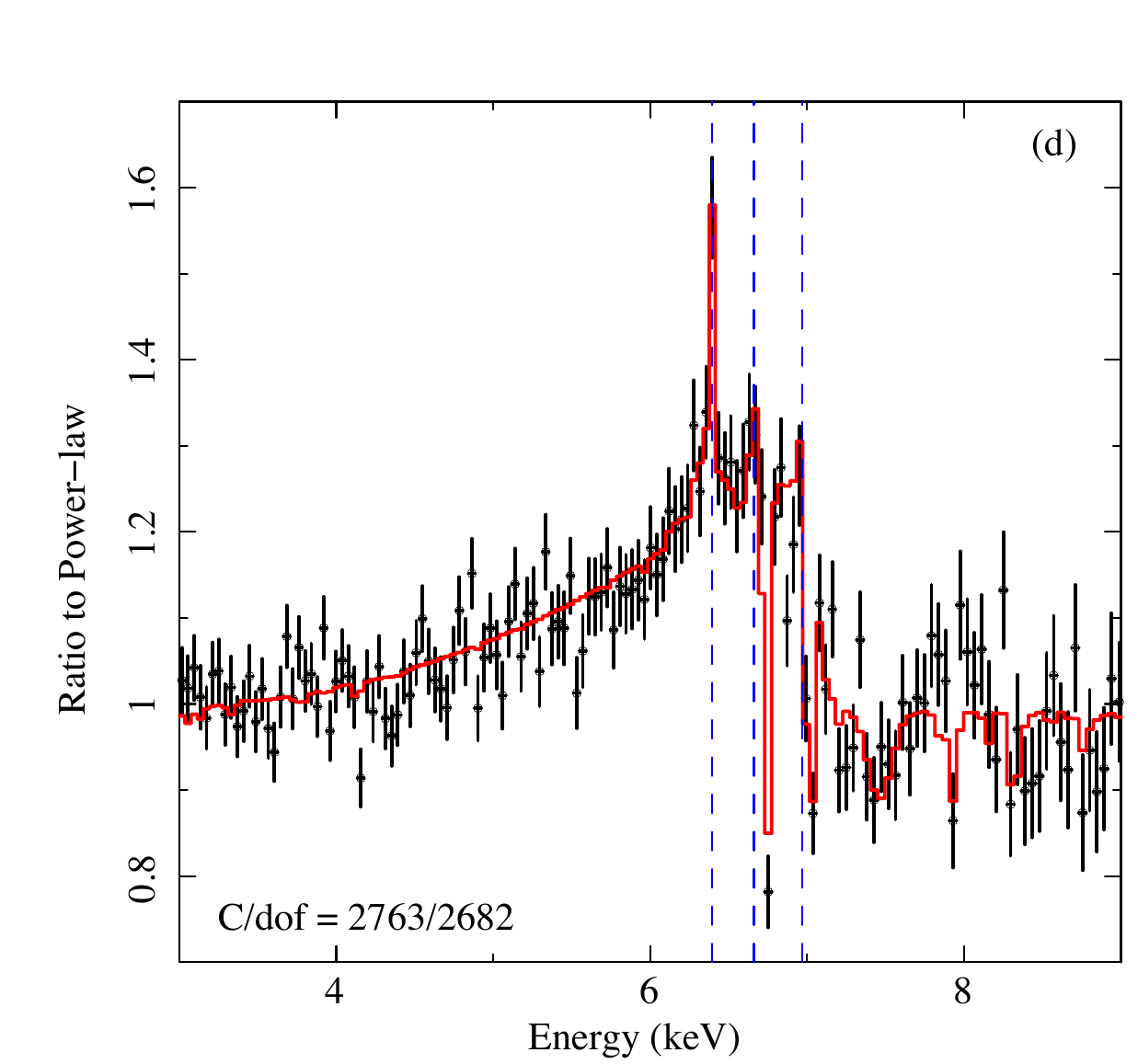}
        }
    }
    \end{center}
    \caption{{\small Ratio of the time-averaged {\it XRISM}/Resolve data to a simple photoabsorbed power-law model, as in Fig.~\ref{fig:rsl_AY_ratplots} (black points).  In each panel, a different, refitted model is plotted in red against the same ratio: (a) the continuum model of the photoabsorbed power-law, as in 
    Fig.~\ref{fig:phpo}; (b) the power-law plus narrow emission, \ie distant neutral reflection and narrow K$\alpha$ lines from Fe\,{\sc xxv} and Fe\,{\sc xxvi}, but no HIO or inner disk reflection; (c) the above model plus the multi-zone HIO; (d) the best-fitting model to the 2024 data, which includes
all of the above components plus relativistic, inner disk reflection via the {\tt relxilllpCp} model.  The continuum and narrow features alone cannot replicate the spectral shape; the broad features of inner disk reflection must be included in order to properly fit the data.  In all panels, the blue hashed vertical lines show, from left to right, the rest-frame positions of the Fe K$\alpha$, Fe\,{\sc xxv} and Fe\,{\sc xxvi} features in emission and/or absorption.}}
\label{fig:rsl_ratplots}
\end{figure}

\begin{figure}
\includegraphics[width=1.0\textwidth]{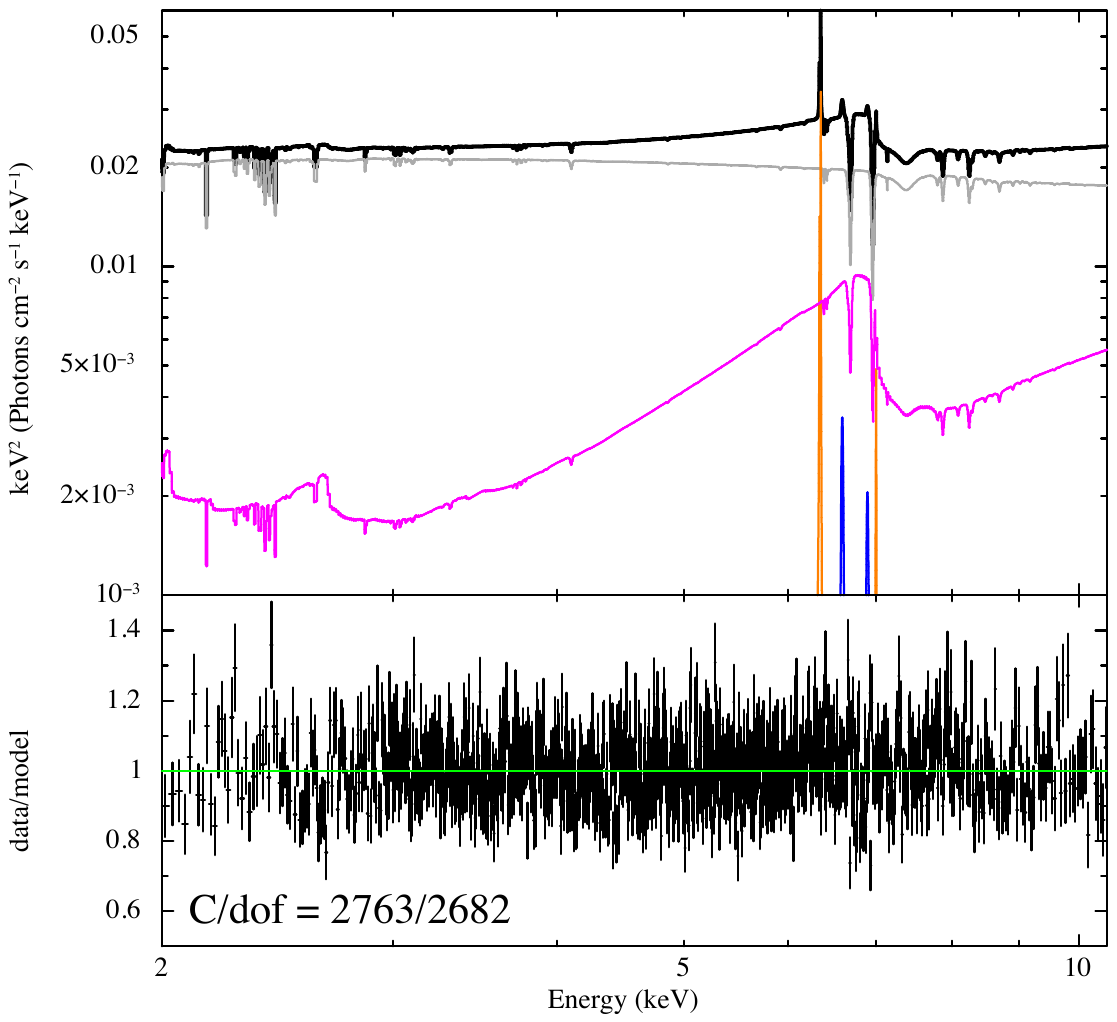}
\caption{{\small {\it Top:} Best-fit model to the $2-11 \keV$ {\it XRISM}/Resolve
    data (total model is in black).  The model consists of a
    three-zone low-ionization warm absorber (two gas zones, one dust zone) plus a two-zone
    high-ionization wind modifying
    the continuum (grey), distant torus reflection via {\tt
      MYtorus} (orange) and inner
    disk reflection via {\tt relxilllpCp} (magenta).  Fe\,{\sc xxv} and Fe\,{\sc xxvi}
    emission lines are also included (blue).  {\it Bottom:} Ratio of the best-fit
    model to the data.  All remaining features are ${<}3\sigma$.}}
\label{fig:rsl_bestfit_eemo_rat}
\end{figure}

The Resolve data ratioed against a power-law continuum and plotted with the best-fitting model described above are
shown in Fig.~\ref{fig:rsl_ratplots}d, with Figs.~\ref{fig:rsl_ratplots}a-c representing the respective contributions from the continuum alone, continuum plus narrow emission, and continuum plus narrow emission and absorption.  Note that {\bf the continuum and narrow features alone cannot replicate the spectral shape; the broad features of inner disk reflection must be included in order to accurately fit the data.}  The final best-fitting model components and
quality of fit across the entire Resolve bandpass are
shown in Fig.~\ref{fig:rsl_bestfit_eemo_rat}, and the corresponding parameter values with $90\%$-confidence uncertainties are listed in Table~\ref{tab:best_fit_tab}.  The final Resolve model syntax is as follows: {\tt constant $\times$ TBabs $\times$ TBvarabs $\times$ WA$_1$ $\times$ WA$_2$ $\times$ HIO$_1$ $\times$ HIO$_2$ $\times$(MYtorus-lines + relxillllpCp + zgauss + zgauss)}.

Notably, {\bf the addition of the HIO components to
the Resolve model does not significantly alter the best-fit parameters of the inner
disk reflection.}  Though Resolve alone cannot statistically constrain the
{\tt relxilllpCp} parameters due to its comparably narrow energy range and
lower effective area
vs. the broadband data, the lack of impact of including this absorption on
the inner disk reflection yields two vital pieces of information: {\bf (1)
{\em XRISM} is uniquely capable of isolating signatures of highly-ionized absorption from the blue wing of the broad Fe K line profile;
(2) both broad Fe K lines {\em and} highly-ionized absorption can, and do,
coexist in AGN.}  

\begin{figure}
\hbox{
  \includegraphics[width=0.5\textwidth, trim = 1cm 0cm 4cm 1cm, clip]{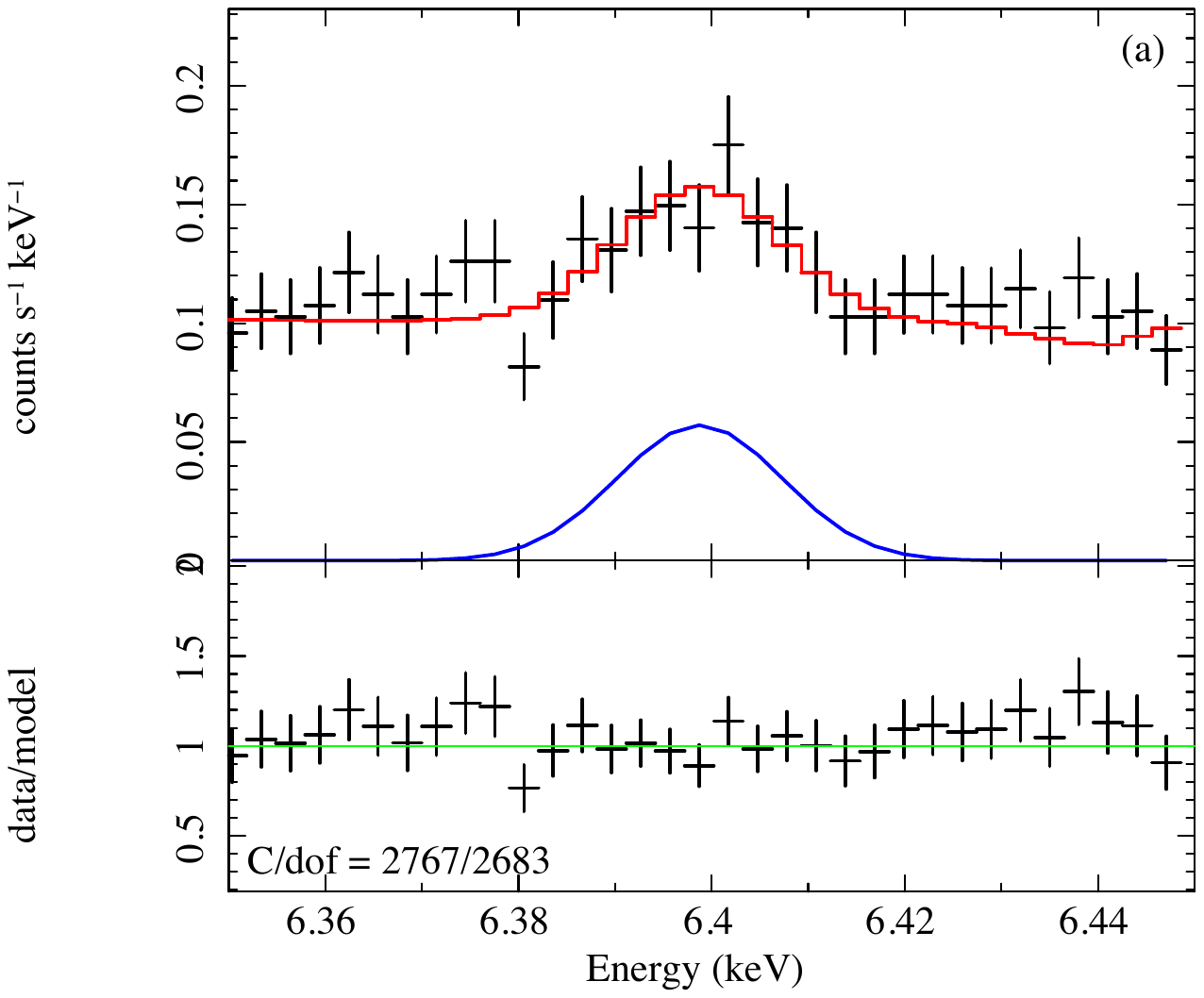}
  \includegraphics[width=0.5\textwidth, trim = 1cm 0cm 4cm 1cm, clip]{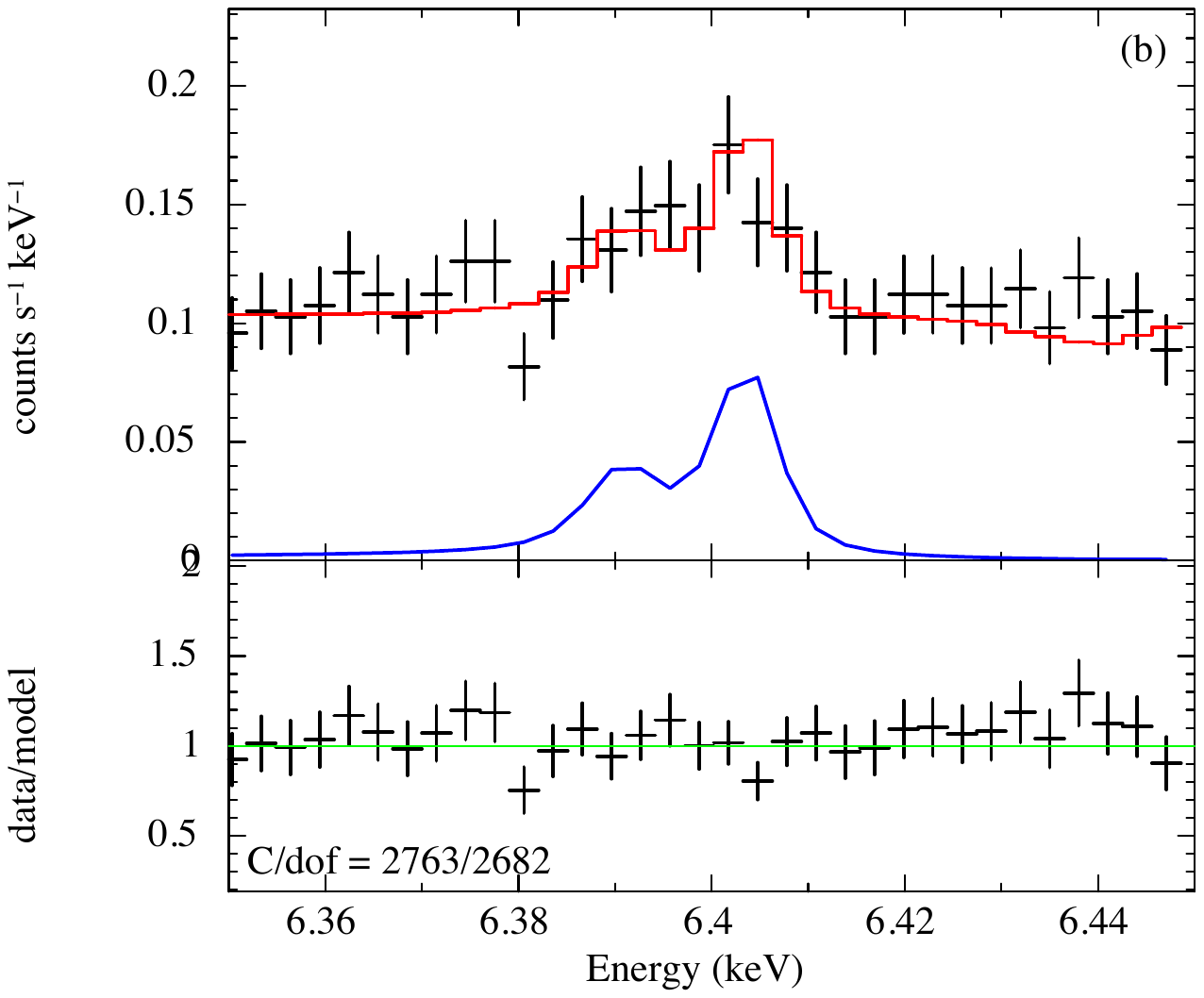}
}
\caption{{\small Comparison of the Resolve data in the narrow Fe K$\alpha$ line region (black points) fit with a
single Gaussian component (left) vs. the {\tt MYtorus} model (right).
{\tt MYtorus} fits the Resolve data slightly better.  The total model is shown in red and the narrow Fe K$\alpha$ line contribution in blue.  Data-to-model ratios are shown in the bottom panel of each plot.}}
\label{fig:myt_v_gauss}
\end{figure}

Historically, the broad Fe K line has been more prominent in the
X-ray spectrum of MCG-6 than its narrow counterpart.  Even so, the
spectral resolution of Resolve provides unique insight into the shape
of the narrow Fe K$\alpha$ feature in emission, as recently reported in
NGC~4151 by \citet{Miller2024}.  Though the lower flux of the narrow
line in MCG-6 prevents us from performing a similarly detailed analysis of
its line shape, we note that the {\tt MYtorus} model, with its Fe
K$\alpha_1$ and K$\alpha_2$ components, provides
a slightly better fit to the narrow Fe K$\alpha$ feature than that obtained
with a simple Gaussian model, as is commonly employed in
lower-resolution data.  Replacing the two Fe K$\alpha$ line
components of {\tt MYtorus} with a single Gaussian results in a marginal worsening of the fit by
${\rm \Delta C/ \Delta dof} = +4/+1$ ($E = 6.399 \pm 0.003 \keV$, $\sigma = 8 \pm
2 \eV$, $EW = 11 \pm 2 \eV$).  This is apparent visually as well (Fig.~\ref{fig:myt_v_gauss}a vs. Fig.~\ref{fig:myt_v_gauss}b).  We also tested a preliminary, high-resolution version of the {\tt xillver} model designed for Resolve data, which has its iron abundance fixed at the solar value (Garcia \etal, in prep.), in place of the {\tt MYtorus} component; this resulted in a marginal improvement in the fit of ${\rm \Delta C/ \Delta dof} = -4/0$ compared with the Gaussian line.  However, in this version of the model we are not yet able to consistently link the {\tt xillver} parameters to those of the {\tt relxilllpCp} component, so we elect not to pursue this model further at this time.

\subsubsection{Joint Spectral Fitting}
\label{sec:joint}

Having characterized the broadband spectral shape with the combination of {\it XRISM}/Xtend, {\it XMM}/pn and {\it NuSTAR}/FPMA and FPMB, then having utilized {\it XRISM}/Resolve to model the HIO and narrow emission features in the Fe K bandpass, we now combine all five time-averaged datasets in a joint spectral analysis.  Our main objective is to simultaneously measure the properties of the continuum and distant reflection, and to get preliminary constraints on the physical properties of all zones of absorption in order to isolate and correctly describe the inner disk reflection.  We begin by assuming the same model used for the best fit to the broadband data with the additional components needed to fit the Resolve data in the Fe K band.  The best-fit parameter values and their $90\%$ uncertainties are shown in Table~\ref{tab:best_fit_tab}.  The joint model syntax is as follows: {\tt constant $\times$ TBabs $\times$ TBvarabs $\times$ WA$_1$ $\times$ WA$_2$ $\times$ HIO$_1$ $\times$ HIO$_2$ $\times$ (MYtorus-lines + relxillllpCp + zgauss + zgauss)}.

{\small
\begin{table}
\hspace{-1.0cm}
\begin{tabular}{|lllll|}\hline\hline
{\bf Component} & {\bf Parameter (units)} & {\bf Xtend+pn+{\em NuSTAR}} & {\bf Resolve} & {\bf All}\\
\hline \hline
{\tt const} (Resolve) &  & --- & $1.017^{+0.004}_{-0.04}$ & $1.017^{+0.004}_{-0.004}$ \\
{\tt const} (Xtend) &  & $1.025^{+0.003}_{-0.003}$ & $1.025^{+0.003}_{-0.003}$ & $1.025^{+0.003}_{-0.003}$ \\
{\tt const} (FPMA) &  & $0.981^{+0.004}_{-0.004}$ & $0.981^{+0.004}_{-0.004}$ & $0.981^{+0.004}_{-0.004}$ \\
{\tt const} (FPMB) &  & $0.982^{+0.004}_{-0.004}$ & $0.982^{+0.004}_{-0.004}$ & $0.982^{+0.004}_{-0.004}$ \\
\hline
{\tt TBabs} & ${\rm log}\,N_{\rm H}\,(\pcmsq)$ & $20.56f$ & $20.56f$ & $20.56f$ \\
{\tt WA}$_1$ & ${\rm log}\,N_{\rm H}\,(\pcmsq)$ & $21.64^{+0.03}_{-0.03}$ & $21.64f$ & $21.54^{+0.02}_{-0.01}$ \\
          & ${\rm log}\,\xi\,(\ergcmps)$ & $1.01^{+0.49}_{-1.01}$ & $1.01f$ & ${\leq}1.05$ \\
          & $v_b\,(\kmps)$ & ${\leq}200*$ & $200f*$ & ${\leq}200*$ \\
          & $v_{\rm out}\,(\kmps)$ & ${\geq}-435*$ & $-435f*$ & ${\geq}-435*$ \\
{\tt WA}$_2$ & ${\rm log}\,N_{\rm H}\,(\pcmsq)$ & $21.85^{+0.03}_{-0.03}$ & $21.85f$ & $22.03^{+0.02}_{-0.02}$ \\
          & ${\rm log}\,\xi\,(\ergcmps)$ & $2.84^{+0.04}_{-0.04}$ & $2.84f$ & $2.81^{+0.04}_{-0.04}$ \\
          & $v_b\,(\kmps)$ & ${\leq}200*$ & $200f*$ & ${\leq}200*$ \\
          & $v_{\rm out}\,(\kmps)$ & ${\geq}-435*$ & $-435f*$ & ${\geq}-435*$ \\
{\tt TBvarabs} & ${\rm log}\,N_{\rm Fe}\,(\pcmsq)$ & $16.78^{+0.36}_{-0.26}$ & $16.78f$ & $17.13^{+0.05}_{-0.04}$ \\
                & $v_{\rm out}\,(\kmps)$ & $0f$ & $0f$ & $0f$ \\
\hline
{\tt HIO}$_1$ & ${\rm log}\,N_{\rm H}\,(\pcmsq)$ & --- & $22.72^{+0.05}_{-0.10}$ & $22.57^{+0.14}_{-0.22}$ \\
          & ${\rm log}\,\xi\,(\ergcmps)$ & --- & $5.01^{+0.06}_{-0.21}$ & $4.75^{+0.24}_{-0.23}$ \\
          & $v_b\,(\kmps)$ & --- & $700^{+150}_{-140}$ & $700^{+150}_{-140}$ \\
          & $v_{\rm out}\,(\kmps)$ & --- & $-2300^{+90}_{-120}$ & $-2310^{+90}_{-120}$ \\
{\tt HIO}$_2$ & ${\rm log}\,N_{\rm H}\,(\pcmsq)$ & --- & $22.72^{+0.22}_{-0.64}$ & $22.87^{+0.25}_{-0.17}$ \\
          & ${\rm log}\,\xi\,(\ergcmps)$ & --- & $5.50^{+0.00}_{-0.70}$ & $5.48^{+0.02}_{-0.38}$ \\
          & $v_b\,(\kmps)$ & --- & $4800^{+1600}_{-1500}$ & $6460^{+1770}_{-2000}$ \\
          & $v_{\rm out}\,(\kmps)$ & --- & $-20100^{+1500}_{-1400}$ & $-19200^{+1120}_{-700}$ \\
\hline
{\tt MYtorus} & ${\rm log}\,N_{\rm H}\,(\pcmsq)$ & $24f$ & $24f$ & $24f$ \\
          & incl$\,(\degmark)$ & $30f$ & $30f$ & $30f$ \\
          & $\Gamma$ & $2.19*$ & $2.19f*$ & $2.11*$ \\
          & $K_{\rm MT}\,(\phpcmsqps)$ & $1.10^{+0.18}_{-0.18} \times 10^{-2}$ & $6.21^{+1.01}_{-0.84} \times 10^{-3}$ & $4.86^{+3.24}_{-1.21} \times 10^{-3}$ \\
\hline
{\tt zgauss} & $E (\keV)$ & --- & $6.66^{+0.05}_{-0.05}$ & $6.70^{+0.02}_{-0.02}$ \\
          & $\sigma (\eV)$ & --- & $12^{+8}_{-4}*$ & $8^{+3}_{-1}*$ \\
          & $K_{\rm line} (\phpcmsqps)$ & --- & $4.00^{+4.65}_{-3.00} \times 10^{-6}$ & $3.92^{+0.94}_{-0.77} \times 10^{-6}$ \\
{\tt zgauss} & $E (\keV)$ & --- & $6.97^{+0.01}_{-0.01}$ & $6.97^{+0.01}_{-0.03}$ \\
          & $\sigma (\eV)$ & --- & $12*$ & $8*$ \\
          & $K_{\rm line} (\phpcmsqps)$ & --- & $1.54^{+0.94}_{-0.77} \times 10^{-5}$ & $4.22^{+0.94}_{-0.77} \times 10^{-6}$ \\          
\hline
{\tt relxilllpCp} & incl$\,(\degmark)$ & $38^{+1}_{-1}$ & $41^{+4}_{-15}$ & $36^{+2}_{-1}$ \\
          & $a$ & ${\geq}0.97$ & ${\geq}0$ & ${\geq}0.65$ \\
          & $h\,(r_{\rm g})$ & $3.22^{+0.10}_{-0.07}$ & $3.80^{+0.35}_{-0.20}$ & $7.59^{+0.62}_{-0.65}$ \\
          & $r_{\rm in}$ & $r_{\rm ISCO}f$ & $r_{\rm ISCO}f$ & $r_{\rm ISCO}f$ \\
          & $r_{\rm out}\,(r_{\rm g})$ & $400f$ & $400f$ & $400f$ \\          
          & $\Gamma$ & $2.19^{+0.03}_{-0.02}$ & $2.19f*$ & $2.11^{+0.01}_{-0.01}$ \\
          & ${\rm log}\,\xi\,(\ergcmps)$ & $1.30^{+0.02}_{-0.04}$ & $1.55^{+1.15}_{-0.05}$ & $1.83^{+0.04}_{-0.02}$ \\
          & ${\rm log}\,n_e\,(\pcmcu)$ & ${\leq}15.02$ & $19.02^{+0.98}_{-0.12}$ & ${\leq}16.1$ \\
          & $A_{\rm Fe}/{\rm solar}$ & $3.53^{+0.11}_{-0.11}$ & $2.97^{+0.98}_{-0.47}$ & $4.45^{+0.04}_{-0.56}$ \\
          & $kT_{\rm e}\,(\keV)$ & $161^{+97}_{-110}$ & $161f$ & $43^{+7}_{-8}$ \\
          & $R$ & $1.20^{+0.31}_{-0.05}$ & $0.40^{+0.02}_{-0.06}$ & $0.94^{+0.13}_{-0.04}$ \\
          & $K_R (\phpcmsqps)$ & $1.15^{+0.22}_{-0.26} \times 10^{-3}$ & $1.14^{+2.21}_{-0.97} \times 10^{-3}$ & $5.29^{+0.41}_{-0.35} \times 10^{-4}$ \\
\hline \hline
Final fit & ${\rm C/dof}$ & $847/540$ & $2763/2682$ & $3603/3222$ \\
\hline \hline
\end{tabular}
\hspace{-1.0cm}
\caption{\small{Best-fit parameters, their values and errors for the model fits to the lower-resolution detectors, Resolve-only, and joint data (Resolve, Xtend, {\it XMM}/pn and {\it NuSTAR}).  Parameters marked
    with an {\it ``f"} are held fixed in the fit, while those marked with an ``*" are
    tied to another parameter (see text).  The column density
    for Galactic absorption \citep[modeled via {\tt TBabs};][]{Wilms2006} is fixed at the value given by weighted average
    from the HI4PI survey \citep{Bekhti2016}.
    The {\tt MYtorus} model for distant reflection and the {\tt
      relxilllpCp} model for primary continuum and inner disk reflection have their photon
    indices tied.  {\tt RelxilllpCp} also
    has its inner disk radius fixed to the ISCO for a given spin.
    Parameters not listed are fixed at their default values.}}
\label{tab:best_fit_tab}
\end{table}
}

The most striking takeaway from the joint fit is that the spin parameter of the SMBH is not very precisely constrained: $a \geq 0.65$.  This is a more precise constraint than that measured from the Resolve data alone ($a \geq 0$), though it is in sharp contrast to the best fit obtained from the broadband CCD-resolution data ($a \geq 0.97$).  The former is not surprising, given the limited bandpass and S/N of the Resolve data, but it is interesting that the addition of these data to the broadband data resulted in a less precise spin constraint than that measured from the broadband data alone.  A likely reason is the AGN's variability during the observation: because the flux varies by ${\sim}3\times$, the time-averaged data effectively represent the superposition of several spectral states.  The Resolve data is most sensitive to this variability in the $2-11 \keV$ band, as changes in the continuum during the observation will have consequences for the morphologies of the residual broad and narrow lines in the Fe K region that other model components are parameterizing.  These line shape changes would not be detectable with the lower-resolution instruments in the joint fit.
Forcing the model to jointly fit all of the data might therefore be expected to result in a fit that ``splits the difference" between the Resolve fit and the lower-resolution broadband fit.  

\begin{figure}
\includegraphics[width=1.0\textwidth]{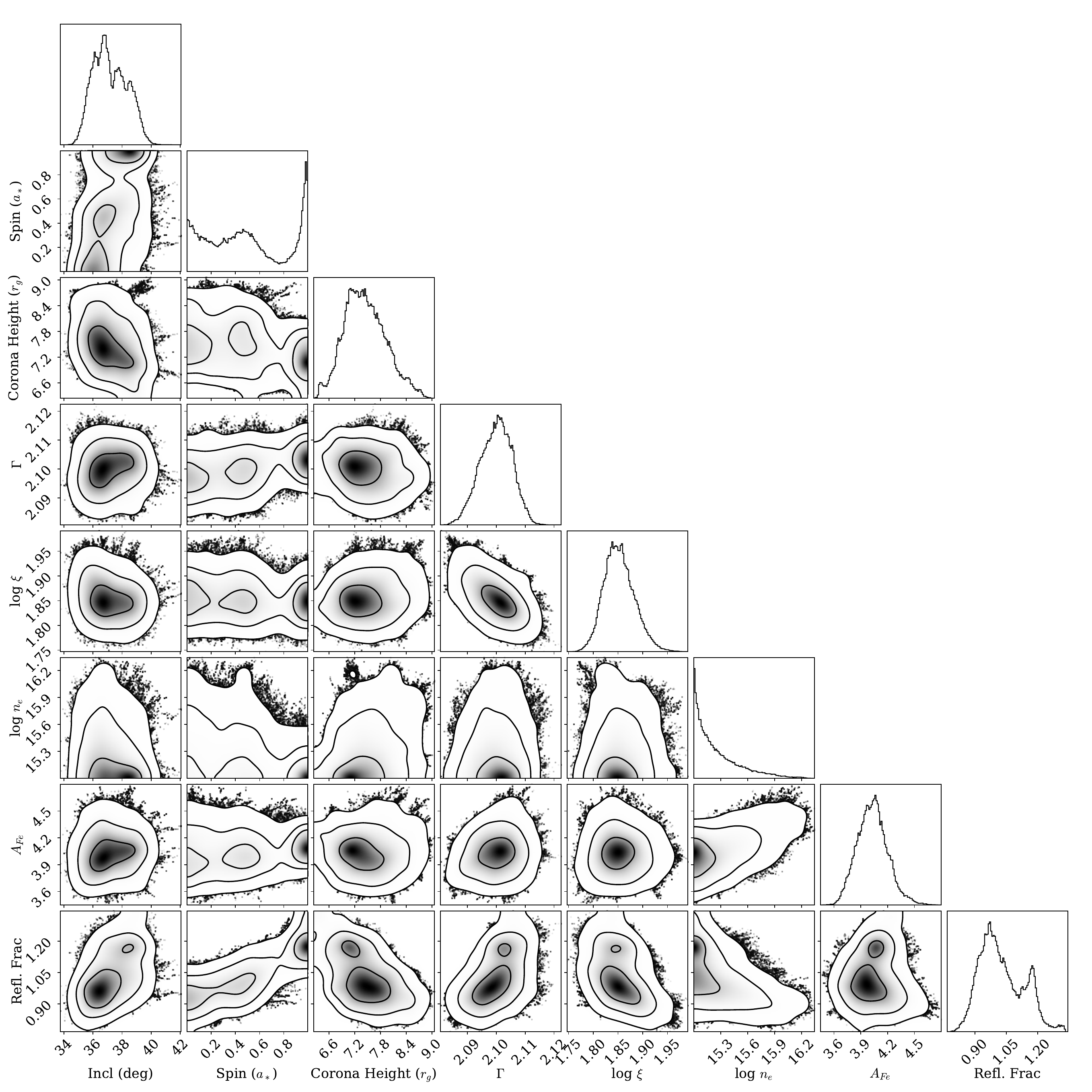}
\caption{{\small Corner plot showing the posterior distributions for the free parameters in the {\tt relxilllpCp} inner disk reflection component of the best-fitting model for the joint fit to the Resolve, Xtend, pn, FPMA and FPMB data.  Single parameter distributions are shown on the ends (y-axes show probability density), as well as plots showing the $1\sigma$, $2\sigma$ and $3\sigma$ contours for the posterior distributions of parameter pairs.}}
\label{fig:corner}
\end{figure}

To probe this hypothesis and to examine the parameter space more thoroughly, we employ a Goodman-Weare Markov Chain Monte Carlo (MCMC) algorithm within {\sc xspec}.  This approach ensures that we settle on global minima for all model parameters rather than getting stuck in local minima.  We run the MCMC algorithm with 100 walkers for $2.5 \times 10^6$ steps, burning (\ie discarding) the first $10^5$ steps to confirm that the parts of the chains we employ have reached equilibrium and do not retain bias due to their starting positions.  The MCMC posterior distributions for each free parameter in the inner disk reflection model, as well as contours of parameter pairs, are shown in a ``corner plot" (Fig.~\ref{fig:corner}).  

Reinforcing the spectral fitting results described above, the spin posterior distribution is not a Gaussian function with a single peak, as it should be if the model describes the time-averaged data accurately.  There is a statistical preference for high prograde values, in keeping with previously published results for MCG-6, but low prograde values are not decisively ruled out.  Likewise, there are multiple peaks in the posterior distributions of two related parameters within the {\tt relxilllpCp} model: the disk inclination angle (Incl) and the fraction of photons reflected from the inner disk (Refl. Frac.).  The results for the spin and disk inclination, particularly, are telling: we know that these values cannot be changing during the course of the observation.  When the corona varies within an observation but only time-averaged data is considered, degeneracies can appear between model parameters representing physical properties of the system that are driven by the corona and those that are inherent to the inner disk/SMBH.  An example is the assumed emissivity profile of the disk (which here we are parameterizing as the height of a point source corona) and the SMBH spin.  If the inner disk is under-illuminated relative to the assumed emissivity profile during part of the observation (\eg if the corona is at a larger height, or is accelerated/beamed away from the inner disk), the model can compensate for this by truncating the inner disk, hence fitting lower values of the black hole spin \citep{Fabian2012,Fabian2014}.  We can resolve this degeneracy in by fitting the time-resolved spectrum, knowing that the spin parameter should not change between the time intervals.  A forthcoming time-resolved spectral analysis will take the spectral variability during the observation into account, linking parameters such as disk inclination and SMBH spin between time intervals in order to obtain precise constraints on these parameters (Wilkins \etal, in prep.). 

Another interesting takeaway from the joint spectral modeling is that no separate component of ``soft excess" emission is needed in order to obtain an adequate fit to the data, in spite of the broadband spectral shape of the source being remarkably similar to the long {\it XMM} observations of 2001 \citep[as shown in, \eg][]{Fabian2002, BR2006} and 2013 \citep{Marinucci2014}.  In contrast to the analysis of \citet{BR2006}, but aligned with the analysis of \citet{Marinucci2014}, the data from the 2024 campaign are well described by only a compact corona illuminating an optically-thick disk that extends down to the ISCO around a rapidly-rotating SMBH, modified by multiple components of absorbing gas along the line of sight.  There is no indication of any additional or extended emission component ${<}2 \keV$; the relativistic reflection model accounts for all of the soft excess present in MCG-6 in a manner that is consistent with the reflection features detected in the higher-energy parts of the spectrum.  There are two likely contributing factors to this difference in model fitting despite the similarity of the spectra: (1) the addition of the {\it NuSTAR} data allows for a more holistic, accurate parametrization of the continuum and reflection components, which impacts the model fit at low energies as well; (2) the {\tt relxilllpCp} model provides a more accurate description of the relativistic reflection component than the older {\tt kerrconv}$\times${\tt reflion} model employed by \citet{BR2006}.

\begin{figure}
\centering
\subfloat{%
    \includegraphics[width=0.7\textwidth, trim = 1cm 0cm 0cm 0cm, clip]{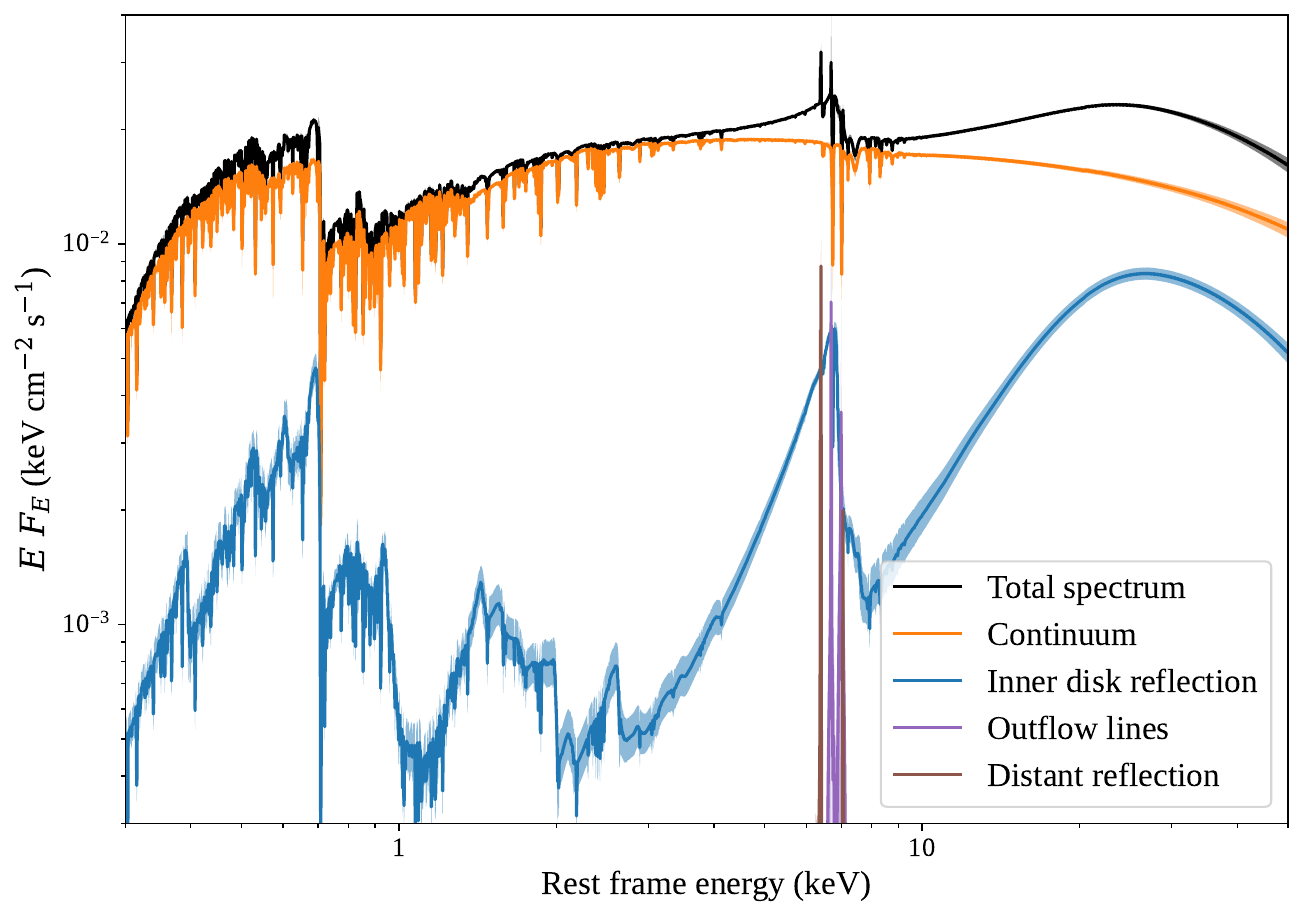}
}
\\
\subfloat{%
    \includegraphics[width=0.8\textwidth, trim = 0cm 0cm 0cm 12.5cm, clip]{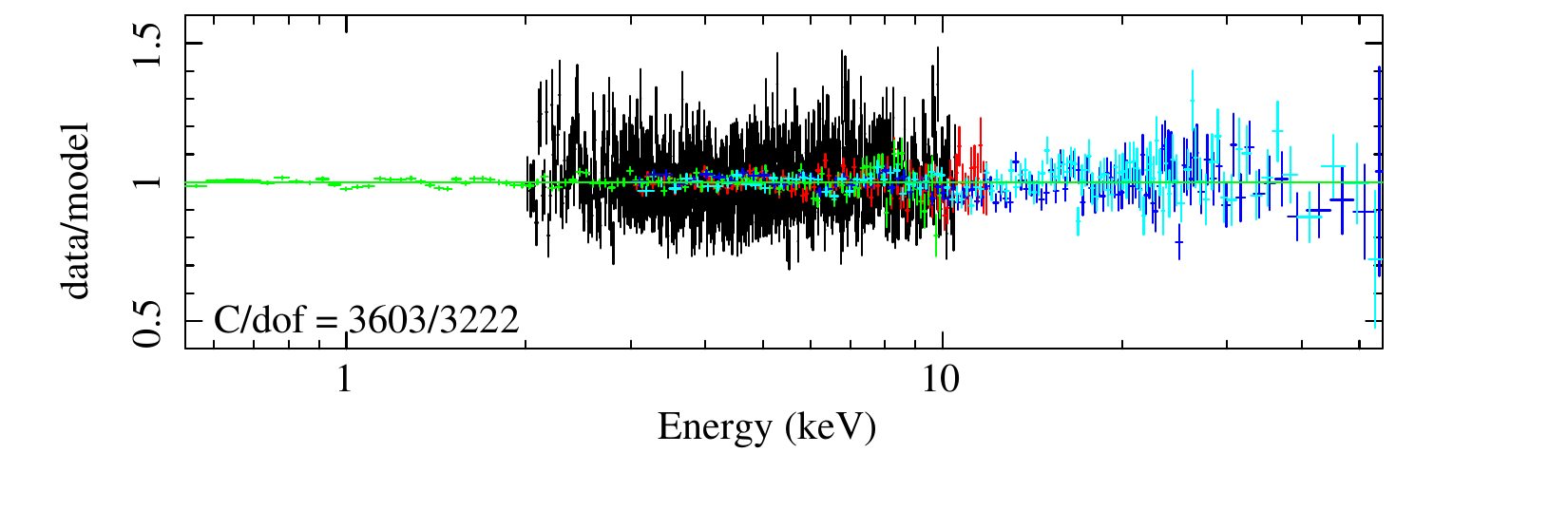}
}
\caption{{\small{\it Top:} Best-fit model to the joint analysis of the Resolve, Xtend, pn, FPMA and FPMB data from $0.5-55 \keV$.  Mean and $1\sigma$-confidence intervals for each of the model components derived from the MCMC are shown, estimated from 10,000 realizations of the spectrum drawn from the MCMC posteriors.  The model consists of a
    three-zone low-ionization warm absorber (two gas zones, one dust zone) plus a two-zone
    high-ionization wind modifying
    the continuum, distant torus reflection via {\tt
      MYtorus} and inner
    disk reflection via {\tt relxilllpCp}.  Fe\,{\sc xxv} and Fe\,{\sc xxvi}
    emission lines, likely re-emission from the outflowing wind, are also detected.  {\it Bottom:} Ratio of the best-fit
    joint model to the data.  {\it XRISM}/Resolve is in black, {\it XRISM}/Xtend is in red, {\it XMM}/pn is in green, {\it
  NuSTAR}/FPMA is in dark blue, and {\it NuSTAR}/FPMB is in light
    blue.}}
\label{fig:5det_bestfit_eemo}
\end{figure}

\begin{figure}
\includegraphics[width=0.25\textwidth, trim = 0cm 0cm 10cm 0cm]{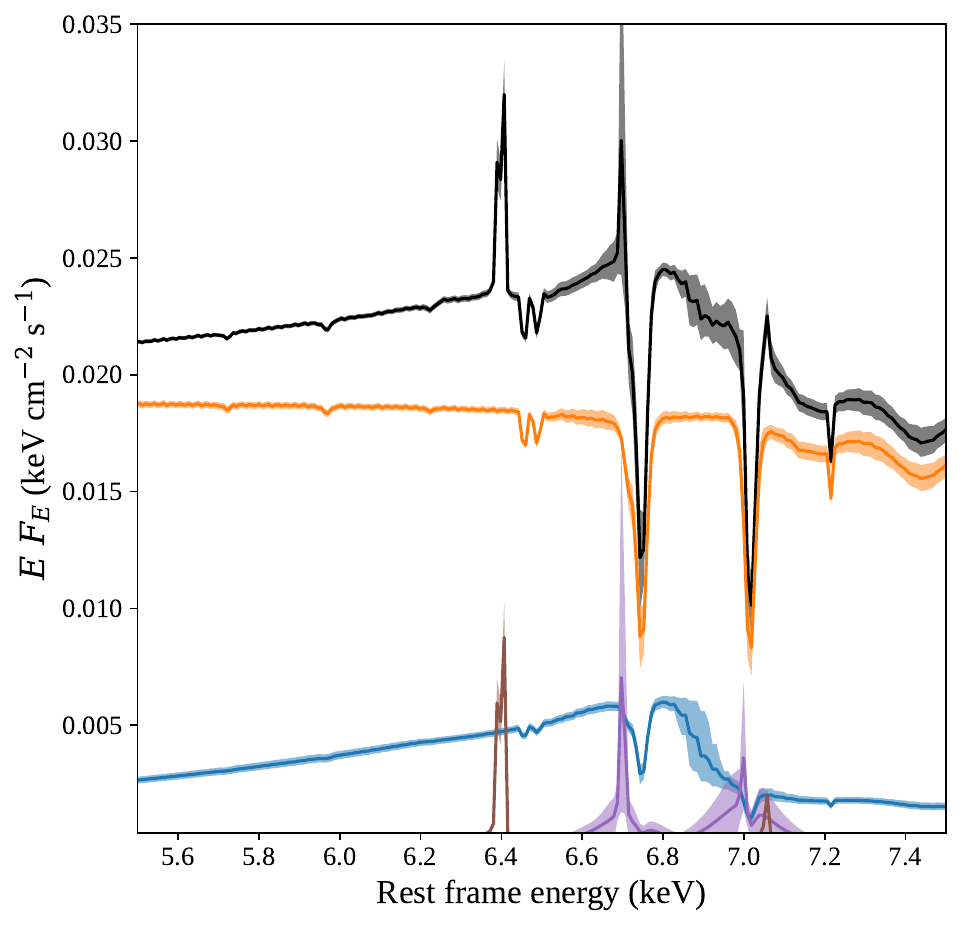}
\caption{{\small Zoom-in of Fig.~\ref{fig:5det_bestfit_eemo} (top) showing the Fe K band with the mean and $1\sigma$-confidence intervals for each of the components of the best-fit model derived from the MCMC.}}
\label{fig:5det_bestfit_eemo_zoom}
\end{figure}

The final best-fit model to the time-averaged data is shown in Fig.~\ref{fig:5det_bestfit_eemo}, which results in a goodness-of-fit of ${\rm C/dof} = 3603/3222$.  A zoom-in of the Fe K band is shown in Fig.~\ref{fig:5det_bestfit_eemo_zoom}. 

\section{Discussion}
\label{sec:discussion}

Though the results of our spectral modeling of the time-averaged data will be further refined by time-resolved analysis (Wilkins \etal, in prep.) as well as detailed modeling of the soft band data using {\it XMM}/RGS (Ogorza{\l}ek \etal, in prep.), our time-averaged spectral analysis of the 2024 observing campaign on MCG-6 has revealed several unique insights that merit further exploration.  We pursue these in the following subsections.

\subsection{Robustness of Relativistic Reflection and SMBH Spin Measurements}
\label{sec:rel_refl}

As discussed in \S\ref{sec:intro}, the inability of CCD-resolution instruments to deconvolve the broad and narrow emission and absorption lines in the Fe K bandpass has created degeneracies between the relativistic reflection model and the absorption-dominated model for AGN with reported broad Fe K$\alpha$ lines, such as MCG-6.  The advent of the Resolve micro-calorimeter opens a new avenue for breaking these modeling degeneracies, particularly when used simultaneously with broadband, higher-S/N X-ray data, as from {\it XMM}/pn and {\it NuSTAR}.

The Resolve data for MCG-6 definitively characterize the narrow emission and absorption lines as well as the signatures of relativistic reflection in the ${\sim}6-8 \keV$ band for the first time.  The narrow Fe K$\alpha$ feature from distant reflection (Gaussian $\sigma = 8 \pm 2 \eV$) is easily separated from the broader component (Gaussian $\sigma = 2785 \pm 173 \eV$, though properly accounting for the skewed, asymmetrical shape of the broad component is necessary in order to obtain a good statistical fit to the Resolve data), and the latter is unmistakably detected at ${>}5\sigma$ (Fig.~\ref{fig:rsl_delchi}a).  Removing the relativistic reflection component from the best-fit model and refitting worsens the overall fit by $\Delta{\rm C}/\Delta{\rm dof} = +485/+1$, quantifying the importance of inner disk reflection in the Resolve spectrum (see Fig.~\ref{fig:rsl_ratplots} for an illustration of the importance of the broad Fe K line to the Resolve fit).  The degradation in fit without inner disk reflection is even more stark in the joint spectral analysis of all five datasets: $\Delta{\rm C}/\Delta{\rm dof} = +1740/+1$.  

Individual absorption lines from multiple kinematic components of Fe\,{\sc xxv} and Fe\,{\sc xxvi} are also isolated and accurately modeled by Resolve with greater spectral resolution in a fraction of the exposure time required by previous observations with {\it Chandra}/HETG.  Not only does this facilitate a precise characterization of these features, but it also enables changes in the properties of the absorbing gas to be tracked in response to continuum flux variability during the {\it XRISM} observation (Ogorza{\l}ek \etal, in prep.; Rogantini \etal, in prep.)

In our time-averaged spectral analysis, {\bf we find that the inclusion of the multi-zone HIO does not significantly affect the best-fit model parameters of the inner disk reflection component.  Though the overall fit to the Resolve data worsens by $\Delta{\rm C}/\Delta{\rm dof} = +154/+8$ when these two zones of absorption are removed and the model is refit, the parameter values of the {\tt relxilllpCp} component remain unchanged within their respective uncertainties.}  This is because Resolve unambiguously separates narrow absorption and emission features from the underlying broad line profile, rather than blending the broad and narrow line components together, as happens at CCD resolution.  Interestingly, removing the two low-ionization absorbers from the best-fit model to the Resolve data also does not yield a significant change in the {\tt relxilllpCp} parameters.  This seems counterintuitive, given that modeling the spectral curvature correctly ${<}3 \keV$ is critical for isolating the SMBH-spin-encoding red wing of the broad Fe K$\alpha$ line \citep{BR2006,Brenneman2013}, but with the gate valve closed, Resolve loses effective area rapidly below $3 \keV$.  A better test of the importance of the low-ionization absorption is to remove these components from the joint fit to all five datasets.  Doing so yields an overall fit that is worsened by $\Delta{\rm C}/\Delta{\rm dof} = +1139/+6$ and results in several significant changes to the coronal properties and inner disk reflection parameters.  Most importantly, this yields unphysical values of the accretion disk density and iron abundance (both pegged at the top of their ranges: ${\rm log}\,n_e = 20 \pcmcu$ and $A_{\rm Fe} = 10\times$ solar), as well as an unconstrained SMBH spin parameter.  These changes are driven by the high count rate and corresponding high S/N in the soft part of the spectrum in the {\it XMM}/pn data.

\begin{figure}
\hbox{
  \includegraphics[width=0.5\textwidth, trim = 1cm 0cm 4cm 1cm, clip]{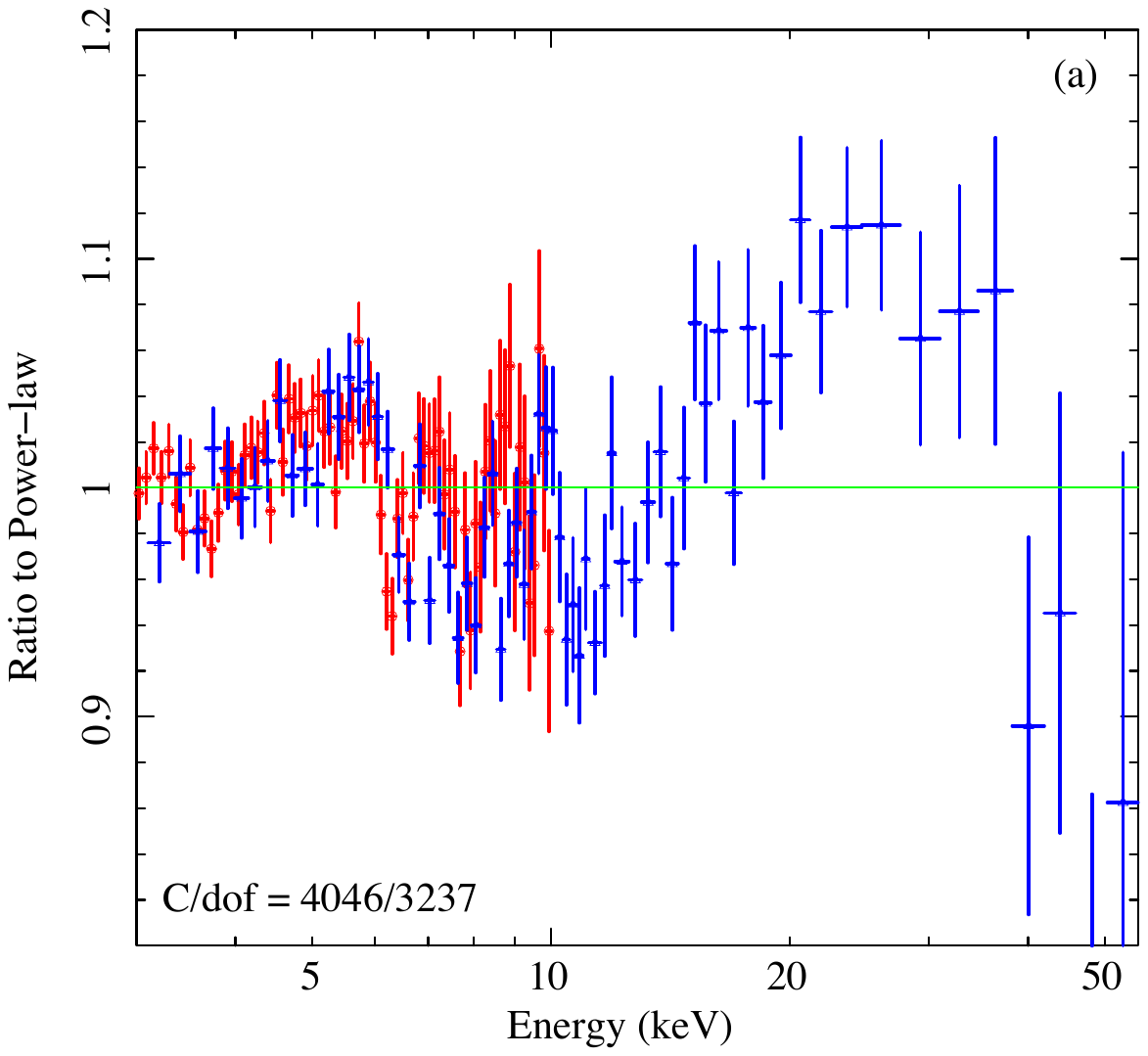}
  \includegraphics[width=0.5\textwidth, trim = 1cm 0cm 4cm 1cm, clip]{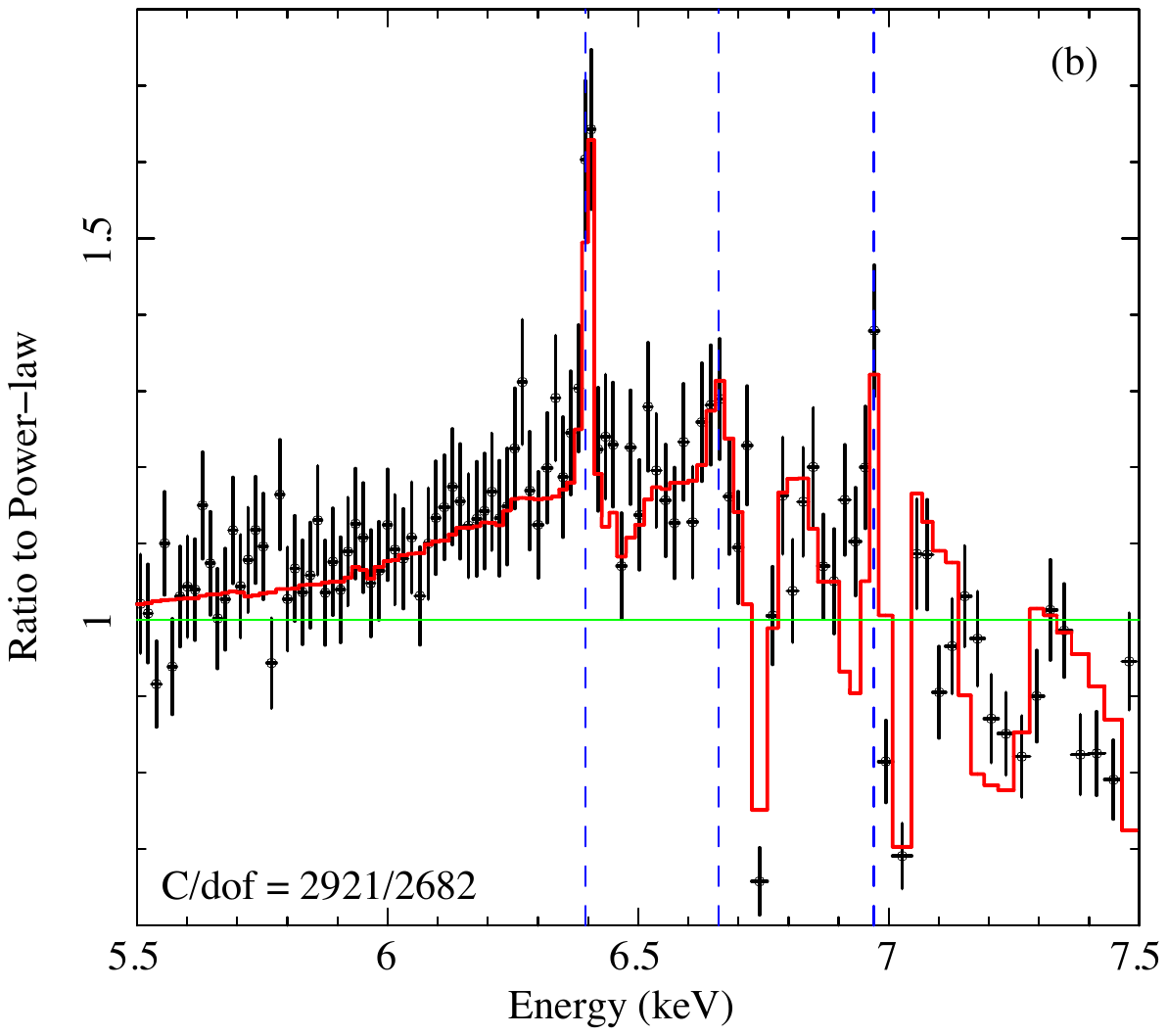}
}
\caption{{\small {\it Left:} Broadband data-to-model ratio showing the poor fit of the disk-wind model of \citet{Matzeu2022} to the joint dataset.  For visual clarity, only the {\it XMM}/pn (red points) and {\it NuSTAR}/FPMB data (blue points) are shown.  The model is clearly inadequate in the Fe K band and ${\geq}10 \keV$.  {\it Right:} Zoom-in of the disk-wind model fit to the Resolve data, which are shown as ratio to a power-law as in Fig.~\ref{fig:rsl_AY_ratplots} (black points).  The model fit begins to worsen ${\geq}7 \keV$, coincident with where the blue wing of the broad Fe K$\alpha$ line from relativistic reflection would appear.  The blue hashed vertical lines show, from left to right, the rest-frame positions of the Fe K$\alpha$, Fe\,{\sc xxv} and Fe\,{\sc xxvi} features in emission and/or absorption.}}
\label{fig:disk_wind}
\end{figure}

The quality of our joint dataset can also test whether absorption-dominated models (\S\ref{sec:intro}) can achieve statistical fits to the data that are comparable with those obtained from inner disk reflection models.  To investigate this question, we replace the {\tt relxilllpCp} component with the ``disk-wind" model of \citet{Matzeu2022} in the joint spectral analysis.  This model improves upon the work of \citet{Sim2008,Sim2010a,Sim2010b} in representing the absorption and scattering of continuum photons by outflowing gas in winds originating from the accretion disk.
We find that the best fit we are able to obtain for the broadband data is worse than that of the inner disk reflection model by $\Delta{\rm C}/\Delta{\rm dof} = +710/+7$, with strong residual features in both the Fe K band and ${>}10 \keV$ where the model does not adequately fit the data, even when the {\tt MYtorus} scattered component is added back in (Fig.~\ref{fig:disk_wind}a).  The fit to the Resolve data alone is also significantly degraded: $\Delta{\rm C}/\Delta{\rm dof} = +158/+7$.  This is most apparent ${\geq}7 \keV$ (Fig.~\ref{fig:disk_wind}b), where the reflection model is most sensitive to the inclination angle of the disk in the blue wing of the Fe K$\alpha$ line.  The disk-wind model does not fit the data as well in this region, meaning that the broad emission feature in the data is not as well replicated by electron scattering from a disk wind as it is by relativistic reflection from the innermost disk.  The best-fit parameters of this wind model feature a launching radius of $r=64\,r_{\rm g}$, mass loss rate \textit{\.m}$ = 4\%$ of the Eddington value, illuminating X-ray luminosity $L_X = 8\%$ of the Eddington value, terminal velocity $v_{\rm out}=-0.06c$ and observer's inclination angle $i=55\degmark$.  Though most of these parameters are broadly consistent with those observed in other aspects of our model fitting in \S\ref{sec:joint}, the inclination angle is substantially higher than what we measure using the inner disk reflection model (which is consistent with values reported previously in the literature), and is also seemingly at odds with MCG-6's classification as a NLSy1 AGN.    

We therefore conclude that inner disk reflection is a necessary component of the X-ray spectrum of MCG-6.  We also conclude that accurate measurements of the properties of the inner accretion disk and SMBH spin can only be obtained by properly characterizing both the discrete spectral features in the Fe K band \textbf{\em and} the spectral curvature induced by absorption from low-ionization gas.  Achieving both aims simultaneously is the only way to definitively isolate the underlying signatures of inner disk reflection.  

A natural consequence of this statement is to question the reliability of previously-published SMBH spin measurements made without the diagnostic benefits of Resolve.  It is therefore interesting that our joint, five-dataset spin measurement for MCG-6 is consistent with those published using reflection modeling of only lower-resolution data \citep[\eg][]{BR2006,Marinucci2014}, even if the posterior distribution is complex for the best fit to the time-averaged data.  There are likely multiple factors that contribute to this finding: 
\begin{itemize}
    \item{The relative prominence of inner disk reflection and breadth of the Fe K$\alpha$ line in MCG-6 vs. other Sy 1 AGN \citep[\eg NGC~4151,][]{Miller2024} makes the SMBH spin easier to measure, even with limited spectral resolution;}
    \item{Gratings data ${<}3 \keV$ from {\it Chandra} and {\it XMM}/RGS used in some previous analyses enabled accurate modeling of the low-ionization absorption components, which can have a significant impact on isolating the red wing of the broad Fe K$\alpha$ line and measuring spin;}
    \item{Careful modeling of broadband, simultaneous data from multiple X-ray telescopes has helped to reliably constrain the properties of the continuum, and multi-epoch spectral analysis based on large archives of data has provided additional constraints on the distant reflection, as this component varies on timescales longer than the inner disk reflection.}
\end{itemize}

It will be important to investigate whether the SMBH spin measurements made in other AGN using {\it XRISM} (particularly in tandem with {\it NuSTAR} and/or {\it XMM}) differ from those obtained using CCD-resolution data alone.    
The HIO lines in the Fe K band in MCG-6 are rather narrow, with moderate column density and small equivalent width, and thus do not significantly impact the spectral fitting of lower-resolution data for this source.  AGN that harbor higher-column HIOs with similar kinematic profiles will likely be more affected, as these absorption lines will appear to remove flux from the relativistic Fe K$\alpha$ line at lower resolution.
Likewise, for spin measurements made during the {\it XRISM} era while the Resolve gate valve remains closed, there is great benefit to obtaining simultaneous data from low-energy X-ray gratings instruments and/or broadband X-ray data from instruments with CCD-like spectral resolution. 

\subsection{Does the Distant Reflection Originate in a Torus?}
\label{sec:torus}

As noted in \S\ref{sec:resolve}, the narrow Fe K$\alpha$ emission feature in MCG-6 is significantly less prominent than the broad emission line.  When both narrow and broad lines in the Resolve data are fit with Gaussian components, their respective equivalent widths are $EW_n = (11 \pm 2) \eV$ and $EW_b = (1044 \pm 101) \eV$.  Fitting the broad line with a more accurate morphology via {\tt relline} ($EW_b = 569 \pm 24 \eV$), its strength relative to the continuum is ${>}50\times$ that of the narrow line.  

In the unified model of AGN \citep{Antonucci1993}, the observer's line of sight to or through the putative molecular ``torus" determines the relative prominence of the narrow and broad Fe K$\alpha$ lines.  Stronger narrow lines are detected in AGN with a preponderance of optically-thick, cold gas and dust along the line of sight at large distances from the central engine, obstructing the view toward the innermost regions of the system and scattering the continuum emission into distant reflection features.  Stronger broad lines are seen in AGN where the line of sight to the inner accretion disk and corona does not pass through as much cold gas and/or dust.  Assuming that the planes of the inner accretion disk and torus are aligned, the relative strengths of the lines in MCG-6 imply that our line of sight to the inner disk does not significantly intersect the torus.  This is supported by the measurement of relatively low disk inclination in the joint fit to all five datasets ($i = 36\degmark$).  

Given that the basic qualitative premise of the unified model seems to be satisfied in MCG-6, the puzzle is why we find a slightly better statistical fit without the scattered continuum of distant reflection in {\tt MYtorus} than we do with this component included.  We first noted this oddity in \S\ref{sec:broadband}, but the joint fit to all five datasets confirms that the fit with the scattered component is worse by $\Delta{\rm C}/\Delta{\rm dof} = +13/0$ (the degrees of freedom do not change, as the three components of the {\tt MYtorus} models are linked).  Early attempts by \citet{GF1991} to holistically model the narrow Fe K$\alpha$ line and its associated Compton hump predict that $EW_n \sim 125 \eV$ for a neutral slab of gas inclined at $i = 35 \degmark$ to the line of sight and irradiated by a power-law with $\Gamma = 2.1$, as is the case in MCG-6.  This is ${>}10\times$ stronger than the narrow feature that we actually observe in the Resolve data: $EW_n = (11 \pm 2) \eV$.  The authors also predict that the albedo of the reflected continuum at $30 \keV$ should exceed that of the $6.4 \keV$ line by ${\sim}7\times$, implying a strong Compton hump, but even reducing the strength of this feature by a factor of ${\sim}10\times$ to match the narrow Fe K$\alpha$ line, we do not detect this scattered continuum from distant material significantly in our data.  The relative strength of this feature could be lessened by several physical factors: \eg a highly inclined angle of emission of the scattered photons away from the line of sight, a low ($N_{\rm H} \ll 10^{24} \pcmsq$) column density of the scattering medium, significant outflow of the corona away from the disk (as in the case of a failed jet, or base of a jet).  The first explanation is inconsistent with the small measured inclination angle of the inner disk, if we assume that it is aligned with the outer disk/torus.  The second explanation would also impact the narrow Fe K$\alpha$ feature, yet we detect it in the data.  Finally, when left free to vary, the parameter describing the coronal outflow within {\tt relxilllpCp} is consistent with zero, and MCG-6 has never been known to harbor a jet.  

\citet{GF1991} find that increasing the iron abundance decreases the prominence of the Compton hump vs. the narrow Fe K$\alpha$ feature by a factor of several.  Our best-fit iron abundance for the inner disk reflection component in the joint fit is $A_{\rm Fe} \sim 4\times$ solar, which is consistent with results reported in more than a dozen AGN over the past two decades \citep[][and references therein]{Garcia2018}.  However, the {\tt MYtorus} model has the iron abundance hardwired at the solar value.  An explanation for the relative weakness of the distant reflection continuum vs. the narrow Fe K$\alpha$ feature could then be that the distant reflector is also comprised of gas and/or dust with super-solar iron abundance.  We test this theory by replacing the {\tt MYtorus} component with a {\tt pexmon} component \citep{Nandra2007}, which relies on out-of-date atomic information and assumes a plane-parallel slab rather than a torus, but offers the ability to vary the iron abundance of the reflector.  With {\tt pexmon}, the joint fit to all five datasets is worsened by $\Delta{\rm C}/\Delta{\rm dof} = +61/-3$.  The iron abundance of the distant reflector finds an unphysically-high best-fit value of $A_{\rm Fe} = 11\times$ solar and is unconstrained.  If super-solar iron abundance is a contributing factor in the weakness of the scattered continuum of the distant reflector, this must be explored with more advanced models and higher S/N data, likely in different AGN with more prominent distant reflection components.

Another potential contributing factor is that the distant reflection could originate not in a traditional torus, as represented by the {\tt MYtorus} model, but in a non-Compton-thick and/or non-uniform medium, such as circumnuclear clouds \citep[\eg][]{Balokovic2018,Buchner2021} or a ``warm wind" \citep{Elvis2000}.  If the scattering medium is relatively uniform, the signatures of distant reflection would be expected to display spectral variability on timescales consistent with the light-crossing time between the central engine and the medium.  While any variability of the narrow Fe K$\alpha$ feature in these 2024 data will be the subject of later papers (Wilkins \etal, in prep.; Rogantini \etal, in prep.), the time-averaged Resolve data when fit with a Gaussian component show a line width of $\sigma = (8 \pm 2) \eV$.  This corresponds to a Keplerian velocity of ${\sim}215 \kmps$ if the line is emitted from a disk-like structure with an inclination of $i = 36 \degmark$ to the line of sight, implying an origin at ${>}1.9 \times 10^6\,r_{\rm g}$ from the SMBH.  The light-crossing time in this scenario would be nearly six months, making variability of the distant reflection detectable between observations, but not within them.  If variability of the narrow Fe K$\alpha$ feature is observed on timescales significantly shorter than this, it likely implies an origin in a non-uniform structure where the spectral variations correspond to changing covering fraction and/or column density of the medium \citep[\eg as in NGC~4151, where the narrow Fe K$\alpha$ line is theorized to originate in BLR clouds, per][]{Zoghbi2019}.  The non-uniform torus scenario could also potentially be investigated using infrared observations with {\it JWST} as a complement to longer X-ray observations with {\it XRISM} and {\it NuSTAR}.

\section{Conclusions}
\label{sec:conclusions}

We have analyzed the time-averaged spectra of {\it XRISM}, {\it XMM}/pn and {\it NuSTAR} from the 2024 observing campaign on MCG-6.  This is the richest X-ray dataset ever obtained for this archetypal NLSy1 AGN.  We summarize our principal conclusions below.
\begin{enumerate}
\item{The signatures of relativistic reflection that have long been cited in MCG-6 with previous instruments are also seen in the {\it XRISM} data.  The revolutionary ${<}5$-eV spectral resolution of the Resolve micro-calorimeter confirms that a strong, broad, asymmetric Fe K$\alpha$ line is required in order to obtain an adequate model fit.  Resolve cleanly separates the broad line from narrow emission and absorption lines in the Fe K band for the first time, and the parameters of the relativistic reflection model are not impacted by the modeling of the high-ionization absorption components.  A disk-driven wind model for this feature is statistically ruled out.}
\item{The best-fitting model for the joint fit to all five datasets is very well described by a reflection from the inner accretion disk illuminated by a compact corona and modified by multi-zone ionized absorption from an outflowing wind along the line of sight.  No additional component is needed in order to reproduce the shape of the spectrum at soft energies (\ie there is no ``soft excess" beyond what is included through relativistic reflection).} 
\item{Though the MCMC-based posterior distribution of the SMBH spin parameter peaks at high prograde spin values, consistent with previously published measurements, the AGN's spectral variability during the observing campaign yields a complex parameter space that represents the superposition of multiple spectral states.  A forthcoming time-resolved spectral analysis will present a much more precise SMBH spin constraint by taking the variability into account.}
\item{While reflection from distant material is required in order to reproduce the observed narrow Fe K$\alpha$ feature at $E = 6.4 \keV$, the contribution from relativistic, inner disk reflection exceeds that of distant reflection by a factor of ${\sim}50\times$.  Further, the scattered continuum component of this distant reflection is not robustly detected, which may imply that the distant reflector is non-uniform in nature.}
\item{We detect five components of the outflowing wind in these data: one dusty zone and four gaseous zones.  We also find tentative evidence for the presence of a very fast component in this wind with $v_{\rm out} \sim -20000 \kmps$, in support of previous studies.}
\end{enumerate}

\begin{center}
    {\bf Acknowledgments}
\end{center}

We thank the anonymous referee for an insightful and constructive review that improved this manuscript.  
We thank Tahir Yaqoob for helpful advice in navigating the {\it XRISM} data reduction during a changing calibration landscape, and Richard Mushotzky for feedback that improved the manuscript.  We are grateful to the directors and scheduling and operations teams of {\it XMM-Newton} and {\it NuSTAR} for coordinated observations that greatly enhance the scientific return of these {\it XRISM} data.  LB thanks Andy Young for stimulating conversations on the {\it Chandra}/HETG data, and gratefully acknowledges funding support from NASA under grant number 80NSSC24K0684.  MM acknowledges support from JSPS KAKENHI Grant Number JP21K13958 and Yamada Science Foundation.  This paper employs a list of Chandra datasets, obtained by the Chandra X-ray Observatory, contained in the Chandra Data Collection (CDC) ADS/Sa.CXO\#obs/04759~\href{https://doi.org/10.25574/cdc.ADS/Sa.CXO\#obs/04759}{doi:10.25574/cdc.ADS/Sa.CXO\#obs/04759}, ADS/Sa.CXO\#obs/04760~\href{https://doi.org/10.25574/cdc.ADS/Sa.CXO\#obs/04760}{doi:10.25574/cdc.ADS/Sa.CXO\#obs/04760},\\ ADS/Sa.CXO\#obs/04761~\href{https://doi.org/10.25574/cdc.ADS/Sa.CXO\#obs/04761}{doi:10.25574/cdc.ADS/Sa.CXO\#obs/04761},\\ ADS/Sa.CXO\#obs/04762~\href{https://doi.org/10.25574/cdc.ADS/Sa.CXO\#obs/04762}{doi:10.25574/cdc.ADS/Sa.CXO\#obs/04762}. 

\bibliography{adsrefs}{}

\begin{thebibliography}{}
\expandafter\ifx\csname natexlab\endcsname\relax\def\natexlab#1{#1}\fi
\providecommand{\url}[1]{\href{#1}{#1}}
\providecommand{\dodoi}[1]{doi:~\href{http://doi.org/#1}{\nolinkurl{#1}}}
\providecommand{\doeprint}[1]{\href{http://ascl.net/#1}{\nolinkurl{http://ascl.net/#1}}}
\providecommand{\doarXiv}[1]{\href{https://arxiv.org/abs/#1}{\nolinkurl{https://arxiv.org/abs/#1}}}

\bibitem[{{Antonucci}(1993)}]{Antonucci1993}
{Antonucci}, R. 1993, \araa, 31, 473, \dodoi{10.1146/annurev.aa.31.090193.002353}

\bibitem[{{Balokovi{\'c}} {et~al.}(2018){Balokovi{\'c}}, {Brightman}, {Harrison}, {Comastri}, {Ricci}, {Buchner}, {Gandhi}, {Farrah}, \& {Stern}}]{Balokovic2018}
{Balokovi{\'c}}, M., {Brightman}, M., {Harrison}, F.~A., {et~al.} 2018, \apj, 854, 42, \dodoi{10.3847/1538-4357/aaa7eb}

\bibitem[{{Bambi} {et~al.}(2021){Bambi}, {Brenneman}, {Dauser}, {Garc{\'\i}a}, {Grinberg}, {Ingram}, {Jiang}, {Liu}, {Lohfink}, {Marinucci}, {Mastroserio}, {Middei}, {Nampalliwar}, {Nied{\'z}wiecki}, {Steiner}, {Tripathi}, \& {Zdziarski}}]{Bambi2021}
{Bambi}, C., {Brenneman}, L.~W., {Dauser}, T., {et~al.} 2021, \ssr, 217, 65, \dodoi{10.1007/s11214-021-00841-8}

\bibitem[{{Brenneman}(2013)}]{Brenneman2013}
{Brenneman}, L. 2013, {Measuring the Angular Momentum of Supermassive Black Holes}, \dodoi{10.1007/978-1-4614-7771-6}

\bibitem[{{Brenneman} \& {Reynolds}(2006)}]{BR2006}
{Brenneman}, L.~W., \& {Reynolds}, C.~S. 2006, \apj, 652, 1028, \dodoi{10.1086/508146}

\bibitem[{{Buchner} {et~al.}(2021){Buchner}, {Brightman}, {Balokovi{\'c}}, {Wada}, {Bauer}, \& {Nandra}}]{Buchner2021}
{Buchner}, J., {Brightman}, M., {Balokovi{\'c}}, M., {et~al.} 2021, \aap, 651, A58, \dodoi{10.1051/0004-6361/201834963}

\bibitem[{{Cash}(1979)}]{Cash1979}
{Cash}, W. 1979, \apj, 228, 939, \dodoi{10.1086/156922}

\bibitem[{{Chen} {et~al.}(2011){Chen}, {Dai}, {Kochanek}, {Chartas}, {Blackburne}, \& {Koz{\l}owski}}]{Chen2011}
{Chen}, B., {Dai}, X., {Kochanek}, C.~S., {et~al.} 2011, \apjl, 740, L34, \dodoi{10.1088/2041-8205/740/2/L34}

\bibitem[{{Chiang} \& {Fabian}(2011)}]{Chiang2011}
{Chiang}, C.-Y., \& {Fabian}, A.~C. 2011, \mnras, 414, 2345, \dodoi{10.1111/j.1365-2966.2011.18553.x}

\bibitem[{{Chiaraluce} {et~al.}(2018){Chiaraluce}, {Vagnetti}, {Tombesi}, \& {Paolillo}}]{Chiaraluce2018}
{Chiaraluce}, E., {Vagnetti}, F., {Tombesi}, F., \& {Paolillo}, M. 2018, \aap, 619, A95, \dodoi{10.1051/0004-6361/201833631}

\bibitem[{{Dabrowski} {et~al.}(1997){Dabrowski}, {Fabian}, {Iwasawa}, {Lasenby}, \& {Reynolds}}]{Dabrowski1997}
{Dabrowski}, Y., {Fabian}, A.~C., {Iwasawa}, K., {Lasenby}, A.~N., \& {Reynolds}, C.~S. 1997, \mnras, 288, L11, \dodoi{10.1093/mnras/288.1.L11}

\bibitem[{{Dauser} {et~al.}(2010){Dauser}, {Wilms}, {Reynolds}, \& {Brenneman}}]{Dauser2010}
{Dauser}, T., {Wilms}, J., {Reynolds}, C.~S., \& {Brenneman}, L.~W. 2010, \mnras, 409, 1534, \dodoi{10.1111/j.1365-2966.2010.17393.x}

\bibitem[{{Elvis}(2000)}]{Elvis2000}
{Elvis}, M. 2000, \apj, 545, 63, \dodoi{10.1086/317778}

\bibitem[{{Fabian} {et~al.}(2015){Fabian}, {Lohfink}, {Kara}, {Parker}, {Vasudevan}, \& {Reynolds}}]{Fabian2015}
{Fabian}, A.~C., {Lohfink}, A., {Kara}, E., {et~al.} 2015, \mnras, 451, 4375, \dodoi{10.1093/mnras/stv1218}

\bibitem[{{Fabian} {et~al.}(2014){Fabian}, {Parker}, {Wilkins}, {Miller}, {Kara}, {Reynolds}, \& {Dauser}}]{Fabian2014}
{Fabian}, A.~C., {Parker}, M.~L., {Wilkins}, D.~R., {et~al.} 2014, \mnras, 439, 2307, \dodoi{10.1093/mnras/stu045}

\bibitem[{{Fabian} {et~al.}(1989){Fabian}, {Rees}, {Stella}, \& {White}}]{Fabian1989}
{Fabian}, A.~C., {Rees}, M.~J., {Stella}, L., \& {White}, N.~E. 1989, \mnras, 238, 729, \dodoi{10.1093/mnras/238.3.729}

\bibitem[{{Fabian} {et~al.}(2002){Fabian}, {Vaughan}, {Nandra}, {Iwasawa}, {Ballantyne}, {Lee}, {De Rosa}, {Turner}, \& {Young}}]{Fabian2002}
{Fabian}, A.~C., {Vaughan}, S., {Nandra}, K., {et~al.} 2002, \mnras, 335, L1, \dodoi{10.1046/j.1365-8711.2002.05740.x}

\bibitem[{{Fabian} {et~al.}(2012){Fabian}, {Wilkins}, {Miller}, {Reis}, {Reynolds}, {Cackett}, {Nowak}, {Pooley}, {Pottschmidt}, {Sanders}, {Ross}, \& {Wilms}}]{Fabian2012}
{Fabian}, A.~C., {Wilkins}, D.~R., {Miller}, J.~M., {et~al.} 2012, \mnras, 424, 217, \dodoi{10.1111/j.1365-2966.2012.21185.x}

\bibitem[{{Ferland} {et~al.}(2013){Ferland}, {Porter}, {van Hoof}, {Williams}, {Abel}, {Lykins}, {Shaw}, {Henney}, \& {Stancil}}]{Ferland2013}
{Ferland}, G.~J., {Porter}, R.~L., {van Hoof}, P.~A.~M., {et~al.} 2013, \rmxaa, 49, 137, \dodoi{10.48550/arXiv.1302.4485}

\bibitem[{{Garc{\'\i}a} {et~al.}(2013){Garc{\'\i}a}, {Dauser}, {Reynolds}, {Kallman}, {McClintock}, {Wilms}, \& {Eikmann}}]{Garcia2013}
{Garc{\'\i}a}, J., {Dauser}, T., {Reynolds}, C.~S., {et~al.} 2013, \apj, 768, 146, \dodoi{10.1088/0004-637X/768/2/146}

\bibitem[{{Garc{\'\i}a} {et~al.}(2018){Garc{\'\i}a}, {Kallman}, {Bautista}, {Mendoza}, {Deprince}, {Palmeri}, \& {Quinet}}]{Garcia2018}
{Garc{\'\i}a}, J.~A., {Kallman}, T.~R., {Bautista}, M., {et~al.} 2018, in Astronomical Society of the Pacific Conference Series, Vol. 515, Workshop on Astrophysical Opacities, 282, \dodoi{10.48550/arXiv.1805.00581}

\bibitem[{{George} \& {Fabian}(1991)}]{GF1991}
{George}, I.~M., \& {Fabian}, A.~C. 1991, \mnras, 249, 352, \dodoi{10.1093/mnras/249.2.352}

\bibitem[{{Guainazzi} {et~al.}(1999){Guainazzi}, {Matt}, {Molendi}, {Orr}, {Fiore}, {Grandi}, {Matteuzzi}, {Mineo}, {Perola}, {Parmar}, \& {Piro}}]{Guainazzi1999}
{Guainazzi}, M., {Matt}, G., {Molendi}, S., {et~al.} 1999, \aap, 341, L27, \dodoi{10.48550/arXiv.astro-ph/9811246}

\bibitem[{{Harrison} {et~al.}(2013){Harrison}, {Craig}, {Christensen}, {Hailey}, {Zhang}, {Boggs}, {Stern}, {Cook}, {Forster}, {Giommi}, {Grefenstette}, {Kim}, {Kitaguchi}, {Koglin}, {Madsen}, {Mao}, {Miyasaka}, {Mori}, {Perri}, {Pivovaroff}, {Puccetti}, {Rana}, {Westergaard}, {Willis}, {Zoglauer}, {An}, {Bachetti}, {Barri{\`e}re}, {Bellm}, {Bhalerao}, {Brejnholt}, {Fuerst}, {Liebe}, {Markwardt}, {Nynka}, {Vogel}, {Walton}, {Wik}, {Alexander}, {Cominsky}, {Hornschemeier}, {Hornstrup}, {Kaspi}, {Madejski}, {Matt}, {Molendi}, {Smith}, {Tomsick}, {Ajello}, {Ballantyne}, {Balokovi{\'c}}, {Barret}, {Bauer}, {Blandford}, {Brandt}, {Brenneman}, {Chiang}, {Chakrabarty}, {Chenevez}, {Comastri}, {Dufour}, {Elvis}, {Fabian}, {Farrah}, {Fryer}, {Gotthelf}, {Grindlay}, {Helfand}, {Krivonos}, {Meier}, {Miller}, {Natalucci}, {Ogle}, {Ofek}, {Ptak}, {Reynolds}, {Rigby}, {Tagliaferri}, {Thorsett}, {Treister}, \& {Urry}}]{Harrison2013}
{Harrison}, F.~A., {Craig}, W.~W., {Christensen}, F.~E., {et~al.} 2013, \apj, 770, 103, \dodoi{10.1088/0004-637X/770/2/103}

\bibitem[{{HI4PI Collaboration} {et~al.}(2016){HI4PI Collaboration}, {Ben Bekhti}, {Fl{\"o}er}, {Keller}, {Kerp}, {Lenz}, {Winkel}, {Bailin}, {Calabretta}, {Dedes}, {Ford}, {Gibson}, {Haud}, {Janowiecki}, {Kalberla}, {Lockman}, {McClure-Griffiths}, {Murphy}, {Nakanishi}, {Pisano}, \& {Staveley-Smith}}]{Bekhti2016}
{HI4PI Collaboration}, {Ben Bekhti}, N., {Fl{\"o}er}, L., {et~al.} 2016, \aap, 594, A116, \dodoi{10.1051/0004-6361/201629178}

\bibitem[{{Igo} {et~al.}(2020){Igo}, {Parker}, {Matzeu}, {Alston}, {Alvarez Crespo}, {F{\"u}rst}, {Buisson}, {Lobban}, {Joyce}, {Mallick}, {Schartel}, \& {Santos-Lle{\'o}}}]{Igo2020}
{Igo}, Z., {Parker}, M.~L., {Matzeu}, G.~A., {et~al.} 2020, \mnras, 493, 1088, \dodoi{10.1093/mnras/staa265}

\bibitem[{{Ishisaki} {et~al.}(2022){Ishisaki}, {Kelley}, {Awaki}, {Balleza}, {Barnstable}, {Bialas}, {Boissay-Malaquin}, {Brown}, {Canavan}, {Cumbee}, {Carnahan}, {Chiao}, {Comber}, {Costantini}, {den Herder}, {Dercksen}, {de Vries}, {DiPirro}, {Eckart}, {Ezoe}, {Ferrigno}, {Fujimoto}, {Gorter}, {Graham}, {Grim}, {Hartz}, {Hayakawa}, {Hayashi}, {Hell}, {Hoshino}, {Ichinohe}, {Ishida}, {Ishikawa}, {James}, {Kenyon}, {Kilbourne}, {Kimball}, {Kitamoto}, {Leutenegger}, {Maeda}, {McCammon}, {Miko}, {Mizumoto}, {Okajima}, {Okamoto}, {Paltani}, {Porter}, {Sato}, {Sato}, {Sawada}, {Shinozaki}, {Shipman}, {Shirron}, {Sneiderman}, {Soong}, {Szymkiewicz}, {Szymkowiak}, {Takei}, {Tamura}, {Tsujimoto}, {Uchida}, {Wasserzug}, {Witthoeft}, {Wolfs}, {Yamada}, \& {Yasuda}}]{Ishisaki2022}
{Ishisaki}, Y., {Kelley}, R.~L., {Awaki}, H., {et~al.} 2022, in Society of Photo-Optical Instrumentation Engineers (SPIE) Conference Series, Vol. 12181, Space Telescopes and Instrumentation 2022: Ultraviolet to Gamma Ray, ed. J.-W.~A. {den Herder}, S.~{Nikzad}, \& K.~{Nakazawa}, 121811S, \dodoi{10.1117/12.2630654}

\bibitem[{{Iwasawa} {et~al.}(1996){Iwasawa}, {Fabian}, {Reynolds}, {Nandra}, {Otani}, {Inoue}, {Hayashida}, {Brandt}, {Dotani}, {Kunieda}, {Matsuoka}, \& {Tanaka}}]{Iwasawa1996}
{Iwasawa}, K., {Fabian}, A.~C., {Reynolds}, C.~S., {et~al.} 1996, \mnras, 282, 1038, \dodoi{10.1093/mnras/282.3.1038}

\bibitem[{{Kaastra} \& {Bleeker}(2016)}]{Kaastra2016}
{Kaastra}, J.~S., \& {Bleeker}, J.~A.~M. 2016, \aap, 587, A151, \dodoi{10.1051/0004-6361/201527395}

\bibitem[{{Kallman} \& {Bautista}(2001)}]{Kallman2001}
{Kallman}, T., \& {Bautista}, M. 2001, \apjs, 133, 221, \dodoi{10.1086/319184}

\bibitem[{{Lee} {et~al.}(1999){Lee}, {Fabian}, {Brandt}, {Reynolds}, \& {Iwasawa}}]{Lee1999}
{Lee}, J.~C., {Fabian}, A.~C., {Brandt}, W.~N., {Reynolds}, C.~S., \& {Iwasawa}, K. 1999, \mnras, 310, 973, \dodoi{10.1046/j.1365-8711.1999.02999.x}

\bibitem[{{Lee} {et~al.}(2000){Lee}, {Fabian}, {Reynolds}, {Brandt}, \& {Iwasawa}}]{Lee2000}
{Lee}, J.~C., {Fabian}, A.~C., {Reynolds}, C.~S., {Brandt}, W.~N., \& {Iwasawa}, K. 2000, \mnras, 318, 857, \dodoi{10.1046/j.1365-8711.2000.03835.x}

\bibitem[{{Lee} {et~al.}(2002){Lee}, {Iwasawa}, {Houck}, {Fabian}, {Marshall}, \& {Canizares}}]{Lee2002}
{Lee}, J.~C., {Iwasawa}, K., {Houck}, J.~C., {et~al.} 2002, \apjl, 570, L47, \dodoi{10.1086/340992}

\bibitem[{{Lee} {et~al.}(2001){Lee}, {Ogle}, {Canizares}, {Marshall}, {Schulz}, {Morales}, {Fabian}, \& {Iwasawa}}]{Lee2001}
{Lee}, J.~C., {Ogle}, P.~M., {Canizares}, C.~R., {et~al.} 2001, \apjl, 554, L13, \dodoi{10.1086/320912}

\bibitem[{{Longair}(2011)}]{Longair2011}
{Longair}, M.~S. 2011, {High Energy Astrophysics}

\bibitem[{{Marinucci} {et~al.}(2014){Marinucci}, {Matt}, {Miniutti}, {Guainazzi}, {Parker}, {Brenneman}, {Fabian}, {Kara}, {Arevalo}, {Ballantyne}, {Boggs}, {Cappi}, {Christensen}, {Craig}, {Elvis}, {Hailey}, {Harrison}, {Reynolds}, {Risaliti}, {Stern}, {Walton}, \& {Zhang}}]{Marinucci2014}
{Marinucci}, A., {Matt}, G., {Miniutti}, G., {et~al.} 2014, \apj, 787, 83, \dodoi{10.1088/0004-637X/787/1/83}

\bibitem[{{Matzeu} {et~al.}(2022){Matzeu}, {Lieu}, {Costa}, {Reeves}, {Braito}, {Dadina}, {Nardini}, {Boorman}, {Parker}, {Sim}, {Barret}, {Kammoun}, {Middei}, {Giustini}, {Brusa}, {Cabrera}, \& {Marchesi}}]{Matzeu2022}
{Matzeu}, G.~A., {Lieu}, M., {Costa}, M.~T., {et~al.} 2022, \mnras, 515, 6172, \dodoi{10.1093/mnras/stac2155}

\bibitem[{{Mehdipour} {et~al.}(2025){Mehdipour}, {Kaastra}, {Eckart}, {Gu}, {Ballhausen}, {Behar}, {Diez}, {Fukumura}, {Guainazzi}, {Hagino}, {Kallman}, {Kara}, {Li}, {Miller}, {Mizumoto}, {Noda}, {Ogawa}, {Panagiotou}, {Tanimoto}, \& {Zhao}}]{Mehdipour2025}
{Mehdipour}, M., {Kaastra}, J.~S., {Eckart}, M.~E., {et~al.} 2025, \aap, 699, A228, \dodoi{10.1051/0004-6361/202555623}

\bibitem[{{Miller} {et~al.}(2008){Miller}, {Turner}, \& {Reeves}}]{Miller2008}
{Miller}, L., {Turner}, T.~J., \& {Reeves}, J.~N. 2008, \aap, 483, 437, \dodoi{10.1051/0004-6361:200809590}

\bibitem[{{Miller} {et~al.}(2009){Miller}, {Turner}, \& {Reeves}}]{Miller2009}
---. 2009, \mnras, 399, L69, \dodoi{10.1111/j.1745-3933.2009.00726.x}

\bibitem[{{Miniutti} {et~al.}(2007){Miniutti}, {Fabian}, {Anabuki}, {Crummy}, {Fukazawa}, {Gallo}, {Haba}, {Hayashida}, {Holt}, {Kunieda}, {Larsson}, {Markowitz}, {Matsumoto}, {Ohno}, {Reeves}, {Takahashi}, {Tanaka}, {Terashima}, {Torii}, {Ueda}, {Ushio}, {Watanabe}, {Yamauchi}, \& {Yaqoob}}]{Miniutti2007}
{Miniutti}, G., {Fabian}, A.~C., {Anabuki}, N., {et~al.} 2007, \pasj, 59, 315, \dodoi{10.1093/pasj/59.sp1.S315}

\bibitem[{{Miyakawa} {et~al.}(2012){Miyakawa}, {Ebisawa}, \& {Inoue}}]{Miyakawa2012}
{Miyakawa}, T., {Ebisawa}, K., \& {Inoue}, H. 2012, \pasj, 64, 140, \dodoi{10.1093/pasj/64.6.140}

\bibitem[{{Moshir} {et~al.}(1990){Moshir}, {Others}, \& {Otherss}}]{Moshir1990}
{Moshir}, M., {Others}, O., \& {Otherss}, O. 1990, IRAS Faint Source Catalogue, 0

\bibitem[{{Murphy} \& {Yaqoob}(2009)}]{Murphy2009}
{Murphy}, K.~D., \& {Yaqoob}, T. 2009, \mnras, 397, 1549, \dodoi{10.1111/j.1365-2966.2009.15025.x}

\bibitem[{{Nandra} {et~al.}(2007){Nandra}, {O'Neill}, {George}, \& {Reeves}}]{Nandra2007}
{Nandra}, K., {O'Neill}, P.~M., {George}, I.~M., \& {Reeves}, J.~N. 2007, \mnras, 382, 194, \dodoi{10.1111/j.1365-2966.2007.12331.x}

\bibitem[{{Noda} {et~al.}(2025){Noda}, {Mori}, {Tomida}, {Nakajima}, {Tanaka}, {Murakami}, {Uchida}, {Suzuki}, {Kobayashi}, {Yoneyama}, {Hagino}, {Nobukawa}, {Uchiyama}, {Nobukawa}, {Matsumoto}, {Tsuru}, {Yamauchi}, {Hatsukade}, {Odaka}, {Kohmura}, {Yamaoka}, {Yoshida}, {Kanemaru}, {Hiraga}, {Dotani}, {Ozaki}, {Tsunemi}, {Sato}, {Takaki}, {Terada}, {Miyazaki}, {Kusunoki}, {Otsuka}, {Yokosu}, {Yonemaru}, {Ichikawa}, {Nakano}, {Takemoto}, {Matsushima}, {Urase}, {Kurashima}, {Fuchi}, {Hayakawa}, {Fukuda}, {Kamei}, {Asahina}, {Inoue}, {Amano}, {Aoki}, {Ito}, {Kamatani}, {Takayama}, {Sako}, {Yoshimoto}, {Shima}, {Higuchi}, {Ninoyu}, {Aoki}, {Tsunomachi}, \& {Hayashida}}]{Noda2025}
{Noda}, H., {Mori}, K., {Tomida}, H., {et~al.} 2025, \pasj, \dodoi{10.1093/pasj/psaf011}

\bibitem[{{Reynolds}(2021)}]{Reynolds2021}
{Reynolds}, C.~S. 2021, \araa, 59, 117, \dodoi{10.1146/annurev-astro-112420-035022}

\bibitem[{{Reynolds} \& {Nowak}(2003)}]{RN2003}
{Reynolds}, C.~S., \& {Nowak}, M.~A. 2003, \physrep, 377, 389, \dodoi{10.1016/S0370-1573(02)00584-7}

\bibitem[{{Reynolds} {et~al.}(1997){Reynolds}, {Ward}, {Fabian}, \& {Celotti}}]{Reynolds1997}
{Reynolds}, C.~S., {Ward}, M.~J., {Fabian}, A.~C., \& {Celotti}, A. 1997, \mnras, 291, 403, \dodoi{10.1093/mnras/291.3.403}

\bibitem[{{Reynolds} {et~al.}(2004){Reynolds}, {Wilms}, {Begelman}, {Staubert}, \& {Kendziorra}}]{Reynolds2004}
{Reynolds}, C.~S., {Wilms}, J., {Begelman}, M.~C., {Staubert}, R., \& {Kendziorra}, E. 2004, \mnras, 349, 1153, \dodoi{10.1111/j.1365-2966.2004.07596.x}

\bibitem[{{Sim} {et~al.}(2008){Sim}, {Long}, {Miller}, \& {Turner}}]{Sim2008}
{Sim}, S.~A., {Long}, K.~S., {Miller}, L., \& {Turner}, T.~J. 2008, \mnras, 388, 611, \dodoi{10.1111/j.1365-2966.2008.13466.x}

\bibitem[{{Sim} {et~al.}(2010{\natexlab{a}}){Sim}, {Miller}, {Long}, {Turner}, \& {Reeves}}]{Sim2010a}
{Sim}, S.~A., {Miller}, L., {Long}, K.~S., {Turner}, T.~J., \& {Reeves}, J.~N. 2010{\natexlab{a}}, \mnras, 404, 1369, \dodoi{10.1111/j.1365-2966.2010.16396.x}

\bibitem[{{Sim} {et~al.}(2010{\natexlab{b}}){Sim}, {Proga}, {Miller}, {Long}, \& {Turner}}]{Sim2010b}
{Sim}, S.~A., {Proga}, D., {Miller}, L., {Long}, K.~S., \& {Turner}, T.~J. 2010{\natexlab{b}}, \mnras, 408, 1396, \dodoi{10.1111/j.1365-2966.2010.17215.x}

\bibitem[{{Str{\"u}der} {et~al.}(2001){Str{\"u}der}, {Briel}, {Dennerl}, {Hartmann}, {Kendziorra}, {Meidinger}, {Pfeffermann}, {Reppin}, {Aschenbach}, {Bornemann}, {Br{\"a}uninger}, {Burkert}, {Elender}, {Freyberg}, {Haberl}, {Hartner}, {Heuschmann}, {Hippmann}, {Kastelic}, {Kemmer}, {Kettenring}, {Kink}, {Krause}, {M{\"u}ller}, {Oppitz}, {Pietsch}, {Popp}, {Predehl}, {Read}, {Stephan}, {St{\"o}tter}, {Tr{\"u}mper}, {Holl}, {Kemmer}, {Soltau}, {St{\"o}tter}, {Weber}, {Weichert}, {von Zanthier}, {Carathanassis}, {Lutz}, {Richter}, {Solc}, {B{\"o}ttcher}, {Kuster}, {Staubert}, {Abbey}, {Holland}, {Turner}, {Balasini}, {Bignami}, {La Palombara}, {Villa}, {Buttler}, {Gianini}, {Lain{\'e}}, {Lumb}, \& {Dhez}}]{Struder2001}
{Str{\"u}der}, L., {Briel}, U., {Dennerl}, K., {et~al.} 2001, \aap, 365, L18, \dodoi{10.1051/0004-6361:20000066}

\bibitem[{{Tanaka} {et~al.}(1995){Tanaka}, {Nandra}, {Fabian}, {Inoue}, {Otani}, {Dotani}, {Hayashida}, {Iwasawa}, {Kii}, {Kunieda}, {Makino}, \& {Matsuoka}}]{Tanaka1995}
{Tanaka}, Y., {Nandra}, K., {Fabian}, A.~C., {et~al.} 1995, \nat, 375, 659, \dodoi{10.1038/375659a0}

\bibitem[{{Tatum} {et~al.}(2013){Tatum}, {Turner}, {Miller}, \& {Reeves}}]{Tatum2013}
{Tatum}, M.~M., {Turner}, T.~J., {Miller}, L., \& {Reeves}, J.~N. 2013, \apj, 762, 80, \dodoi{10.1088/0004-637X/762/2/80}

\bibitem[{{Thorne}(1974)}]{Thorne1974}
{Thorne}, K.~S. 1974, \apj, 191, 507, \dodoi{10.1086/152991}

\bibitem[{{Verner} {et~al.}(1996){Verner}, {Ferland}, {Korista}, \& {Yakovlev}}]{Verner1996}
{Verner}, D.~A., {Ferland}, G.~J., {Korista}, K.~T., \& {Yakovlev}, D.~G. 1996, \apj, 465, 487, \dodoi{10.1086/177435}

\bibitem[{{Verner} \& {Yakovlev}(1995)}]{Verner1995}
{Verner}, D.~A., \& {Yakovlev}, D.~G. 1995, \aaps, 109, 125

\bibitem[{{Walton} {et~al.}(2014){Walton}, {Risaliti}, {Harrison}, {Fabian}, {Miller}, {Arevalo}, {Ballantyne}, {Boggs}, {Brenneman}, {Christensen}, {Craig}, {Elvis}, {Fuerst}, {Gandhi}, {Grefenstette}, {Hailey}, {Kara}, {Luo}, {Madsen}, {Marinucci}, {Matt}, {Parker}, {Reynolds}, {Rivers}, {Ross}, {Stern}, \& {Zhang}}]{Walton2014}
{Walton}, D.~J., {Risaliti}, G., {Harrison}, F.~A., {et~al.} 2014, \apj, 788, 76, \dodoi{10.1088/0004-637X/788/1/76}

\bibitem[{{Ward} {et~al.}(1987){Ward}, {Elvis}, {Fabbiano}, {Carleton}, {Willner}, \& {Lawrence}}]{Ward1987}
{Ward}, M., {Elvis}, M., {Fabbiano}, G., {et~al.} 1987, \apj, 315, 74, \dodoi{10.1086/165115}

\bibitem[{{Wilkins} \& {Gallo}(2015)}]{Wilkins2015}
{Wilkins}, D.~R., \& {Gallo}, L.~C. 2015, \mnras, 448, 703, \dodoi{10.1093/mnras/stu2524}

\bibitem[{{Wilms} {et~al.}(2000){Wilms}, {Allen}, \& {McCray}}]{Wilms2000}
{Wilms}, J., {Allen}, A., \& {McCray}, R. 2000, \apj, 542, 914, \dodoi{10.1086/317016}

\bibitem[{{Wilms} {et~al.}(2006){Wilms}, {Juett}, {Schulz}, \& {Nowak}}]{Wilms2006}
{Wilms}, J., {Juett}, A., {Schulz}, N., \& {Nowak}, M. 2006, in AAS/High Energy Astrophysics Division, Vol.~9, AAS/High Energy Astrophysics Division \#9, 13.60

\bibitem[{{Wilms} {et~al.}(2001){Wilms}, {Reynolds}, {Begelman}, {Reeves}, {Molendi}, {Staubert}, \& {Kendziorra}}]{Wilms2001}
{Wilms}, J., {Reynolds}, C.~S., {Begelman}, M.~C., {et~al.} 2001, \mnras, 328, L27, \dodoi{10.1046/j.1365-8711.2001.05066.x}

\bibitem[{{Xiang} {et~al.}(2025){Xiang}, {Miller}, {Behar}, {Boissay-Malaquin}, {Brenneman}, {Buhariwalla}, {Byun}, {Done}, {Gallo}, {Gerolymatou}, {Hagen}, {Kaastra}, {Paltani}, {Porter}, {Mushotzky}, {Noda}, {Mehdipour}, {Minezaki}, {Tashiro}, \& {Zoghbi}}]{Xiang2025}
{Xiang}, X., {Miller}, J.~M., {Behar}, E., {et~al.} 2025, arXiv e-prints, arXiv:2507.09210, \dodoi{10.48550/arXiv.2507.09210}

\bibitem[{{XRISM Collaboration}(2024)}]{Miller2024}
{XRISM Collaboration}. 2024, arXiv e-prints, arXiv:2408.14300, \dodoi{10.48550/arXiv.2408.14300}

\bibitem[{{Yaqoob}(2024)}]{Yaqoob2024}
{Yaqoob}, T. 2024, \mnras, 527, 1093, \dodoi{10.1093/mnras/stad3257}

\bibitem[{{Young} {et~al.}(2005){Young}, {Lee}, {Fabian}, {Reynolds}, {Gibson}, \& {Canizares}}]{Young2005}
{Young}, A.~J., {Lee}, J.~C., {Fabian}, A.~C., {et~al.} 2005, \apj, 631, 733, \dodoi{10.1086/432607}

\bibitem[{{Zoghbi} {et~al.}(2019){Zoghbi}, {Miller}, \& {Cackett}}]{Zoghbi2019}
{Zoghbi}, A., {Miller}, J.~M., \& {Cackett}, E. 2019, \apj, 884, 26, \dodoi{10.3847/1538-4357/ab3e31}

\end{thebibliography}
\bibliographystyle{aasjournal}


\end{document}